\documentclass[structabstract]{aa} % for double columns
\usepackage{amsmath}
\usepackage{rotating}
 \usepackage{color}
\usepackage{graphicx} 
\usepackage{natbib}
\usepackage{overpic}
\usepackage{mathptmx}
\usepackage{anyfontsize}

\usepackage{appendix}
\usepackage{lscape} \usepackage{longtable, needspace} \usepackage{ulem}
\def\new#1 {{\bf #1} } \def\cut#1 {\sout{#1}}

\usepackage{subfigure}
\usepackage{txfonts}
\usepackage{tabu}
\usepackage{multirow}
\usepackage{url}
\usepackage{xcolor}
\definecolor{darkgreen}{RGB}{0 128 128}
\usepackage{booktabs}
\usepackage[flushleft]{threeparttable}
\usepackage[labelfont=bf,singlelinecheck=false]{caption}
\usepackage{lipsum}

\begin{document}

\title{Inferring the evolutionary stages of the internal structures of NGC\,7538\,S and  IRS1 from chemistry } \author{S. Feng\inst{1,2},
H. Beuther\inst{1}, D. Semenov\inst{1}, Th. Henning\inst{1}, H. Linz\inst{1}, E. A. C. Mills\inst{3}, R. Teague\inst{1}}
\institute{1. Max-Planck-Institut f\"ur Astronomie,
K\"onigstuhl 17, D-69117, Heidelberg \\
2. Max-Planck-Institut f\"ur  Extraterrestrische Physik,
Giessenbachstrasse 1, 85748, Garching, syfeng@mpe.mpg.de\\
3. National Radio Astronomy Observatory, 1003 Lopezville Road, Socorro, NM 87801, USA}
\offprints{syfeng@mpe.mpg.de} 
\date{\today} 
\abstract {Radiative feedback of young (proto)stars and gas dynamics including gravitational collapse and outflows are important in high-mass star-forming regions (HMSFRs), for the reason that they may leave footprints on the gas density and temperature distributions, the velocity profile, and the chemical abundances. }
{We unambiguously diagnose the detailed physical mechanisms and the evolutionary status of HMSFRs. }
{We performed 0.4\arcsec ($\sim1000$ AU) resolution observations at 1.37\,mm towards two HMSFRs, NGC\,7538\,S and  IRS1, using the Plateau de Bure Interferometre (PdBI). The observations covered abundant molecular lines, including tracers of gas column density, hot molecular cores, shocks, and complex organic molecules. We present a joint analysis of the 1.37\,mm continuum emission and the line intensity of 15 molecular species (including 22 isotopologues). Assuming local thermal equilibrium (LTE), we derived molecular column densities and molecular abundances for  each internal gas substructure that is spatially resolved. These derived quantities are compared with a suite of 1-D gas-grain models.}
 {NGC\,7538\,S is resolved into at least three dense gas condensations. Despite the comparable continuum intensity of these condensations, their differing molecular line emission is suggestive of  an overall chemical evolutionary trend from the northeast to the southwest. Line emission from MM1 is consistent with a chemically evolved hot molecular core (HMC), whereas MM3 remains a  prestellar candidate that only exhibits emission of lower-excitation lines. The condensation MM2,  located  between MM1 and MM3, shows {{an}} intermediate chemical evolutionary status. Since these three condensations are embedded within the same parent gas core, their differing chemical properties are  most likely due to the different {{warm-up histories}}, rather than the different dynamic timescales. 
{{Despite remaining spatially unresolved, in IRS1 we detect abundant complex organic molecules}} (e.g. $\rm NH_2CHO$, $\rm CH_3OH$, $\rm HCOOCH_3$, $\rm CH_3OCH_3$), indicating that IRS1 is the most chemically evolved HMC presented here. We observe a continuum that is dominated by absorption features with at least three strong emission lines,  potentially from $\rm CH_3OH$. The $\rm CH_3OH$ lines which are purely in emission have higher excitation than the ones being purely in absorption.
Potential reasons for this difference are discussed.}
{This is the first comprehensive comparison of observations of the two high-mass {cores} NGC\,7538\,S and IRS1 and {a chemical model}. We have found that different chemical evolutionary stages can coexist in the same natal gas core. Our achievement illustrates the strength of chemical analysis for {{understanding}} HMSFRs.
}
\titlerunning{} 
\authorrunning{Feng et al.}
 \maketitle

%______________________________________________________________

\section{Introduction}
The ubiquitous multiplicity of high-mass star-forming regions (HMSFRs)  leads to abundant dense gas cores and associated (proto-)stellar outflows, shocks, and disk-like structures. Understanding how  dense gas structures and cores form and evolve in such {clustered} environments, and indeed how such massive stellar clusters come into the existence, {requires} high angular resolution millimetre and submillimetre interferometric observations.\\

{However,} the interpretation of resolved gas structures and kinematics in high angular resolution images is often ambiguous due to the blending of structures, and the confusion of motions from, for example, infall, outflows and rotation. {Determination of the chemical composition may provide independent constraints, aiding in distinguishing} distinct gas components (e.g. \citealt{liu15}) and potentially inferring the evolutionary status or evolutionary history (e.g. \citealt{vD09}). For example, a number of recent observations of HMSFRs have been made on both large scales ($\rm >10^5$ AU; e.g. \citealt{foster11,sanhueza12,hoq13,jackson13,gerner14}) and small scales ($<2000$ AU; e.g. Orion-KL within $\rm 1\arcmin$, \citealt{feng15}; AFGL 2591, \citealt{jimenez12}; infrared dark clouds (IRDC) G11.11-0.12 \citealt{wangk14}) with the aim of classifying their different evolutionary phases. While observations on the larger spatial scales are useful {to characterize the general properties of the entire clump}, high angular resolution observations ($\lesssim$2000~AU scales) can {probe in more detail} the processes associated with HMSFRs, {such as} hierarchical fragmentation and chemical evolution.\\

The focus of the present paper is a comprehensive study of the internal chemical segregation in two HMSFRs: NGC\,7538\,S and IRS1. {Both sources
have been the subject of a kinematical study (\citealp{beuther12}, see Section \ref{target}). }
In the following discussion {we adopt the nomenclature of \citet{zhang09}, where the authors refer to} gas structures according to their characteristic spatial scales. A gas ``clump" is defined as a structure that has  $\rm >1\,pc$ size (e.g. the entire region of NGC\,7538); a ``core" refers to a structure on the size scale of $\rm \sim0.1$ pc, which may contain or will form a small stellar group (e.g. NGC\,7538\,S and IRS1); and ``condensations" are internal substructures in a gas core, which have $\rm \sim0.01$ pc size and will likely form a single star or multiple-star system.\\

Several evolutionary schemes have been suggested to describe the early phases of HMSFRs (e.g. \citealt{beuther07b,zinnecker07,tan14}). These schemes, however, are mainly based on the study of the physical conditions (density, temperature, spectral energy distribution, kinematics, etc.). Our work {focusses on} the interpretation of the observed chemical composition and {a comparison with chemical modelling to improve upon the HMSFR evolutionary sequence proposed by \citet{gerner14}. We note that the observations of \citet{gerner14} have a $\rm 10^5$ AU spatial resolution, thus each region in their sample corresponds to an unresolved core.} Therefore, chemical properties (e.g. spectral line feature, temperature, and molecular abundance) on that scale are dominated by the most chemically evolved, embedded condensation(s), {i.e. NGC\,7538\,S and IRS1 show chemical properties dominated by hot molecular cores (HMCs).}  In the following discussion, {we consider high-mass protostellar objects (HMPOs) and HMCs as condensations.}\\

Characteristics of each evolutionary stage are briefly summarized as follows:

\begin{itemize}
\item[(1)]  {Prestellar objects are formed in cold and dense molecular cores. Examples of these cores are usually detected in Bok globules or  IRDCs. 
At this particular stage, many molecules are forming or have already formed via rapid ion-molecule gas-phase reactions, followed by slower molecular freeze-out and surface processes. From chemical point of view, it makes sense to designate  the onset time of molecular formation as the starting point $t=0$, which likely precede the formation of the dense cores themselves. This particular time moment is very difficult to estimate for any given observed environment but, given relatively short timescale of high-mass star formation (e.g. \citealt{russeil10,tackenberg12}), it  should not be much longer than the typical IRDC lifetime  of $\sim 10-5\times10^4$ yrs. }\\

\item[(2)] Gravitational collapse leads to the formation of the HMPOs and subsequently compact HMCs. The masses {of these embedded protostars have increased to over} $\rm 8 ~M_\odot$ and exhibit a  temperature of hundreds of Kelvin {even at the scale of 0.01\,pc. The chemical complexity of these regions, characterized by the presence of complex organic molecules (COMs\footnote{Carbon-bearing molecules containing $\ge 6$ atoms, which are theoretically sublimated from grain to the gas phase or produced via ion-molecular reaction in the gas phase as temperature increases \citep{garrod08}.}), is driven predominantly by the hot gas component. Although a clear chemical distinction between the HMPO and HMC phases has not yet been observed, the number and intensity of molecular transitions are higher in HMCs than in HMPOs. Furthermore, outflows and shocks associated with HMPOs and HMCs would be expected to enhance the abundance of certain molecules (e.g. sulfur-bearing species) which would provide for an observational discriminant. }\\

\item[(3)] Formation of  more complex molecules continues as the HMC evolves until the molecules are photo-dissociated by the energetic UV-radiation of young stars. {HMSFRs} at this stage may start to exhibit ultra-compact HII (UCHII) regions. 
\end{itemize}

In this paper, some background on the target sources is provided in Section \ref{target}. A brief summary of our observations is given in Section \ref{obs} and the full-bandwidth spectra
are presented in Section \ref{obs-res}. Analysis of chemical abundances and the results of 1-D radiation transfer modelling are presented in Sections \ref{cal} and \ref{model}, respectively. Finally, the results are discussed in Section \ref{dis}, and the conclusions are presented in Section \ref{conclusion}.
%%%%%%%%%%%%%%%%%%%%%%%%%%%%

\section{Targets}\label{target}
{The molecular cloud NGC\,7538 is located in the Perseus arm of our Galaxy at a distance of 2.65 kpc \citep{moscadelli09}. 
Associated with an optically visible HII region, this region harbors several clusters of infrared \citep{martin73,wynn74} and radio continuum sources \citep{campbell84b}. At least three dense gas cores (IRS1, IRS9, and  S) hint at the presence of (proto-)stellar accretion disks which are driving powerful molecular outflows (e.g. \citealt{de05,kraus06,sandell10,barentine12,beuther12,beuther13}).\\
}

Previous studies proposed a sequential evolution of these dense cores, where the ``older" IRS1 lies in the northwest, and the ``younger" S lies in the southeast \citep{elmegreen77,werner79,mcCaughrean91,ojha04, balog04,qiu11}. Therefore, a comparison of the chemical compositions of  IRS1 and  S (hereafter, NGC\,7538\,S in case of confusion) regions, is important to demonstrate a  chemical evolutionary sequence. In addition, our high angular resolution observations may further unveil the sequential chemical evolution of their internal structures. \\

{\bf IRS1} is the brightest millimetre (mm) continuum  source in the molecular cloud NGC\,7538 \citep{scoville86}. 
Its systemic velocity is $\rm V_{lsr} = -57.4~ km~s^{-1}$ \citep{vandertak00, sandell09}.  
A hyper-compact HII (HCHII) region is  present, which may be ionized by a O6 / O7.5 type ($\rm 8 \times10^4~L_\odot$, 30 $\rm M_\odot$, \citealt{werner79,campbell84,akabane05}) host star. Previous millimetre observations of $\rm HCO^+$ and HCN line emission in the central $\rm 1\arcmin$ area \citep{pratap89} have revealed a  $\rm 10\arcsec\text{--}20\arcsec$ scale shell-like structure ($\rm V_{lsr} \sim-56.5~ kms^{-1}$).
Observations of infrared absorption by dust grains and icy mantles have probed an exterior cool gas envelope on $\sim$72,000 AU scale \citep{willner76,willner82,vandertak00,zheng01,gibb04}. Observations of $\rm NH_3$ absorption ($\rm V_{lsr} \sim-60~ kms^{-1}$) against the HCHII region have detected warmer gas in the inner region \citep{wilson83,henkel84,vandertak00,goddi15a}. Numerous maser sources have been detected within the central 1,000 AU: some outline two small-scale knotty outflows (e.g. OH, $\rm H_2O$, $\rm NH_3$, $\rm CH_3OH$, and $\rm H_2CO$;  \citealt{gaume91,gaume95,hutawarakorn03,hoffman03,kurtz04}), while the rest may trace the rotational motion ($\rm CH_3OH$, \citealp{pestalozzi04,pestalozzi09,beuther12,beuther13}). \\

 {\bf NGC\,7538\,S} is an extended ($\sim$0.2 pc) source located 80\arcsec south of IRS1. It is less luminous than IRS1 ($\rm 1.3\times10^4 ~L_\odot$, \citealt{sandell03}) and is considered to be the ``youngest" star-forming core in NGC\,7538. Weak mm free-free emission \citep{sandell04, pestalozzi06,zapata09}, $\rm H_2O$ masers, OH masers,  $\rm CH_3OH$ masers, and optically thick {thermal} lines have been detected in the high gas column density region \citep{sandell10}. 
Mapping the molecular lines and dust continuum emission has inferred an embedded, $\rm 10-100~M_\odot$ young star or star cluster (\citealt{corder09,wright12,naranjo12, beuther12}), which is associated with the gas accretion flow or disk, and the driving of a compact (0.2\,pc), molecular outflow \citep{beuther12,wright14}. High angular resolution observations of SiO line emission have resolved at least two collimated  bipolar outflows \citep{corder09, naranjo12}.\\
%%%%%%%%%%%%%%%%%%%%%%%%%%%%%
%%%%%%%%%%%%%%%%%%%%%%%%%%%%%
\section{Observations} \label{obs}
{Here were summerize the observational parameters which are directly relevant to the following chemical study. For a full details of the PdBI 1.37\,mm observations and data reduction, we direct the reader to \citet{beuther12}.}\\

The observations were carried out in the A and B array configurations on Jan. 26, 2011 and Feb. 10, 2011, respectively. {The baselines range from 88 m to 760 m with 6 antennas, filtering out structures with an extent $>5,100$\,AU.} The phase referencing centers of our target sources are $\rm \alpha_{2000}= 23^h13^m45.36^s$, $\rm \delta_{2000.0}=
61^o28^{'}10.55^{''}$ (IRS1), and $\rm  \alpha_{2000}= 23^h13^m44.86^s$, $\rm \delta_{2000.0} = 61^o28^{'}48.10^{''}$ (NGC\,7538\,S). Observations of target sources were interleaved with the observations of quasars 2146+608, 0059+581, and 0016+731 for the gain phase and amplitude calibrations. We observed 3C345, 3C273, and MWC349 for passband calibration and absolute flux referencing. {We note that} the typical absolute flux accuracy of PdBI observations is $\sim$20\%.\\

{We configured the WIDEX correlator to cover in two polarizations with a the frequency range 217.167-- 220.836\,GHz} and an uniform spectral resolution of 1.9 MHz  (2.66~km\,s$^{-1}$). This spectral resolution is  coarse for kinematical studies; nevertheless is adequate for our primary purpose of {probing} the chemical evolution. We extracted the continuum data of NGC\,7538\,S from the spectral line-free channels.\\

IRS1 exhibits extremely rich spectral line emission, such that the continuum data derived from the spectral line-free channels have a poor S/N. Thus, the continuum image of IRS1 was produced by averaging over all available spectral channels. {A quantitative comparison of the continuum images of IRS1, with and without line contamination shows little difference, suggesting that spectral line emission in IRS1 does not seriously contaminate the continuum image. Spectral line contamination is discussed in \citet{beuther12}.} \\

The continuum image of NGC\,7538\,S achieves an $1\sigma$ rms noise level of 0.94\,mJy\,beam$^{-1}$ while for IRS1 it is dynamic range limited with a $1\sigma$ rms noise level of 22\,mJy\,beam$^{-1}$. The achieved $1\sigma$ rms noise levels for each 2.66~km\,s$^{-1}$ wide spectral channel, are 2.8\,mJy\,beam$^{-1}$ and 26\,mJy\,beam$^{-1}$ for NGC7538\,S and IRS1, respectively.\\

Data calibration and imaging was carried out using the CLIC\footnote{http://www.iram.fr/IRAMFR/GILDAS/doc/html/clic-html/clic.html} and MAPPING\footnote{http://www.iram.fr/IRAMFR/GILDAS/doc/html/map-html/map.html} software packages. The images were generated adopting a ``robust'' weighting scheme, and the Clark algorithm \citep{clark80}. The  synthesized beams are $\sim0.40''\times 0.36''$ for IRS1 and  $\sim0.57''\times 0.37''$ for NGC\,7538\,S, which correspond to a spatial resolution of $\rm \sim1,100$ AU at the assumed source distance of 2.65 kpc.
%%%%%%%%%%%%%%%%%%%%%%%%%%%%%
%%%%%%%%%%%%%%%%%%%%%%%%%%%%%
\section{Observational results}\label{obs-res}

\subsection{Continuum emission}\label{continuum}
\begin{figure}[htb]
\centering
\begin{tabular}{p{0.1cm}p{8.9cm}}
\multirow{2}{*}{
 \begin{sideways}
\small Dec [J2000]
 \end{sideways}
}
&\includegraphics[width=9.cm]{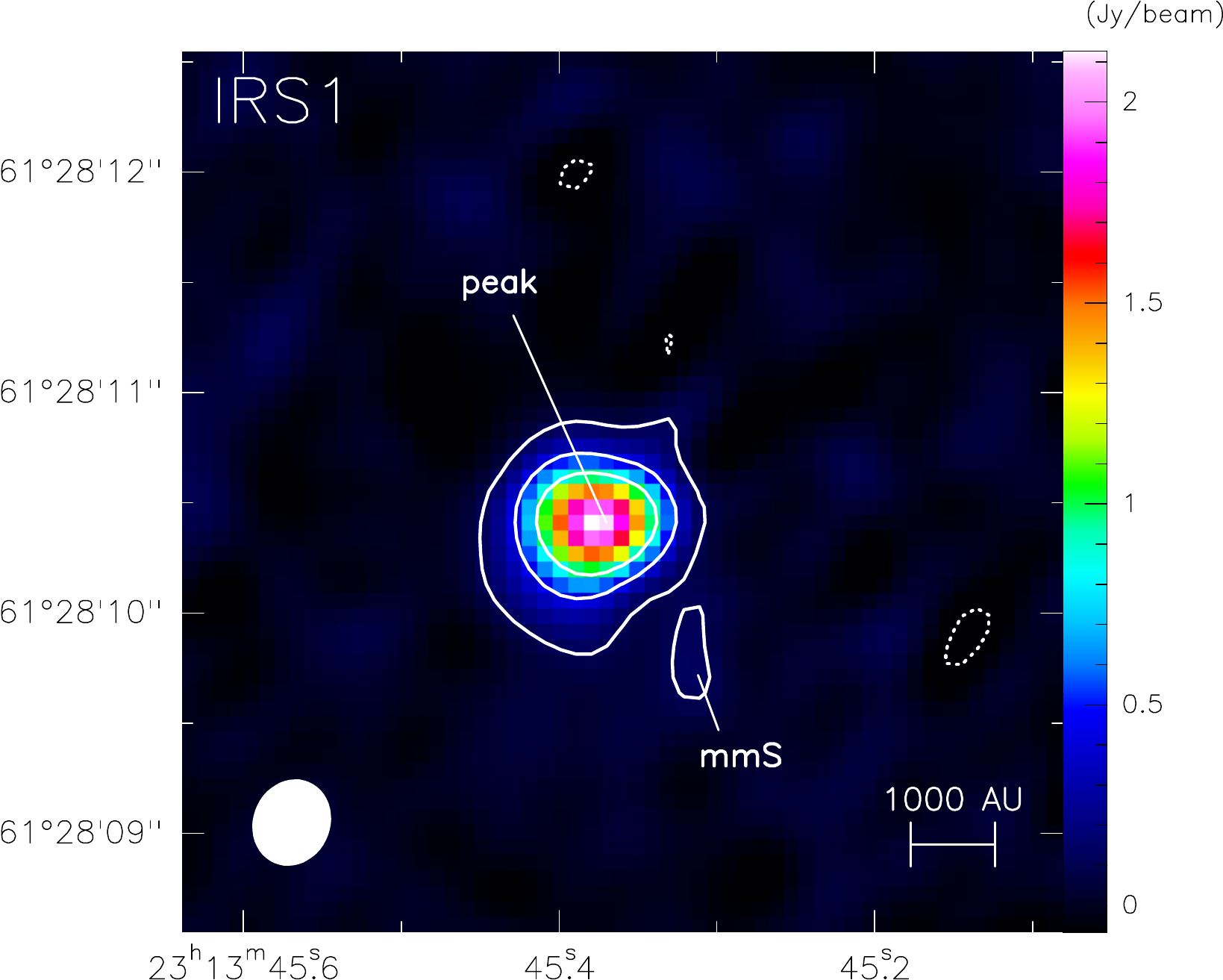}\\
&\includegraphics[width=9.cm]{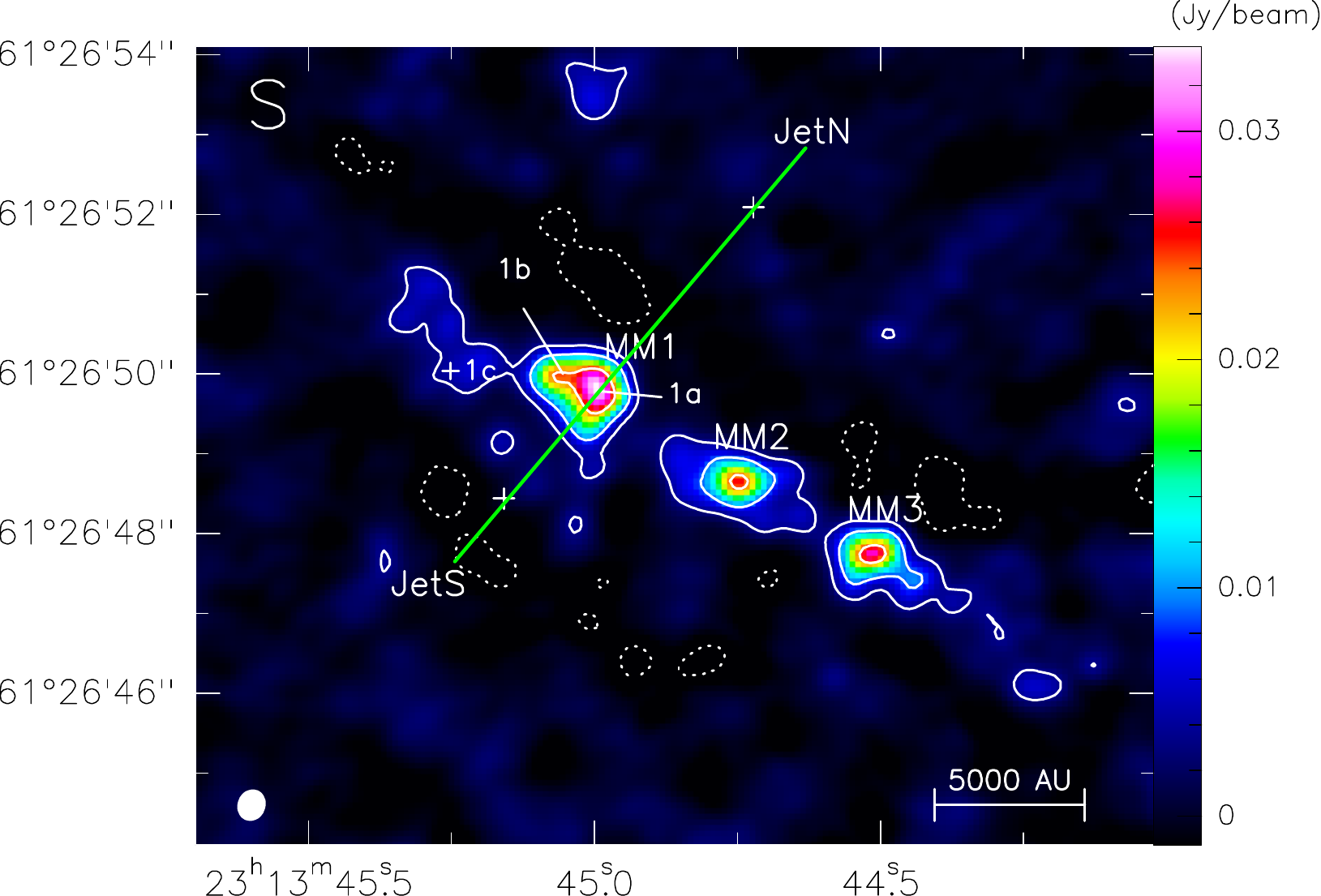}\\
\end{tabular}
\begin{tabular}{c}
\small RA [J2000]
\end{tabular}
\caption{Colormaps and contours of the continuum emission obtained with the PdBI  at 1.37\,mm. Line emission was removed for NGC\,7538\,S and shown to be negligible for IRS1 (see text). For  IRS1 ({\it upper panel}), the white solid contours start at 5$\sigma$  and continue in 20$\sigma$ steps ($\rm \sigma=22~mJy~beam^{-1}$). For NGC\,7538\,S ({\it lower panel}), the contour levels start at 4$\sigma$  and continue in 10$\sigma$ steps ($\rm \sigma=0.94~m~Jy~beam^{-1}$). Dashed contours represent negative emission at the same level as the solid (positive) contours. The synthesized beams are shown at the bottom left of each panel. The direction of the studied outflow is shown with a green line. Labels mark individual substructures we identified.
\label{fig:conti}}
\end{figure}

Figure~\ref{fig:conti} presents the 1.37\,mm continuum images of NGC\,7538\,S and IRS1.
We have resolved several compact substructures, consistent with previous observations (e.g. \citealt{naranjo12,wright12,beuther12,zhu13,wright14}). The nominal absolute positions,  peak specific intensity per beam of the internal substructures, {and the projected size of the internal condensations from 2-D Gaussian fits}, are summarized in Table~\ref{source}.\\

\begin{table*}
\centering
\caption{Properties of continuum substructures in IRS1 and NGC\,7538\,S, identified in Figure~\ref{fig:conti}.  
 \label{source}}
  \scalebox{0.85}{
\begin{tabular}{ll|cccccccc|c}
\hline
\hline
Source & &R.A. &Dec.  &Peak Flux $\rm I_\nu$  &Size ($\theta^{\it d}$)&$\rm T_{rot}$     &$\rm N_{H_{2,1}}^{\it e,g}$  &$\rm N_{H_{2,2}}^{\it f,g}$  &$\rm M^{\it m}$ &Alternative\\
&&[J2000] &[J2000] &($\rm mJy/beam$) &($\rm \arcsec\times\arcsec$, $\rm PA^\circ$) &(K)                  &($\rm 10^{23}cm^{-2}$)   &($\rm 10^{23}cm^{-2}$) &($\rm M_{\odot}$)&Designations\\
&&$\rm 23^h13^m$  &$\rm 61^{\circ}$  & & & & &\\
\hline
\multirow{2}{*}{IRS1} &IRS1-peak     &$\rm 45^s_.4$      &$\rm 28^{'}10^{''}_.4$   &$1945.1\pm182.3$  &$\rm 0.49\arcsec\times0.43\arcsec$, $\rm 13^\circ$   &$214\pm66^h$  &$--^k$   &{\color{black}$44.07\pm31.51$}  &$--^k$ \\
&IRS1-mmS     &$\rm 45^s_.3$      &$\rm 28^{'}09^{''}_.8$   &$126.4\pm3.1$  &$\rm 0.61\arcsec\times0.37\arcsec$, $\rm 87^\circ$   &$162\pm14^i$  &$--^k$  & $--^l$  &$--^k$  &SW extension$^a$\\
            
\hline
\hline
 &    &    &  &   &   &    &    &    & &{\tiny 1a:SMA3$^b$,CARMA $\rm S_{A1}$$^c$}\\
\multirow{5}{*}{S}&MM1      &$\rm 45^s_.0$      &$\rm 26^{'}49^{''}_.8$    &$32.2\pm1.1$  &$\rm 0.87\arcsec\times0.64\arcsec$, $\rm 153^\circ$ &$172\pm23^i$   &$16.90\pm2.52$    & $\rm 16.50\pm1.91$   &$3.28\pm0.52$  &{\tiny 1b:SMA4$^b$,CARMA $\rm S_{A2}$$^c$}\\
 &                                                                                                                               &      &   &   &  &     &   &  & &{\tiny 1c:SMA5$^b$}\\
\cline{2-11}
&MM2    &$\rm 44^s_.7$      &$\rm 26^{'}48^{''}_.7$  &$25.7\pm0.8$   &$\rm 1.08\arcsec\times0.59\arcsec$, $\rm 166^\circ$ &$137\pm14^i $  &$17.05\pm2.11$       &$\rm 2.70\pm0.19$    &$2.54\pm0.30$    &{\tiny 2:SMA2$^b$, $\rm S_B$$^c$}\\
                                                                                                                          
\cline{2-11}
&MM3   &$\rm 44^s_.5$      &$\rm 26^{'}47^{''}_.8$   &$27.0\pm1.1$ &$\rm 0.68\arcsec\times0.49\arcsec$, $\rm 170^\circ$    &$50^j$     &$52.49\pm2.14$  &$\rm 0.29\pm0.01$  &$\rm\le6.58$ &{\tiny 3:SMA1$^b$, $\rm S_C$$^c$}\\
                                                                                                                       
\cline{2-11}
&JetN     &$\rm 44^s_.7$      &$\rm 26^{'}52^{''}_.1$   &$\sigma=0.12$   &$--$ &$150^j$   &$\le0.27 $    &$\rm 2.54\pm 0.01$  &$--$\\
\cline{2-11}
&JetS     &$\rm 45^s_.2$      &$\rm 26^{'}48^{''}_.4$    &$\sigma=0.27$  &$--$  &$150^j$    &$\le0.61$   &$\rm 3.33\pm 0.03$   &$--$\\
\hline
\hline
\end{tabular}
}
\vspace{0.5em}
\begin{tabular}{lp{16cm}}
\footnotesize{\bf Note.}
&\footnotesize{{\it a.}  \citet{zhu13}. }\\
&\footnotesize{{\it b.}   \citet{naranjo12}.} \\
&\footnotesize{{\it c.}  \citet{wright12}.}\\
&\footnotesize{{\it d.}   Projection size is fitted with 2-D Gaussian}\\
&\footnotesize{{\it e.} $\rm H_2$ column densities are calculated from the continuum using Eq.~\ref{gas}. The uncertainties come from the  measurement of  the continuum fluxes and the calculation of  temperatures. }\\
&\footnotesize{{\it f.} $\rm H_2$  column densities are calculated from $\rm C^{18}O$ conversion. The uncertainties come from the  gaussian fit to $\rm C^{18}O~(2\rightarrow1)$ and the excitation  temperatures.}\\ 
&\footnotesize{{\it g.} The column density estimates in MM1--MM3 from dust continuum are thought more reliable than that derived from $\rm C^{18}O$ conversion  ({see Appendix~\ref{appendix:nh2})} .}\\
&\footnotesize{{\it h.}  Temperature lower limit is taken by assuming the same as in IRS1-mmS, and upper limit is taken from \citet{goddi15a} (see details in Section~\ref{tem}). }\\
&\footnotesize{{\it i.} Temperature derived from rotation diagram of $\rm CH_3CN$ in Figure~\ref{rotation}.}\\ 
&\footnotesize{{\it j.} Temperature assumed based on estimation from \citet{sandell10}. }\\
&\footnotesize{{\it k.} Continuum is dominated by free-free emission {at 219 GHz}. }\\
&\footnotesize{{\it l.} Column density could not be obtained because  $\rm C^{18}O~(2\rightarrow1)$ shows {P-Cygni profile.}}\\
&\footnotesize{{\it m.} {Mass on scale of 0.01\,pc is derived from Eq. \ref{gasmass}, {based on total flux reported in \citet{beuther12}, which is corrected for free-free contribution and with $\rm >90\%$ missing flux.}
}}\\
\end{tabular}
\end{table*}

{NGC\,7538\,S} is resolved into three compact condensations ({hereafter} MM1, MM2, and MM3; see Figure \ref{fig:conti}) along the NE-SW direction, {which have nearly identical continuum peak specific intensities. }The most extended condensation, MM1, may be {resolved into} multiple internal substructures, tentatively identified as 1a, 1b, and 1c. MM1-1a and 1b have been detected by  previous 1.2\arcsec resolution 1.3\,mm observations using the Submillimetre Array (SMA,  \citealp{sandell10,naranjo12}), and the previous 1.4 mm/3 mm observations using the CARMA/BIMA array (\citealt{wright12}; see Table~\ref{source} for corresponding observations). MM1-1c has been reported by the previous SMA observations in \citet{naranjo12}, but was undetected by the CARMA observations in \citet{wright12}, {most likely because of insufficient sensitivity.}
We note that MM1-1a and 1b, with an 0.5\arcsec projected separation, are associated with a heavily obscured mid-IR counterpart (Spitzer IRAC and IRS data, \citealt{wright12}), suggesting they may be more ``evolved" than the other compact-mm condensations.\\

In addition, MM1 was suggested to be the driving source of a 0.1--0.4\,pc scale, ionized bipolar outflow \citep{sandell10}\footnote{The outflow has been detected by previous VLA observations \citep{sandell10} coninciding with an OH maser \citep{argon00}, a Class II $\rm CH_3OH$ maser \citep{pestalozzi06}, and a cluster of $\rm H_2O$ maser spots \citep{kameya90}.},
which is perpendicular to the major axis of the proposed edge-on disk \citep{sandell03}. However, this ionized outflow shows no clear continuum detection in our observations, likely because the emission is primarily over extended angular scales and therefore is filtered out due to missing short-spacing data. We mark this ionized outflow as JetN and JetS in Figure~\ref{fig:conti}.\\

At 1.37\,mm, the continuum peak flux specific intensity of {IRS1 }is $\sim$60 times higher than that of NGC\,7538\,S. Our observations, with a spatial resolution of $\sim$1,100 AU, do not find internal substructures towards IRS1, however we do detect a companion in the southwest (IRS1-mmS, hereafter) {with a S/N $\rm >5$}.\\

IRS1-mmS has also been detected by previous 0.7\arcsec resolution mm continuum observations using the SMA and CARMA \citep{zhu13}. {However, the source was detected neither at 1.4\,cm by e.g. \citet{moscadelli09}, nor in high angular resolution observations (0.2\arcsec, $\sim500$ AU) at 843 $\rm \mu m$ by \citet{beuther13}, which have resolved at least three substructures embedded within IRS1.}\\

%%%%%%%%%%%%%%%%%%%%%%%%%%%%%

\subsection{Spectral line}\label{spectra}

\subsubsection{Line identification}\label{id}
\begin{figure*}[h] 
\centering
\includegraphics[width=17.5cm]{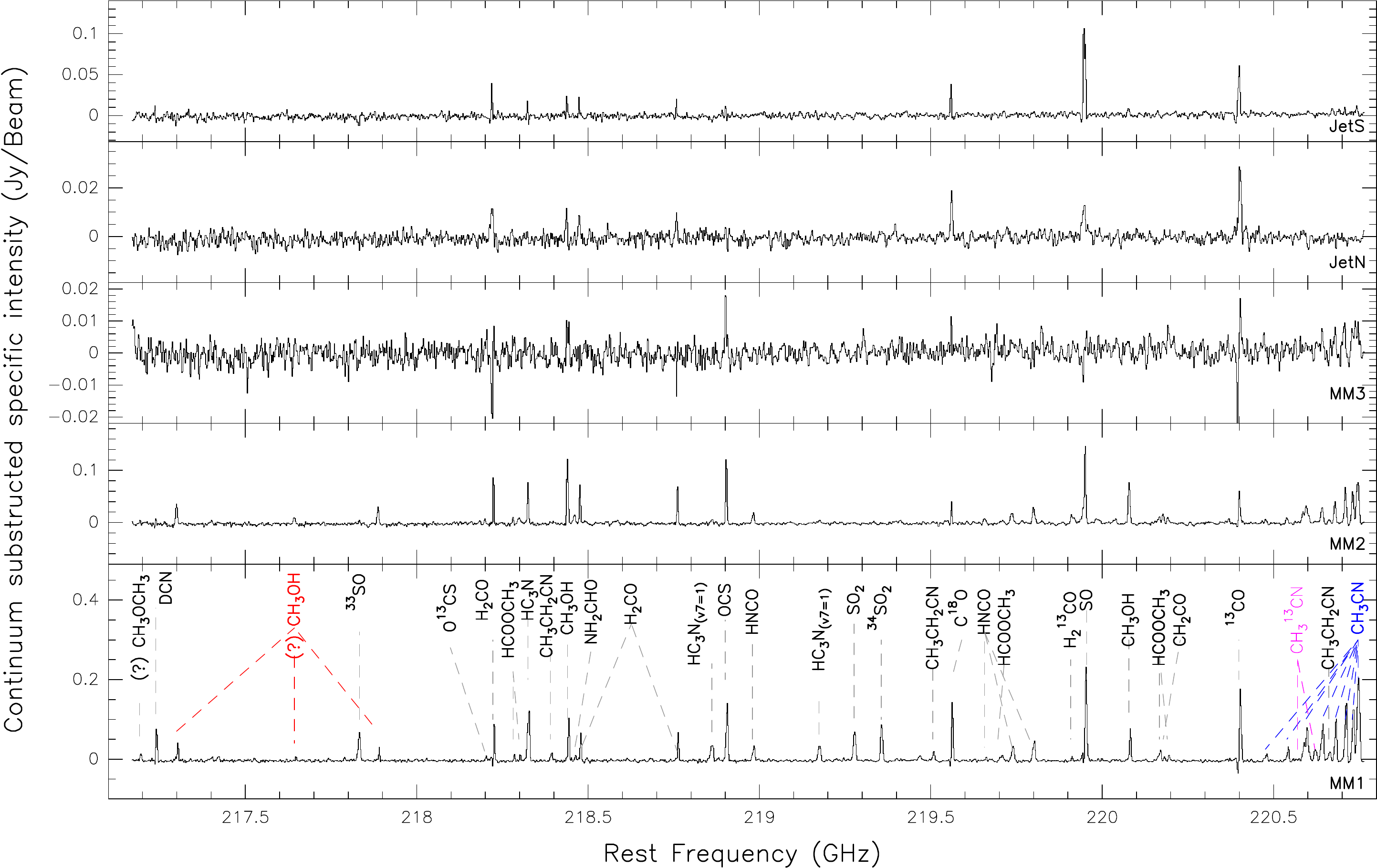}
\includegraphics[width=17.5cm]{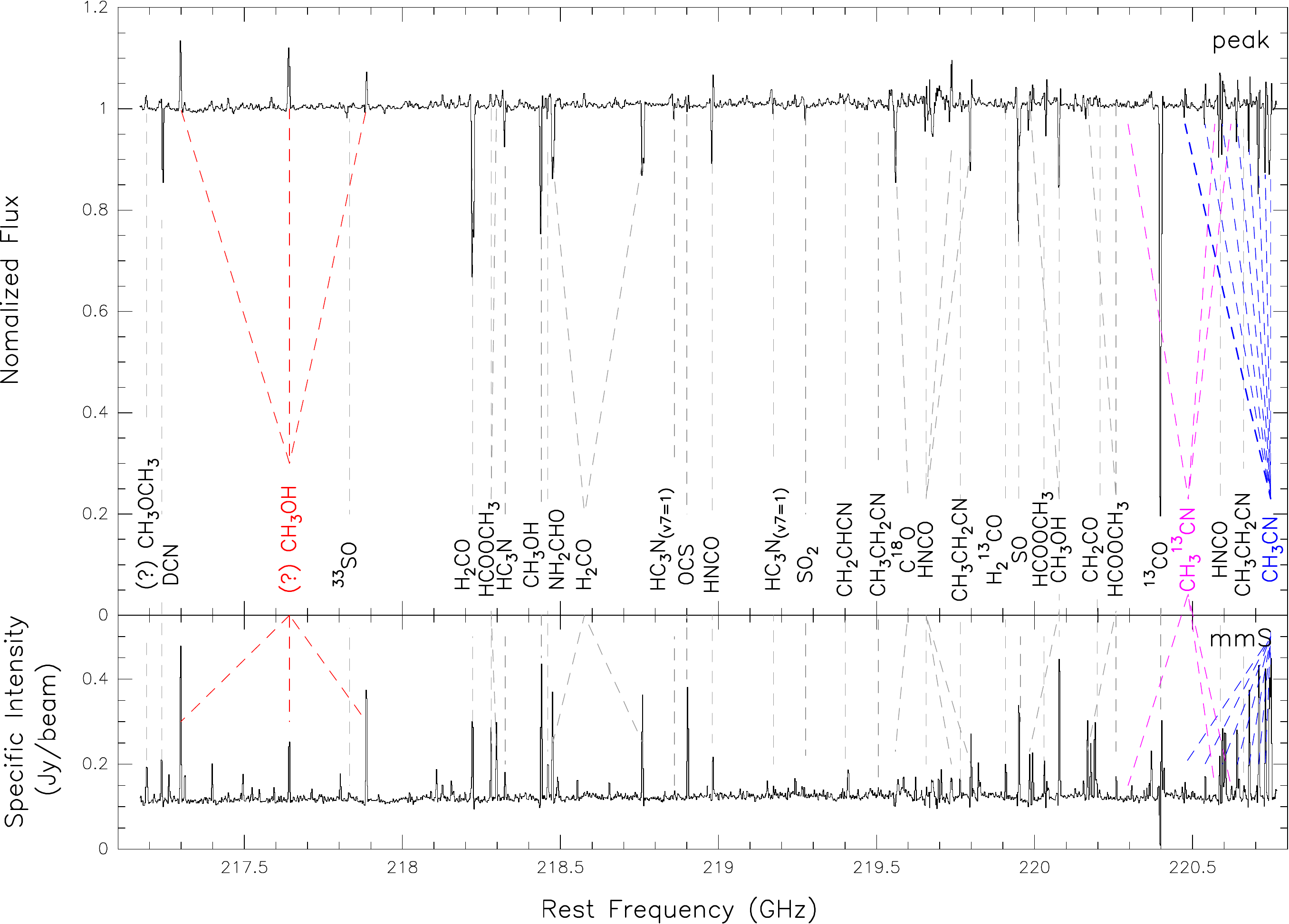}
\caption{Spectra extracted from  positions {denoted}  in Figure~\ref{fig:conti} after imaging the whole data cube. 
All  strong lines are labeled; { the identification of lines marked with``?" is tentative (see Section \ref{dif})}. Continuum-subtracted spectra from different substructures in NGC\,7538\,S are shown in the {\it upper panel}, while the {\it lower panel} presents continuum-unsubstracted spectra from two positions in IRS1 (the spectrum of IRS1-mmS shows the continuum level). All of the detected $\rm HCOOCH_3$ transitions are shown in Figure~\ref{hcooch3}.
}\label{spec}
\end{figure*}

We produced image cubes covering the full available bandwidth ($\sim$4 GHz). The beam-averaged spectra at the peaks of the gas substructures we identified in the 1.37\,mm continuum images (Table~\ref{source}), are given in Figure~\ref{spec}.\\

We identify molecular species based on the spectra of MM1 and the IRS1-peak, which exhibit the strongest line emission. We caution that our line identifications are limited by the blending of multiple velocity components, the broad linewidths and the limited ($\rm \sim2.7~km~s^{-1}$) velocity resolution of our observations. We have cross-compared our results with higher velocity resolution ($\rm \sim1.2~km~s^{-1}$)  spectra of Orion-KL \citep{feng15} and have verified that the identified molecular species are indeed expected. We have used a synthetic spectral fitting program \citep{sanchez11, palau11} {incorporating} molecular data from the JPL\footnote{http://spec.jpl.nasa.gov  \citep{pickett98}} and CDMS\footnote{http://www.astro.uni-koeln.de/cdms/catalog  \citep{muller05}} catalogues, to simultaneously fit multiple lines. The fits are based on the assumptions that all transitions are optically thin, are in local thermodynamic equilibrium (LTE), and have the same linewidth\footnote{The synthetic fitting program we use here is for line detection, but not for finding the best fit of column density or temperature. The optically thin assumption may break down for certain molecular transitions, which leads to some {line intensity deviations} between the synthetic spectra and the observational results.}. This approach is particularly {suited} for self-consistently identifying multiple transitions of a certain molecular species (e.g. Fig.~\ref{hcooch3}).\\

We have identified over 90 molecular lines, which are summarized in Table ~\ref{tab:line}. We mark tentative identifications of some weak lines ($\sim3\sigma$ rms) with symbol ``*", and mark the brightest transition of each species with ``$\dag$". We find large spectral variations within both NGC\,7538\,S and IRS1. \\

{NGC\,7538\,S} shows numerous molecular lines. The three internal condensations, separated by less than 2\arcsec, {display nearly identical continuum peak specific intensities} however show  differences in their line spectra. {MM1 exhibits the strongest intensity for majority of line emission.} We only detect weak OCS ($\rm O^{13}CS$), $\rm ^{13}CO$ ($\rm C^{18}O$), and $\rm CH_3OH$  line emission (where we consider a ``detection" if the line emission is  $\geq4\sigma$ rms) from MM3.\\

In addition, we have detected outflow/shock tracers, such as $\rm H_2CO$ ($\rm H_2^{13}CO$),  $\rm CH_3OH$, SO, OCS, and $\rm C^{18}O$ ($\rm ^{13}CO$),  around the MM1-outflow(s) (i.e. JetN, JetS), despite the low local gas column density. We note that these lines are brighter around the outflows than  MM3.\\

Unlike NGC\,7538\,S, the spectrum of {IRS1-peak} exhibits numerous absorption lines. {Some of these lines (e.g. $\rm HC_3N$, HNCO, $\rm SO_2$, and $\rm C^{18}O$) display P-Cygni profiles, suggesting they may be associated with jets or outflows. Additionally, other lines (e.g. $\rm CH_3OH$, OCS, and $\rm NH_2CHO$) show inverse P-Cygni profiles, indicative of gas infall \citep{beuther12}.} We also detect four strong emission lines marked as ``$\rm ?CH_3OCH_3$" and ``$\rm ?CH_3OH$". We discuss the tentative identification of these lines and their unique properties in Section \ref{dif}.\\

We detect more lines in IRS1-mmS than in MM1--MM3. In particular, multiple lines of $\rm HCOOCH_3$ are detected in IRS1-mmS but are absent in {MM1--MM3 or JetN(S)} (Figure~\ref{hcooch3}). {Moreover, in IRS1-mmS $\rm C^{18}O$ and $\rm ^{13}CO$ are observed in absorption, but the other lines are observed in emission (Figure~\ref{spec}).}

%\onecolumn

\twocolumn

\subsubsection{Line Imaging}\label{image}
\begin{figure*}[htb]
  \small
\begin{center}
\begin{tabular}{lll}
&\multicolumn{2}{c}{\includegraphics[width=13cm]{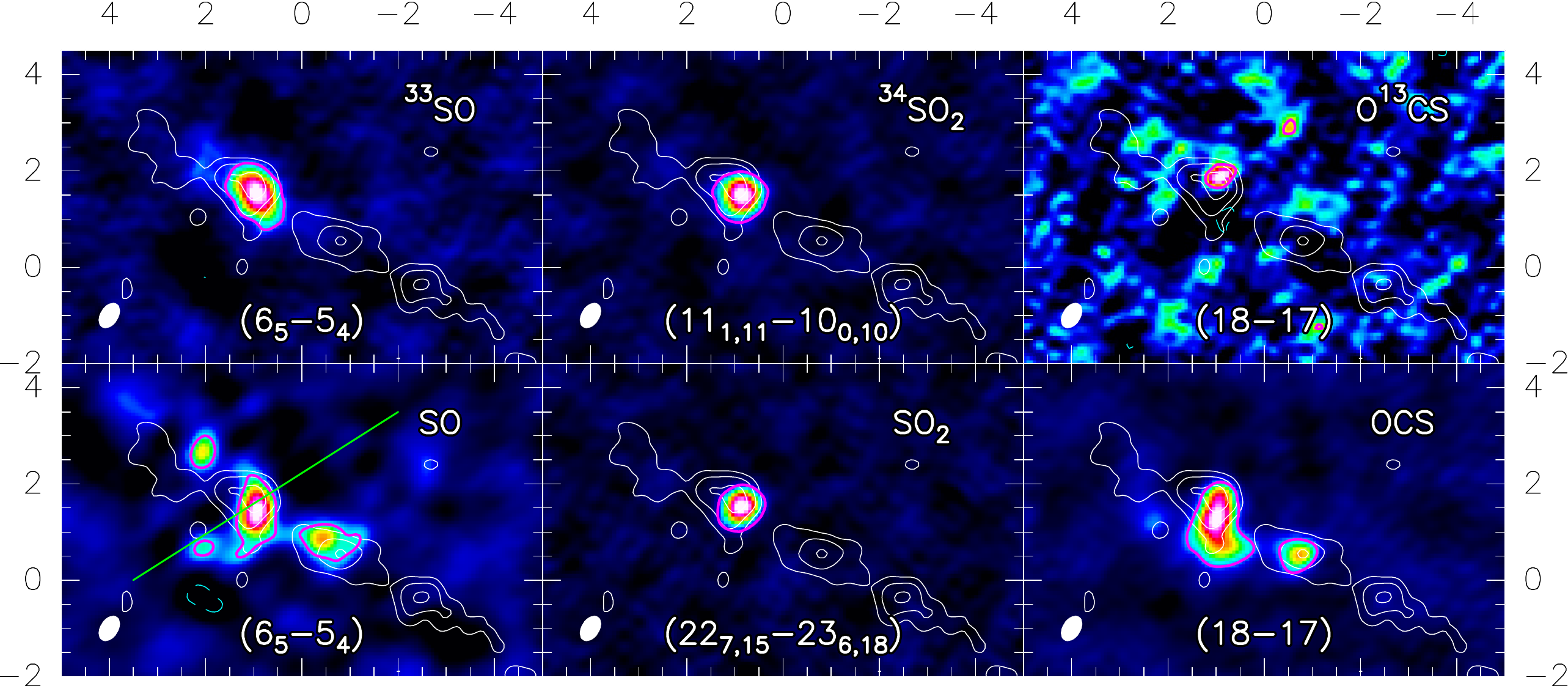}}\\
\multirow{2}{*}{
 \begin{sideways}
\Large$\triangle \delta (\arcsec)$ 
 \end{sideways}
}
&\includegraphics[width=8.5cm]{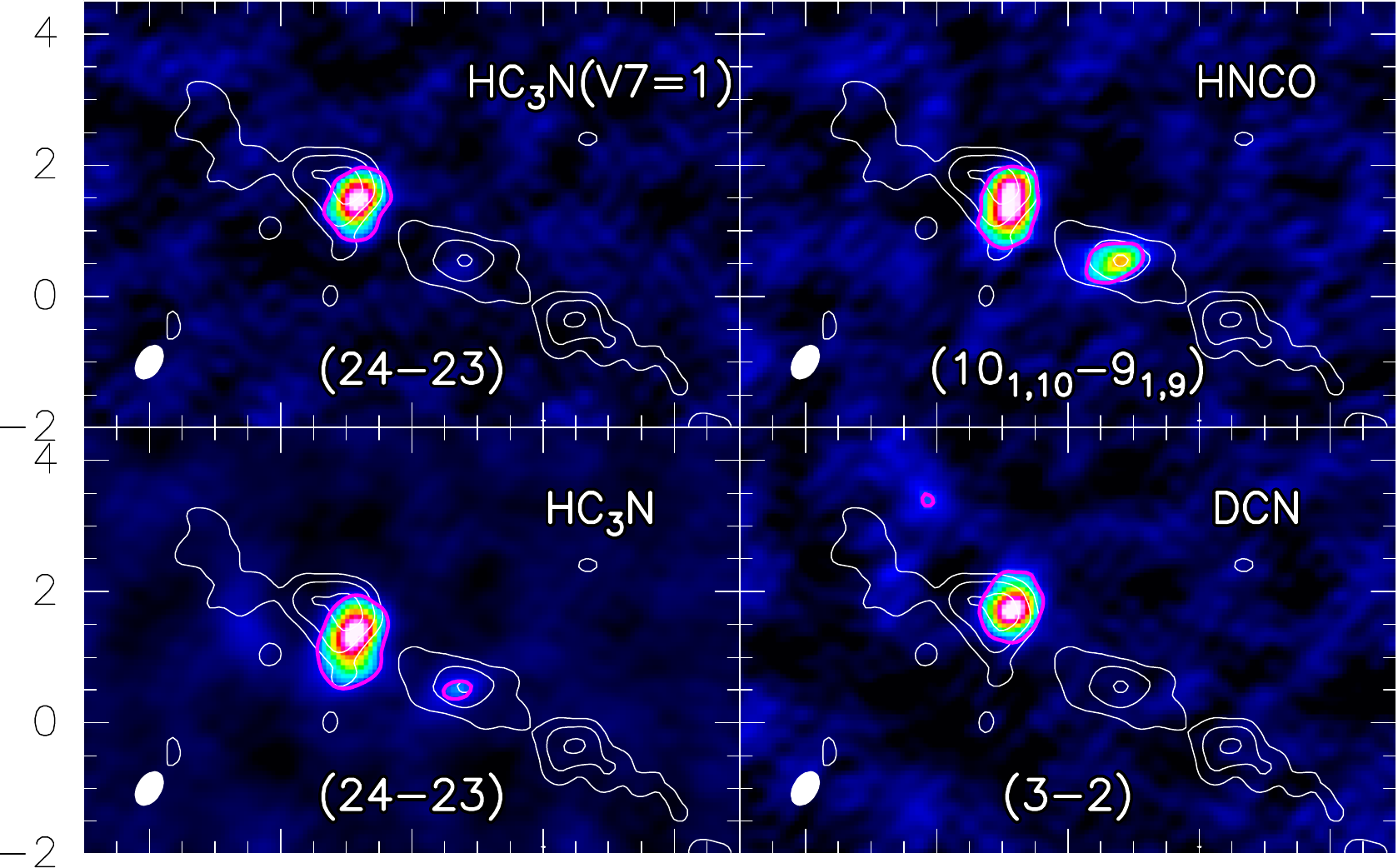}
&\includegraphics[width=8.5cm]{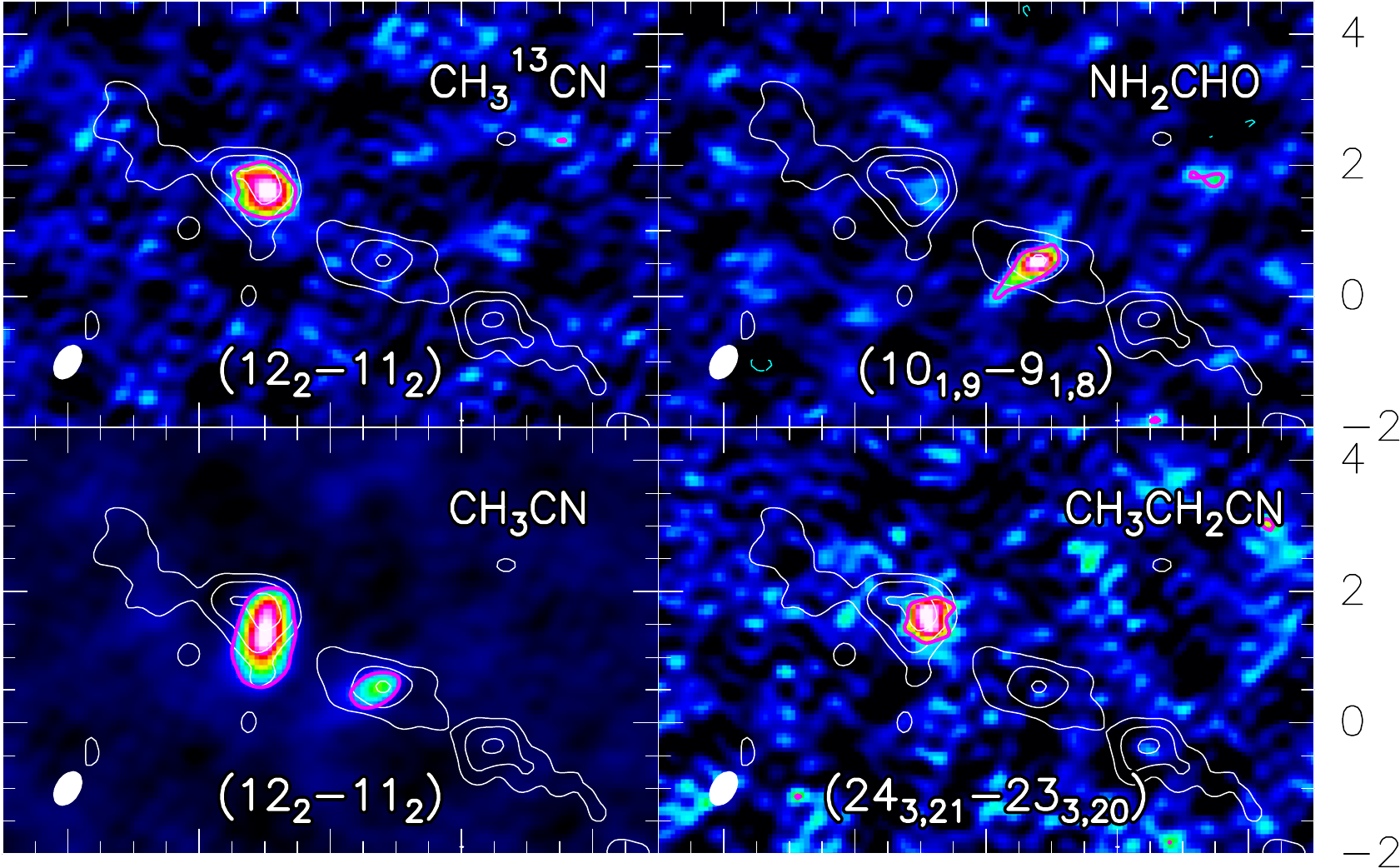}\\
&\includegraphics[width=8.5cm]{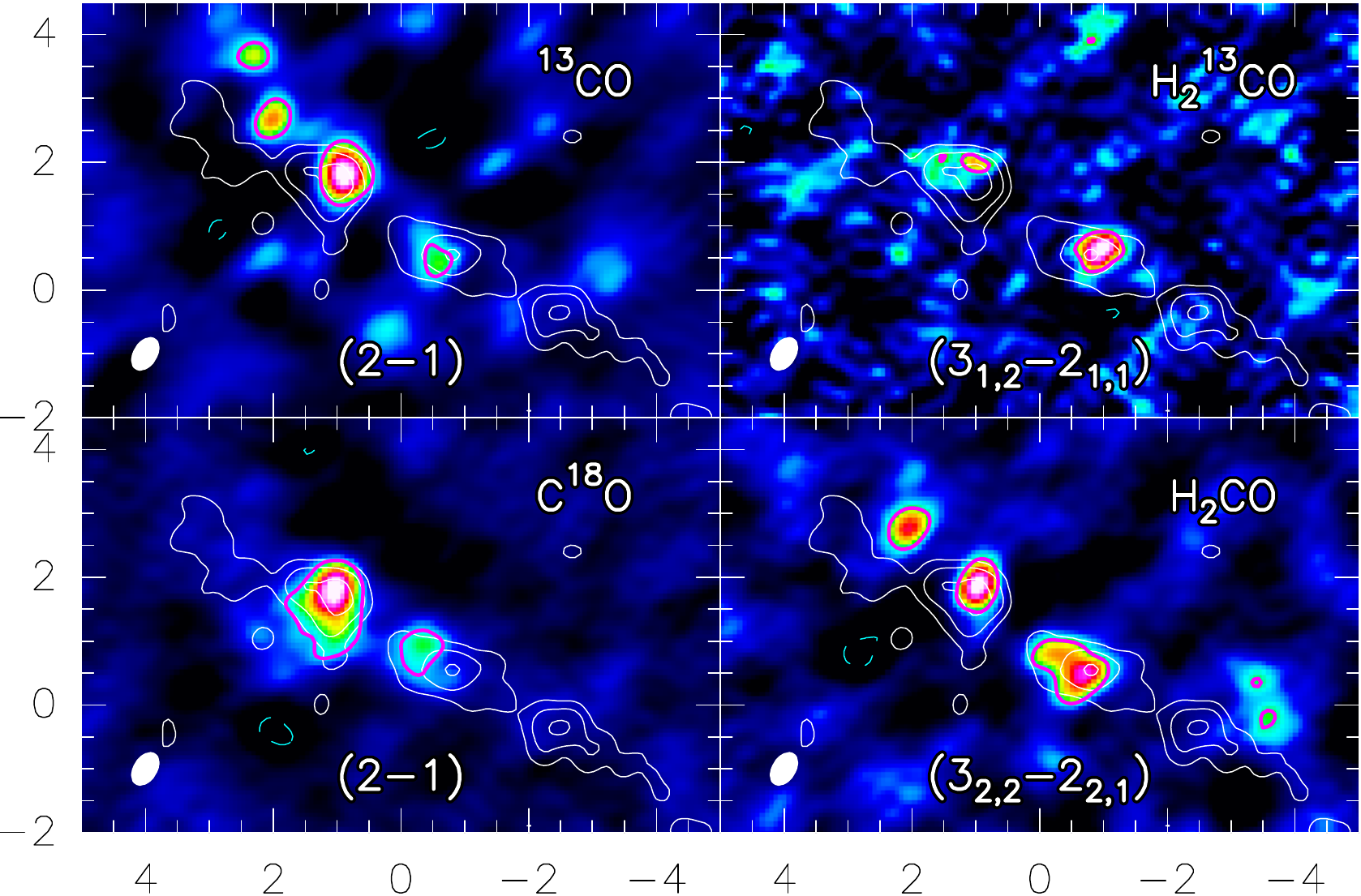}
&\includegraphics[width=8.5cm]{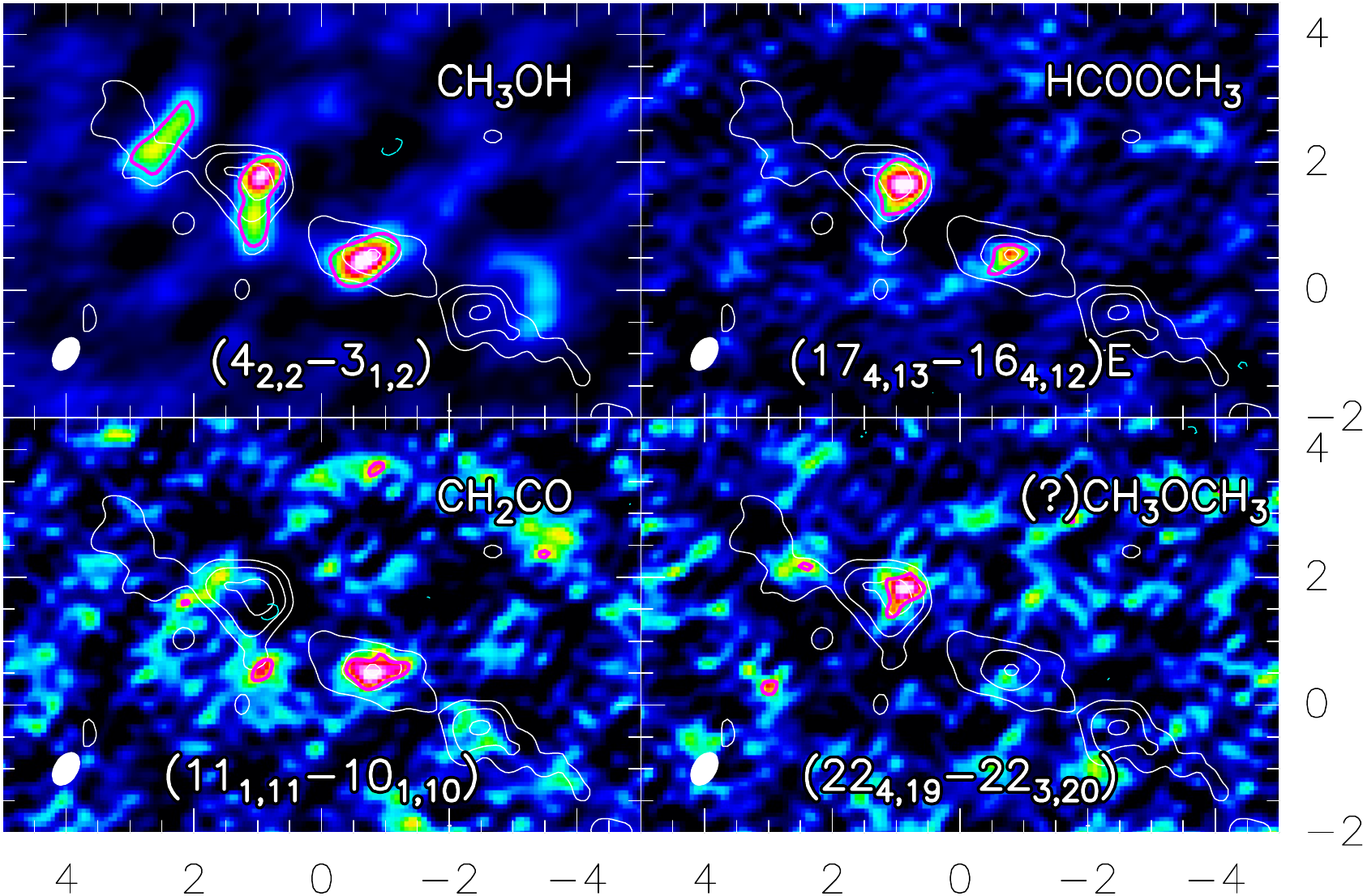}\\
\multicolumn{3}{c}{\Large$\triangle \alpha (\arcsec)$ }
\end{tabular}
\caption{{Maps of the strongest line from each isotopologue} detected in PdBI observations at 1.37\,mm. The intensities have been derived by integrating the line emission over the velocity range shown in Figure~\ref{velpro}. The synthetic beam is shown at the bottom left of each panel. White contours show the continuum  (at 4$\sigma$,  14$\sigma$, and  24$\sigma$ levels with $\rm \sigma=0.94~m~Jy~beam^{-1}$), purple contour reveals the region where molecular specific intensity $\rm >3\sigma$, and the dashed green contour reveals the negative specific intensity beyond $-3\sigma$ because of the interferometric side lobe effect. Green lines in the SO  map sketch  the outflow directions.  All images have different color scales  (in $\rm mJy~ beam^{-1} ~km~ s^{-1} $), increasing from black to white, which are optimized to emphasize the features in the distribution of each line.
}\label{into}
\end{center}
\end{figure*}

Identified molecular lines listed in Table~\ref{tab:line} can be attributed to 15 different species (and 22 isotopologues thereof). In the following discussion, molecular lines observed from IRS1-mmS is beyond the scope of this paper, given its uncertain nature in 1.37\,mm continuum emission. 
For NGC\,7538\,S, we present the strongest, unblended transitions of each isotopologue (marked as ``$\dag$" in Table~\ref{tab:line}) as intensity maps in Figure~\ref{into}. The intensities are integrated over a velocity range\footnote{This range is slightly larger than the FWHM of each line in order to account for the small variation in the peak velocity between different substructures.} as marked in Figure~\ref{velpro},  to enhance the contrast of the images.\\

Different molecular species  have distinct spatial distributions (as in Figure ~\ref{into}), although most  display emission peaks around MM1 and (or) MM2. We note that our observations are not sensitive to emission on angular scales larger than $\sim2 \arcsec$ due to filtering and therefore cannot recover all of the line emission from the extended structures, e.g. outflows.\\

Most of the isotopologues (19 out of 22) peak at MM1 {with line center velocities around  $\rm V_{lsr}\sim-60~km~s^{-1}$. Of these,} SO, $\rm H_2CO$, $\rm CH_3OH$, and $\rm ^{13}CO$ aditionally exhibit strong emission even in MM1-1c. Of all these species peaking towards MM1, several COMs (e.g. $\rm CH_3CN$, $\rm CH_3^{13}CN$, and $\rm CH_3CH_2CN$) are indicative of high temperatures. Furthermore, the presence of vibrationally-excited  $\rm HC_3N~( v_7=1)$ lines indicates mid-IR pumping, and hence suggests the existence of a hot embedded IR-emitting source. {Combined,} these molecular distribution peaks suggest that MM1 is a HMC.\\

{MM2 shows  $\rm >4\sigma$  emission  from $\rm C^{18}O$ and $\rm ^{13}CO$ lines, as well as  from only the main isotopologue lines of the other species with a $\rm V_{lsr}=\sim-58~km~s^{-1}$. These lines have typically lower specific intensities observed in MM2 than  in MM1. Most of these species evaporate from the grain surface early in the warm-up phase, such as $\rm CH_3OH$, $\rm HCOOCH_3$, and HNCO, and our observations did not cover their rare isotopologue lines. Minority of these species are OCS, SO, and $\rm CH_3CN$; their rare isotopologue lines are observed but exhibit $\rm <4\sigma$ emission.}\\

Unlike the majority of species described above, lines of $\rm CH_2CO$  and $\rm NH_2CHO$ are stronger in MM2 than in MM1. {In addition, the only $\rm H_2^{13}CO$ line ($\rm 3_{1,2}\rightarrow2_{1,1}$) exhibits stronger emission in MM2 than in MM1, which is on contrary to the MM1-dominated emission of the $\rm H_2CO$ lines ($\rm 3_{2,2}\rightarrow2_{2,1}$, $\rm 3_{2,1}\rightarrow2_{2,0}$, and $\rm 3_{0,3}-2_{0,2}$).}

%%%%%%%%%%%%%%%%%%%%%%%%%%%%%

 \section{Source temperatures and chemical {variations}}\label{cal}

We derive the spatial variations of gas temperature, molecular column density, and molecular abundance in order to diagnose the chemical evolution at core scale from IRS1 to NGC\,7538\,S, and at condensation scale with the condensations MM1--MM3.\\

\subsection{Temperature}\label{tem}

A comprehensive chemical model should not only reproduce the observed chemical composition, but should also be based on realistic assumptions of gas temperature and density.\\

Our sources show high gas number densities (see discussion in Section~\ref{continu}), so collisions are sufficient to thermalize rotational levels, thereby making the rotational excitation temperature ($\rm T_{rot}$) a good approximation to the kinetic temperature. {\it Rotational diagrams} provides the simplest method to deriving gas temperatures, when multiple ($\rm >3$) transitions with different upper state energy levels $\rm E_u/k$ of an isotopologue have been observed.
This method cannot be simply applied to IRS1-peak due to the confusion with the absorption lines, however we have derived the rotational diagrams of multiple species for MM1, MM2, and IRS1-mmS from their emission lines, as discussed below:
\color{black}

\begin{itemize}
\item  $\rm \bf CH_3CN$ line forests are regarded as good thermometres for dense molecular gas \citep{boucher80, wright95}.  However, if the emission lines are optically thick then the estimated $\rm T_{\rm rot}$ can significantly exceed $\rm T_{\rm kin}$. This can be circumvented using optically thin lines of less abundant isotopologues, e.g. $\rm CH_3^{13}CN$. In our observations most $\rm CH_3^{13}CN$ transitions are blended with the adjacent molecular transitions, except for $\rm CH_3^{13}CN~(12_2\rightarrow12_2)$. Therefore, we estimate an opacity corrected $\rm T_{\rm rot}$ using the optically thick $\rm CH_3CN$ lines for MM1, MM2, and IRS1-mmS (panels 3--5 of Fig.~\ref{rotation}). See Appendix~\ref{ap-temp} (also in \citealt{feng15}) for more details on this iterative approach. \\

  \begin{figure*}[htb]
  \small
\begin{center}
\begin{tabular}{ll}
\includegraphics[width=8cm]{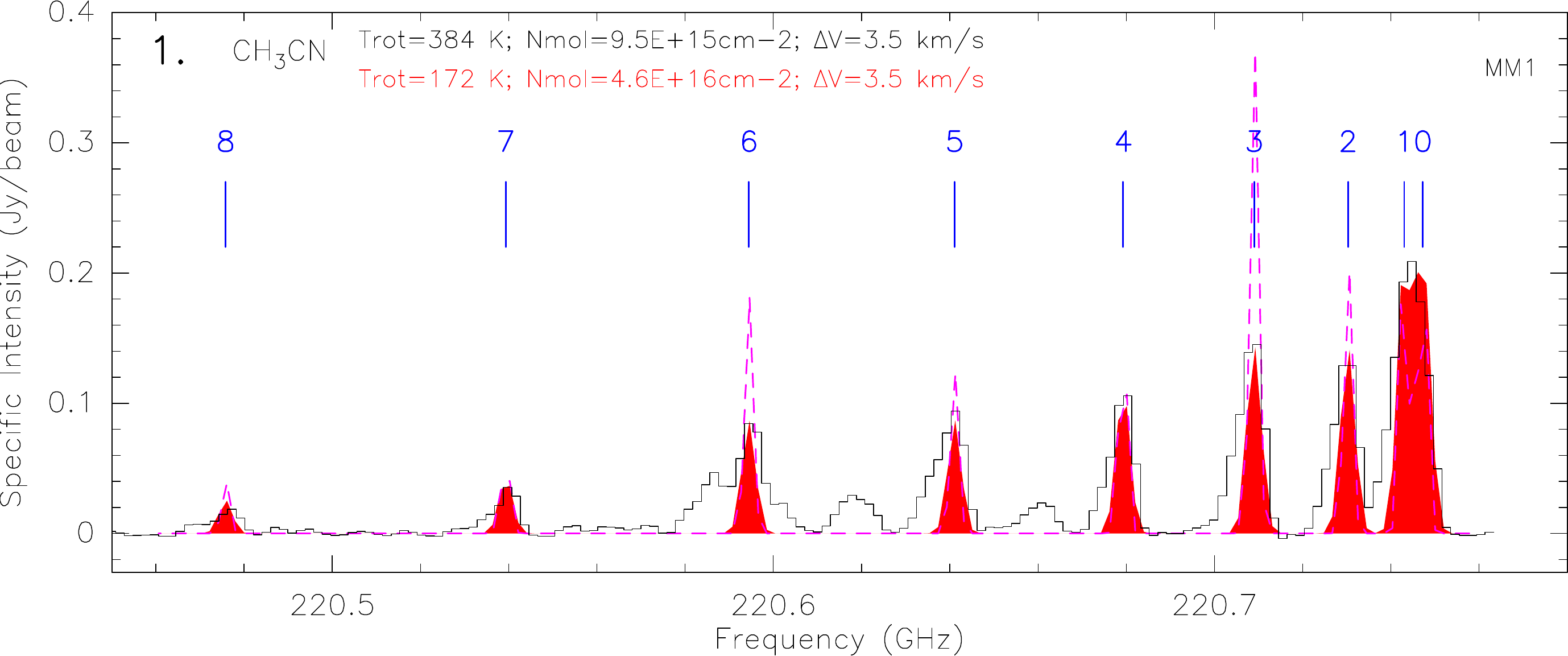}
&\includegraphics[width=8cm]{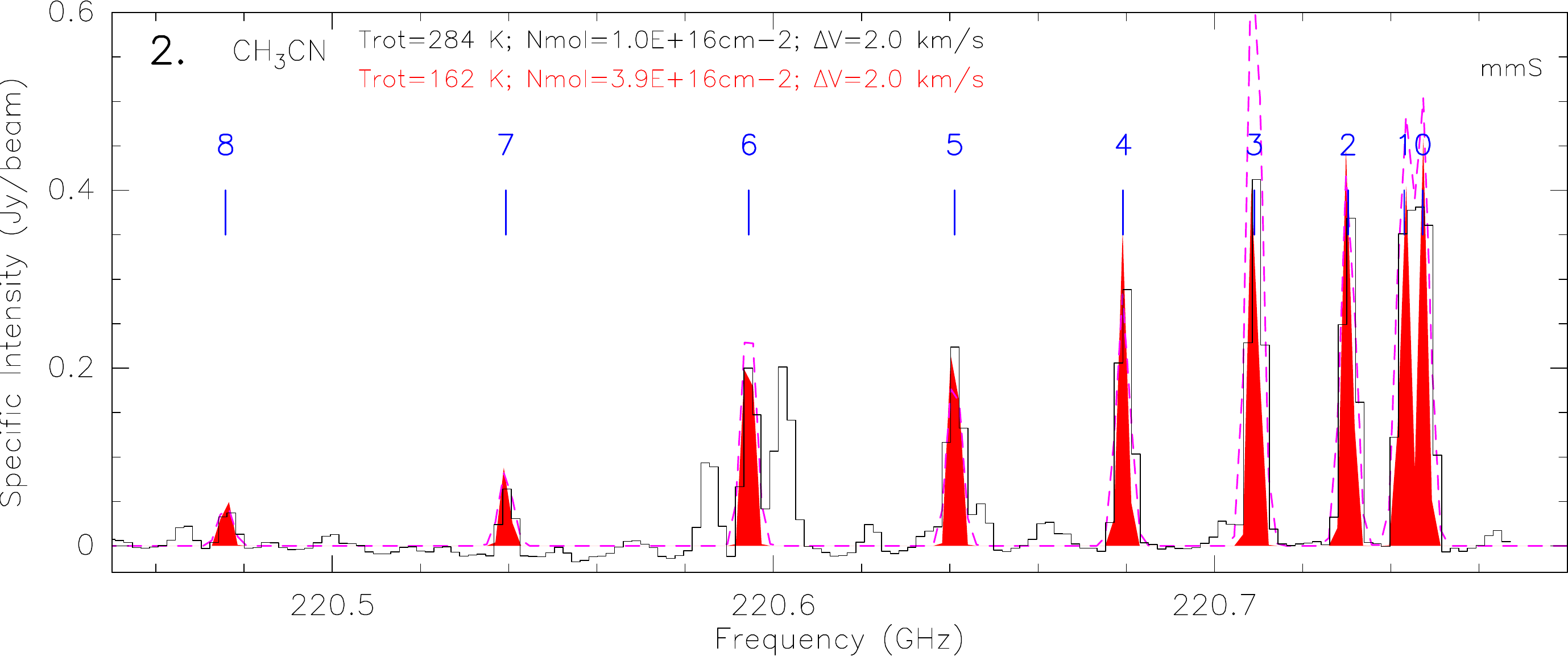}
\end{tabular}
\begin{tabular}{lll}
\includegraphics[width=5cm]{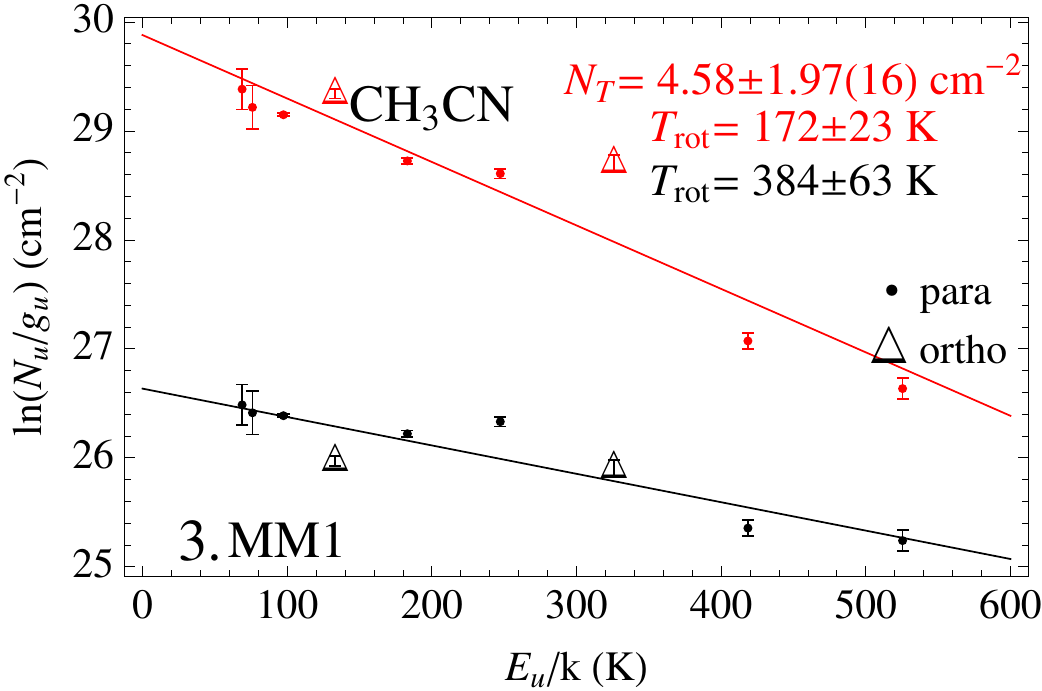}
&\includegraphics[width=5cm]{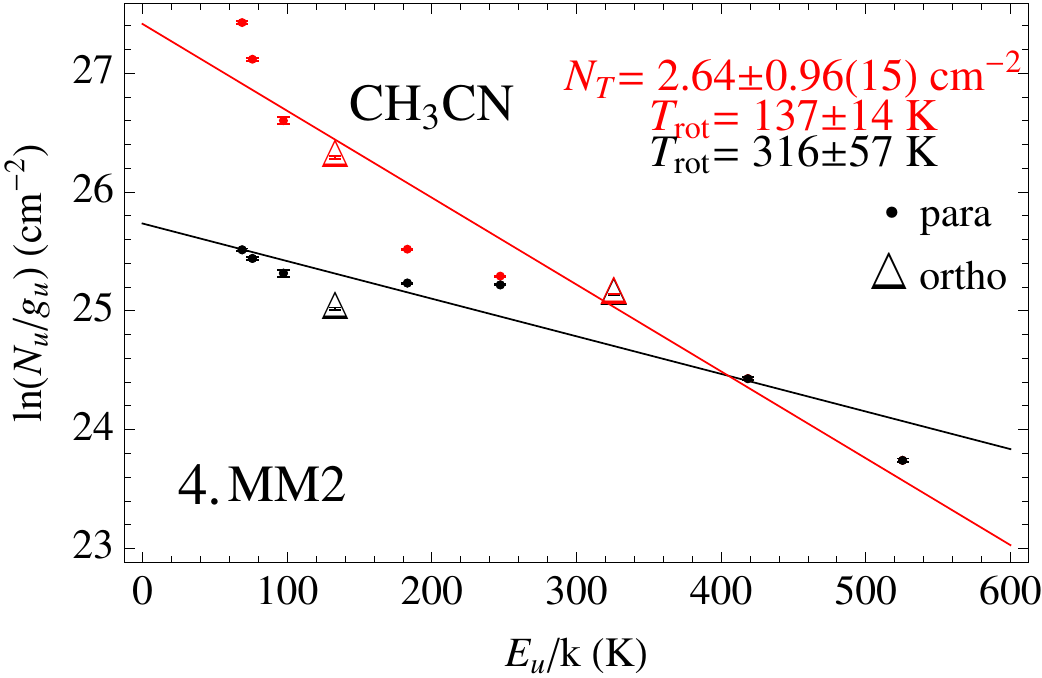}
&\includegraphics[width=5cm]{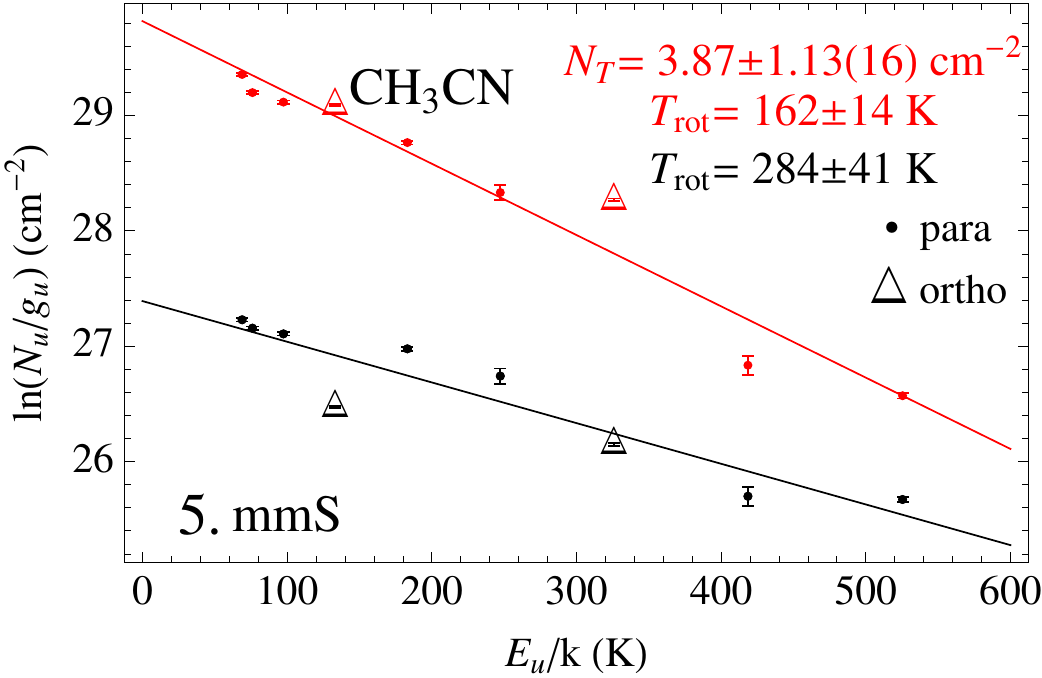}
\end{tabular}
\begin{tabular}{ll}
\includegraphics[width=8cm]{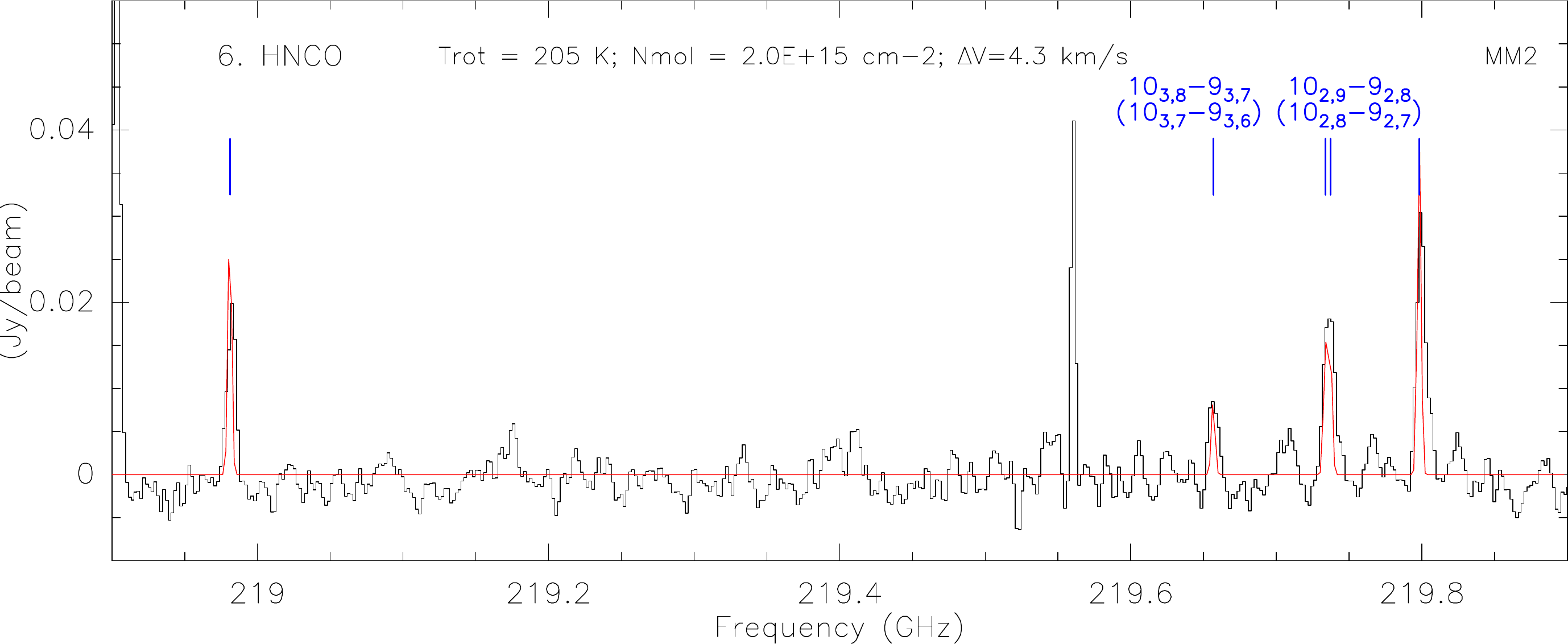}
&\includegraphics[width=8cm]{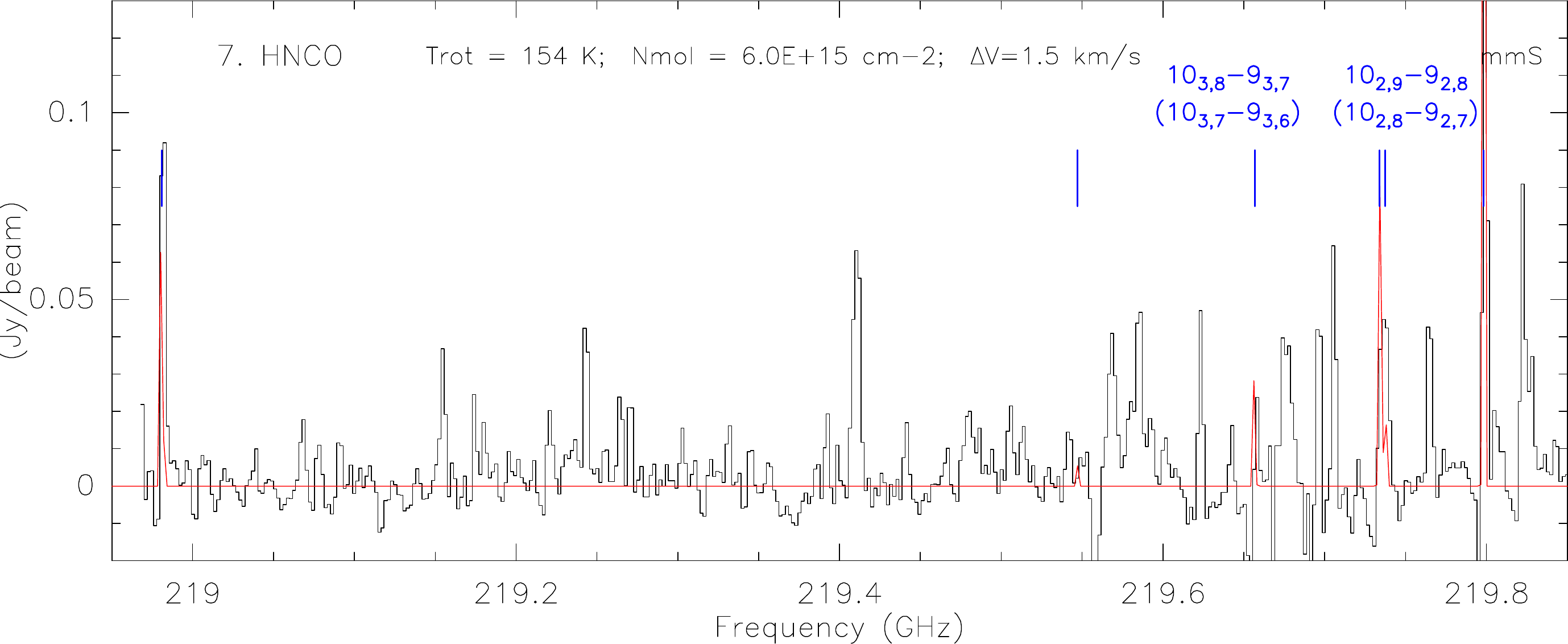}
\end{tabular}
\begin{tabular}{lll}
\includegraphics[width=5cm]{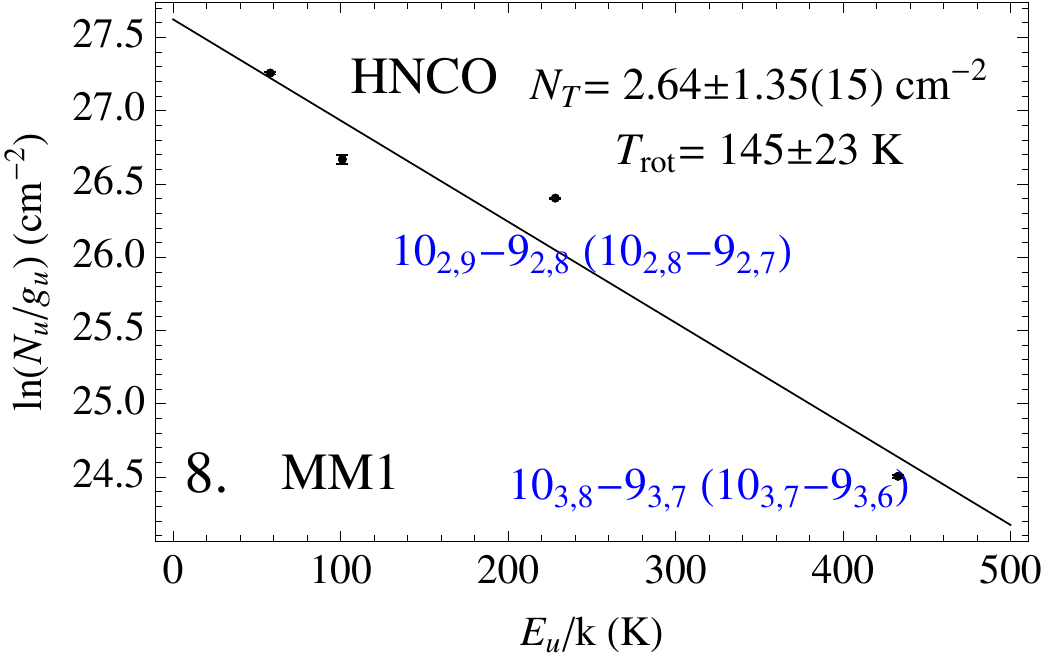}
&\includegraphics[width=5cm]{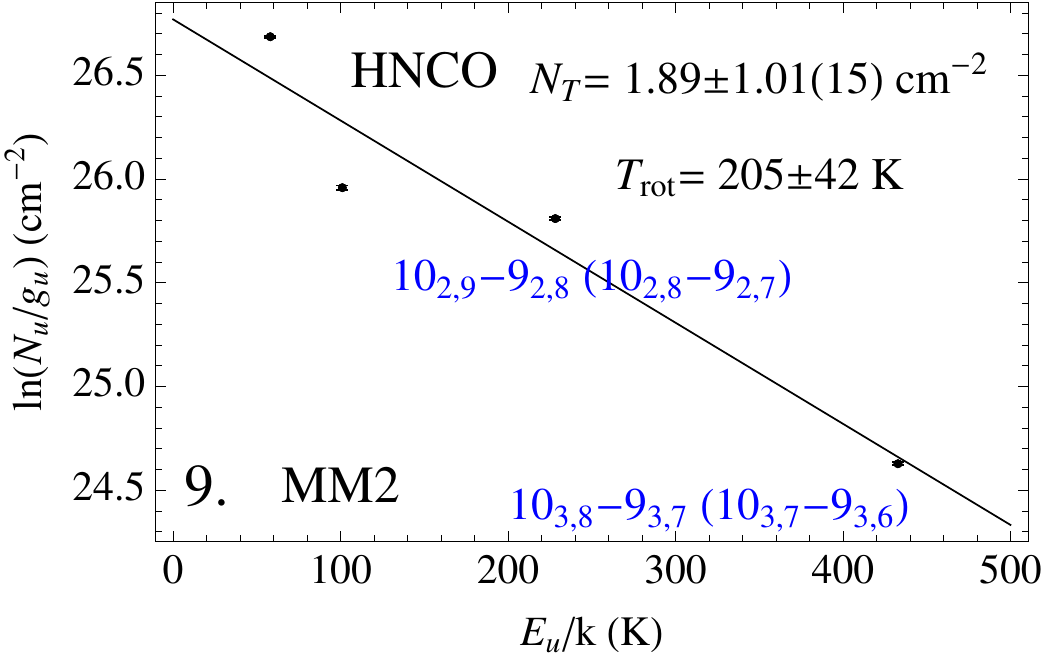}
&\includegraphics[width=5cm]{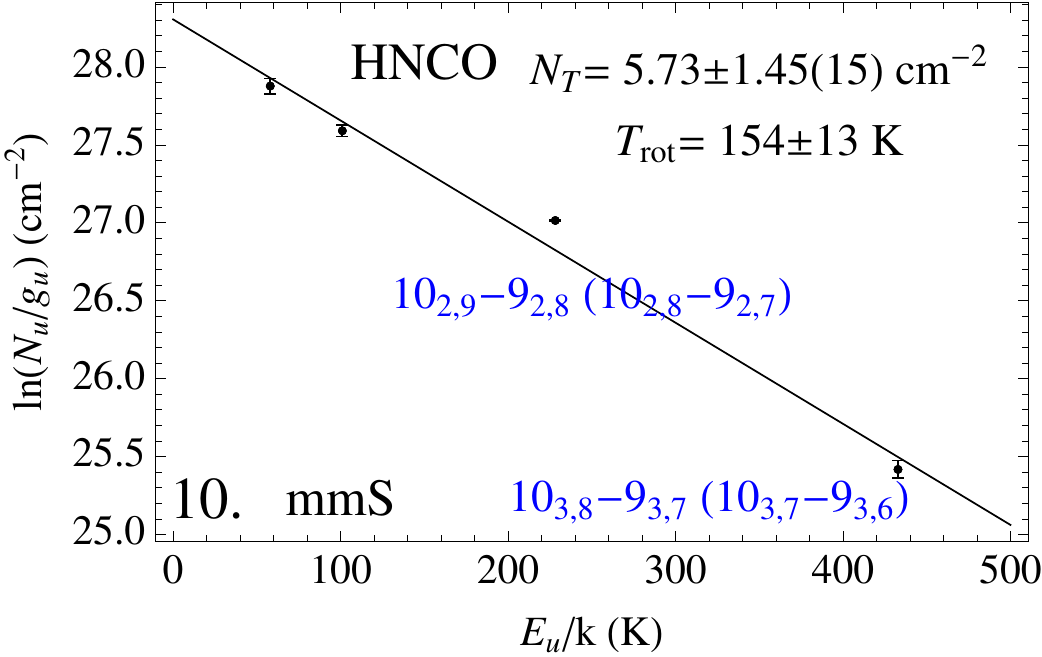}
\end{tabular}
\begin{tabular}{lll}
\includegraphics[width=8cm]{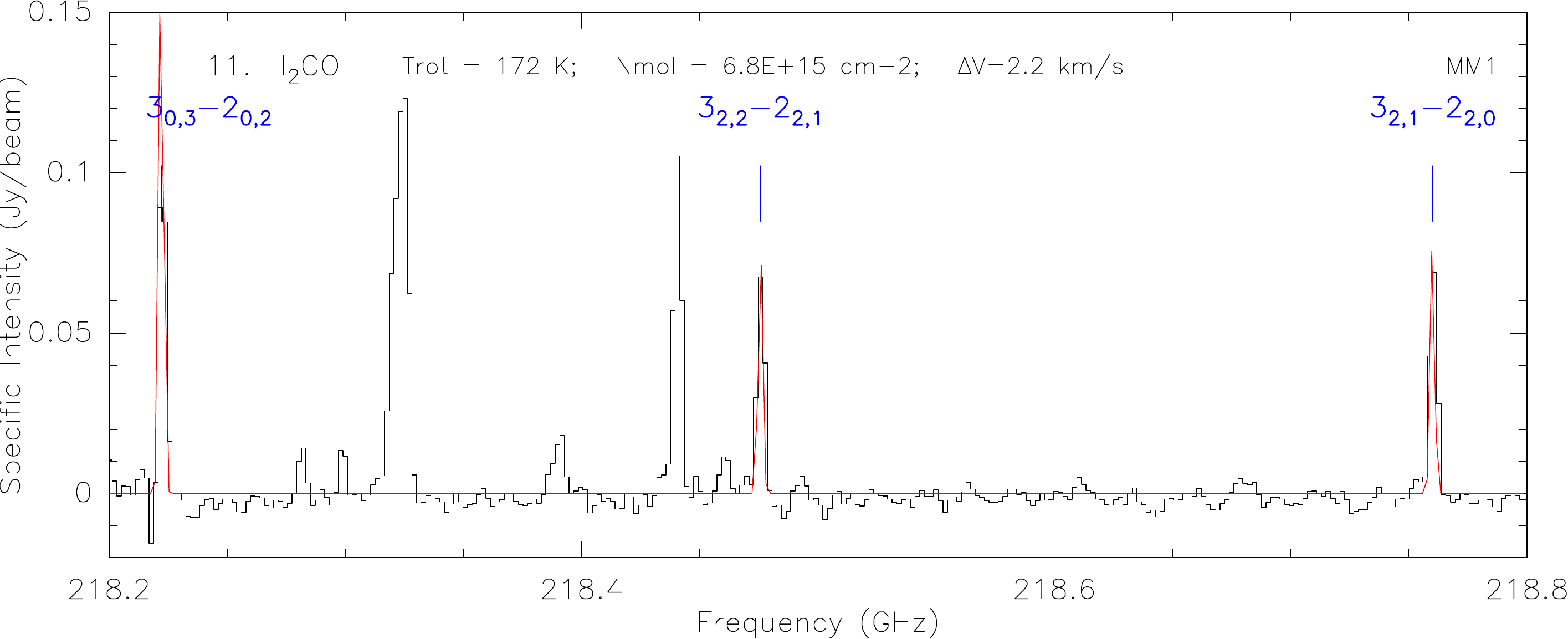}
&\includegraphics[width=5cm]{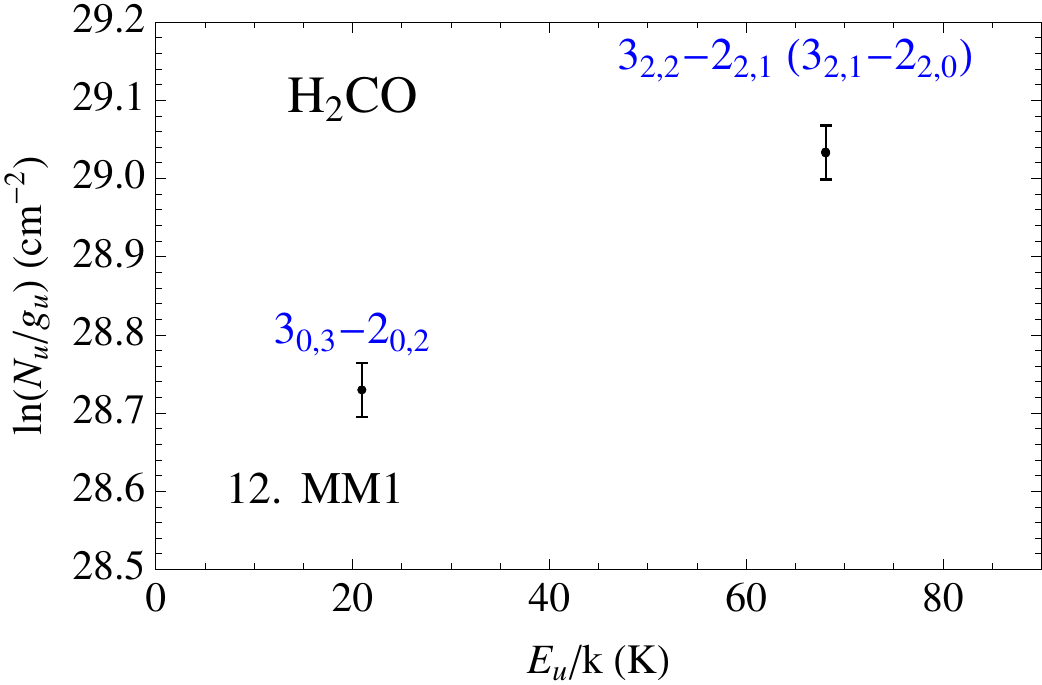}
\end{tabular}
\caption{Rotation diagrames and synthetic spectrum fits of $\rm CH_3CN$, HNCO, and $\rm H_2CO$ in MM1 (or MM2) and IRS1-mmS. Panels 1--2 present the $\rm CH_3CN$ spectra towards MM1 and IRS1-mmS  in black,  fitting with optically thin assumption in dashed purple, and fitting with opacity correction in filled red.  Panels 3--5 show the rotation diagrams of $\rm CH_3CN$ derived for MM1, MM2, and IRS1-mmS, with black dots and fits corresponding to the optically thin assumption, whereas red dots and fits  {include}  correction for opacity.  Panels 6--7 present the HNCO {(assumed optically thin)} synthetic spectrum fits towards MM2 and IRS1-mmS, with blended lines marked {(fit in red)}. Panels 8--10 show the rotation diagrams of HNCO  {(assumed optically thin)}. Panel 11 presents the $\rm H_2CO$ synthetic spectrum fits towards MM1. Panels {12} shows the data points of $\rm H_2CO$ lines in MM1, which  cannot be fitted in the rotation diagram.  The estimated $\rm T_{rot}$ and molecular column density from each rotation diagram is  indicated in each panel. Errors are derived from scatter in the data points.  
}
\label{rotation}
\end{center}
\end{figure*}

We assume that all $\rm K=1-8$ components have the same linewidth in each individual substructure. Although K=0, 1 lines are blended, they have similar relative intensity and $\rm E_u/k$ and therefore, weighting the measured intensity relative to the expected relative intensity for each component yields the individual integrated-intensities of each line. Solid and dashed lines in Panels 1--2 of Figure~\ref{rotation} show the best synthetic spectrum fits with or without, respectively, opacity correction for $\rm CH_3CN$ ladder in MM1 and IRS1-mmS. Panels 3 and 5 show the corresponding rotation diagrams. We find $\rm T_{rot}$ derived without opacity correction is overestimated by a factor of 1.5--2, leading to an underestimation of the $\rm CH_3CN$ column densities by a factor of 3--5.\\

\item {\bf HNCO} has four detected transitions in our observations, with $\rm E_u/k$ ranging from 50 K-- 450 K. The rotational temperatures derived from HNCO are consistent with those derived from $\rm CH_3CN$ in MM1 and IRS1-mmS, however slightly higher than from $\rm CH_3CN$ in MM2. Such  differences in $\rm T_{rot}$ between $\rm CH_3CN$ and HNCO is {also observed at larger scales} (e.g. \citealt{bisschop07a}). Possible explanations include: {(1) The excitation of HNCO may be dominated  by radiative processes rather than collisional \citep{churchwell86}; (2) Our frequency resolution does not allow us to resolve blended hyperfine components\footnote{Two pairs of twin lines $\rm 10_{3,8}\rightarrow9_{3,7}$--$\rm 10_{3,7}\rightarrow9_{3,6}$ and $\rm 10_{2,9}\rightarrow9_{2,8}$--$\rm 10_{2,8}\rightarrow9_{2,7}$ are blended and contain further hyperfine splittings. Each pair of lines {has} the same relative intensity and $\rm E_u/k$. Therefore, the rotation diagram is based on the  assumption that each line have the same contribution to the blending intensity integration.} which may not be in LTE  (Panel 9 of Figure~\ref{rotation}); 
 (3) Limited by angular resolution, we cannot distinguish whether HNCO and $\rm CH_3CN$ trace the same gas components in MM1 -- MM3; and (4) The optically thin assumption may break down for HNCO, in particular in MM2.}\\

\item  {\bf  $\rm H_2CO$}  is  routinely used as a thermometre for dense gas \citep{mangum93,ao13}. {However, as shown in Panel 11 of Figure~\ref{rotation}, the data are not consistent with optically thin emission. Without any clear opacity correction, we cannot use this transitions for further temperature estimates.}\\

\item{\bf Other species}: 
The three $\rm CH_3CH_2CN$ transitions observed have approximately the same $\rm E_u/k$ and therefore cannot provide accurate constraints for $\rm T_{rot}$. Several transitions of $\rm HCOOCH_3$ were covered as well, however, were mostly only marginally detected (Table.~\ref{tab:line}). We detected five $\rm CH_3OH$ transitions. Three of the $\rm CH_3OH$ may be masing (see Section~\ref{dif}), including two in the torsionally excited state (Table~\ref{tab:ch3oh}), such that they cannot constrain $\rm T_{rot}$. \\

\end{itemize}

In the following, we assume that the $\rm CH_3CN$ rotational temperatures in MM1--MM2 and IRS1-mmS {are equal to} the gas kinetic temperature. Given that $\rm CH_3CN$ in MM3 and the outflow regions are observed at $\rm <3\sigma$, and each detected species in these regions present only one line, we are unable to directly measure the gas temperature. Therefore, we assume that temperatures in JetN and JetS are $\sim150$ K following \citet{sandell10} ($\rm 147\pm40~K$). In MM3 the gas temperature must be large enough to excite $\rm >3\sigma$ line emission from $\rm CH_3OH~(4_{2,2}\rightarrow3_{1,2}$, $\rm E_u/k=46$ K), while also being lower than in MM1--MM2 because of the smaller number of lines and their lower intensities. We assume the gas temperature of MM3 to be $\rm \sim50\,K$ which is comparable with that of the associated, more extended gas envelope as constrained by the previous observations of $\rm H^{13}CN$, DCN, and $\rm CH_3CN$  ($\rm 52\pm10~K$, \citealp{sandell10}). Toward IRS1-peak, {we take the gas temperature measured from IRS1-mmS as the  temperature lower limit,  and that measured by \citet{goddi15a} at 0.15\arcsec--0.3\arcsec resolution as the temperature upper limit}.\\

\subsection{Column densities}\label{col}

\subsubsection{Spectral line:} \label{spl}   
We assume that all transitions are in LTE to calculate column densities from integrated intensities of the strongest transition for each isotopologue (see Appendix \ref{ap-abun}).\\

{For several species where we have additionally detected their rare isotopologue lines}, e.g.  $\rm SO ~(6_5\rightarrow5_4)$- $\rm ^{33}SO ~(6_5\rightarrow5_4)$,  $\rm OCS ~(18\rightarrow17)$- $\rm O^{13}CS ~(18\rightarrow17)$,  $\rm ^{13}CO ~(2\rightarrow1)$- $\rm C^{18}O ~(2\rightarrow1$) and  $\rm CH_3^{13}CN ~(12_2\rightarrow11_2)$- $\rm CH_3CN ~(12_2\rightarrow11_2)$, we have derived the molecular column densities assuming an opacity correction (Eqs~\ref{emi}--\ref{eq:tau}). We find $\rm \tau=3\text{--}45$ (see Table \ref{tab:correction}), indicating that the lines are very optically thick. 
For the remaining species we assume that the observed transitions are optically thin, an assumption discussed in Appendix \ref{appendix:molcol}.\\
\color{black}

%%\onecolumn
  \begin{figure*}[ht]
\centering
\includegraphics[width=6cm,angle=-90]{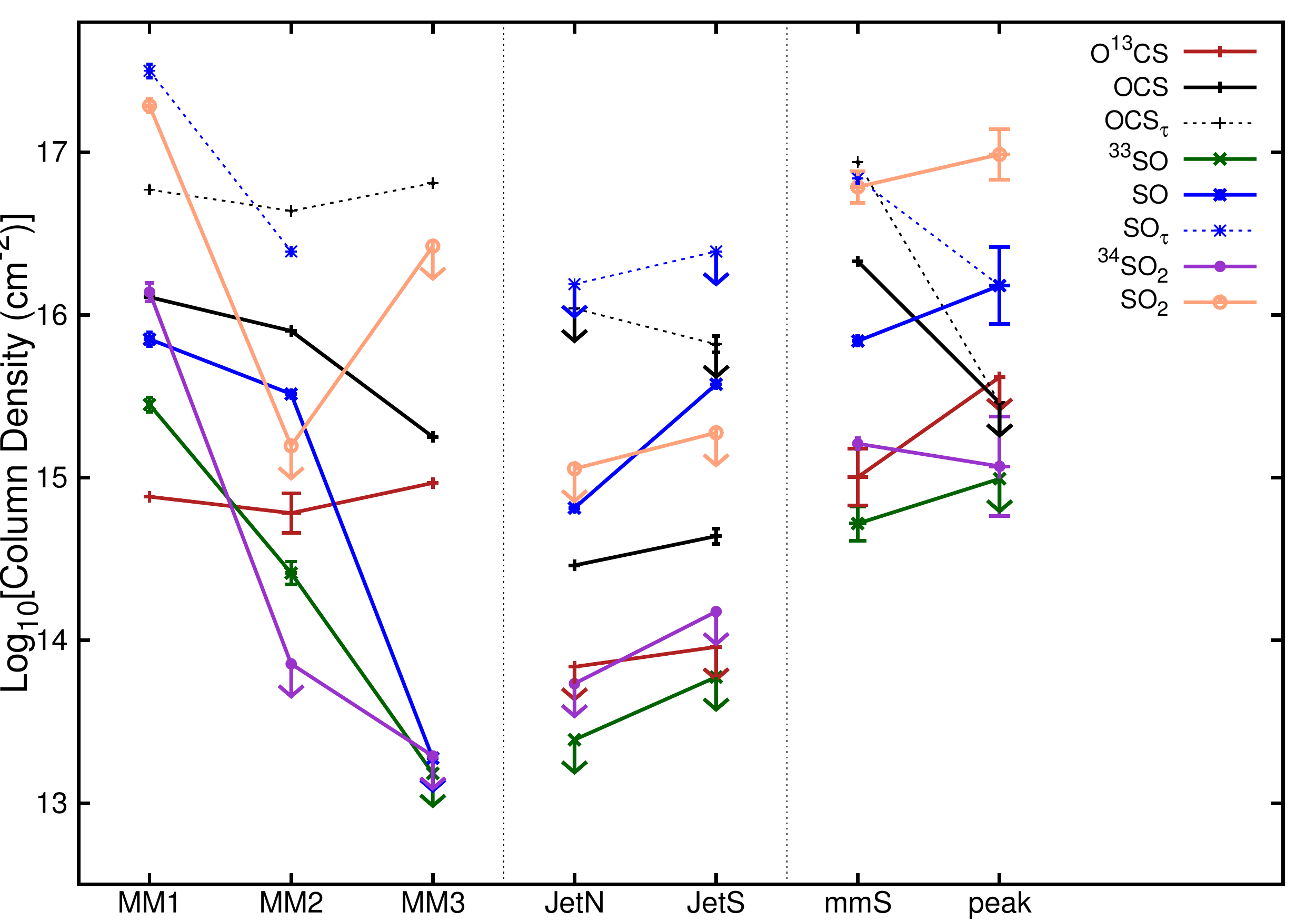}\\
\includegraphics[width=6cm, angle=-90]{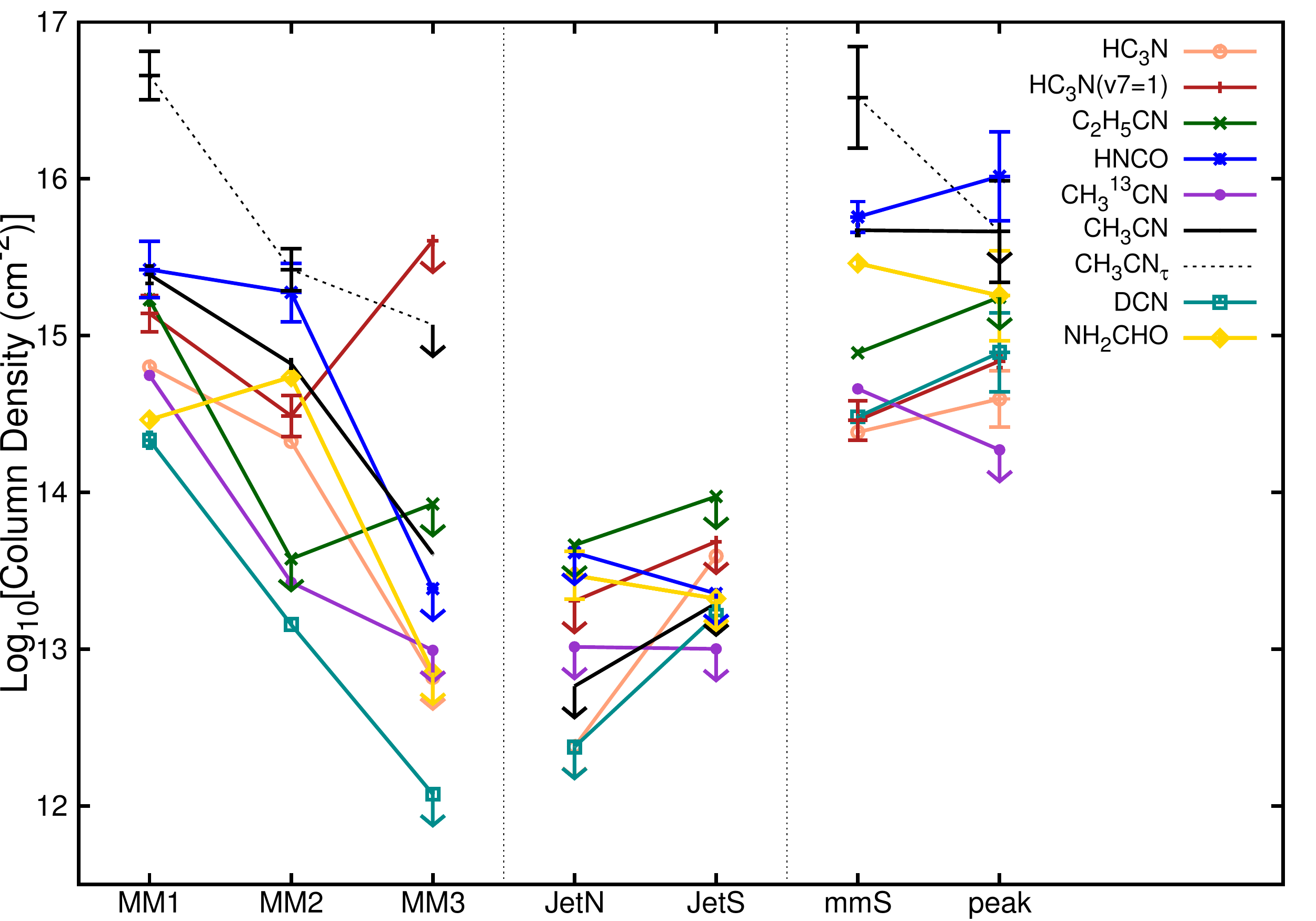}\\
\includegraphics[width=6cm, angle=-90]{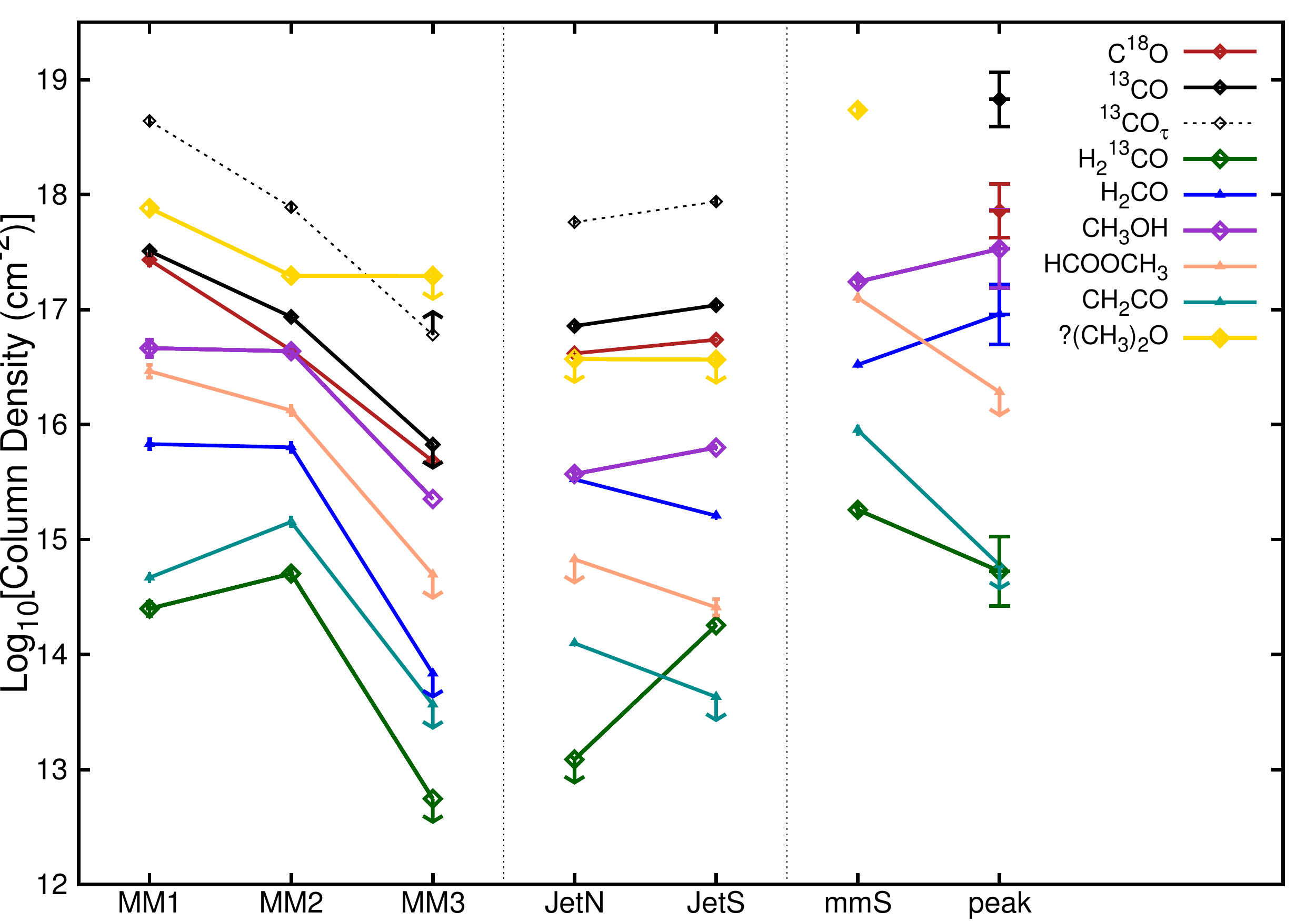}\\

\caption{Molecular column densities toward 5 positions in NGC\,7538\,S and 2 positions in IRS1. Colored solid lines show the {variation} for each isotopologue calculated with the optically thin assumption (arrows highlight the upper  and lower limits), {while dashed lines ($\rm OCS_{\tau}$, $\rm SO_{\tau}$, $\rm CH_3CN_{\tau}$, $\rm ^{12}CO_{\tau}$) show the   variation} calculated with opacity correction from Eq.~\ref{eq:tau}.  The uncertainties at each position are determined from $\rm T_{rot}$, partition function $\rm Q(T_{rot})$, Gaussian fit to  $\rm \int T_B(\upsilon)d\upsilon$, or the scatter of the multi-transitions on the rotation diagrames ($\rm CH_3CN$ and HNCO). For  species (e.g.  HNCO, $\rm SO_2$)   we have not measured the optical depths, the column densities could be underestimated by a factor of 5-10.
}\label{abun_line}
\end{figure*}

Tables \ref{col-Obearing}--\ref{col-Sbearing} and Figure~\ref{abun_line} summarize the derived {beam-averaged} column densities of 22 isotopologues. Column densities  vary significantly between the  substructures. Most present a  decreasing trend from MM1 to MM3, with the exceptions of $\rm NH_2CHO$, $\rm H_2^{13}CO$, and $\rm CH_2CO$.  Given the similar sizes and the 1.37\,mm continuum emission intensities of MM1--MM3, the observed decreasing  column densities are likely because of chemical segregation. 

\subsubsection{Continuum:}\label{continu}
 In NGC\,7538\,S, the 1.37\,mm continuum emission is dominated by thermal dust emission. Observations of \citet{wright12} have shown that the contribution from the free-free emission is negligible. Assuming that the dust emission at mm wavelength is optically thin and that the temperature of the dust and gas temperatures are equal, we are able to calculate the $\rm H_2$ column densities and masses for different substructures. Following \citet{hildebrand83} and \citet{schuller09},

     \begin{eqnarray}
  \rm N_{H_2}=\frac{I_{\nu}R}{B_\nu(T_{dust})\Omega_a \kappa_{\nu}\mu m_H} ~~~~~~~~~(cm^{-2})\label{gas}\\
   \rm M=\rm \frac{S_{\nu}R{\it D}^2}{B_\nu(T_{dust}) \kappa_{\nu}} ~~~~~~~~~~~~~~~~~~~~~~~~~~~~~(g)\label{gasmass}
   \end{eqnarray}

    where $\rm I_\nu$ is the peak specific intensity  for each substructure   (in Jy/beam, taken to be the $3\sigma$ upper limit for the  outflow regions),  {$\rm S_\nu$ is the total flux {(in Jy, free-free contribution excluded)}, measured within $\rm 4\sigma$ contours \citep{beuther12},} R is the gas to dust mass ratio (taken to be 150 from \citealt{draine11}),  $\rm B_\nu(T_{dust})$ is the Planck function for a dust temperature $\rm T_{dust}$,  $\rm \Omega_a$ is the solid angle of the PdBI beam (in $\rm rad^2$),  $\mu$ is the mean molecular weight of the ISM, assumed to be 2.33,   $\rm m_H$ is the mass of an hydrogen atom ($\rm 1.67 \times 10^{-24}$ g), {and $\it D$ is the source distance}. We assume a model of agglomerated grains with thin ice mantles\footnote{When $\rm T_{dust}>100$ K, water ice starts to sublimate. However, along the line of sight, temperature profiles at different radii of the  protostar envelope are unknown. Therefore, we assume the dust absorption coefficient according to the Òthin ice mantleÓ case, based on the modelling result in \citet{gerner14}. In the most extreme case, if  the ice  is completely sublimated in MM1 or if MM3 still have larger water ice mantle, our current estimation leads to 2--4 times overestimation/underestimation of the $\rm H_2$ column density.\label{icemantle}}
    for densities $\rm 10^8~cm^{-3}$ in MM1--MM3\footnote{Table~\ref{source} shows that the projected sizes of MM1--MM3 from 2-D Gaussian fits are larger than the synthetic beam, so filling factors  can be  as one. Thus the number densities of MM1--MM3 are $\rm >10^8~cm^{-3}$.}, and  $\rm 10^6~cm^{-3}$ in JetN and JetS\footnote{Jet/outflow-induced shocks may not only strip the ices mantles, but also partly destroy the grains.}, yielding a dust absorption coefficient at 1.37\,mm of  {\color{black} $\rm \kappa_{\nu}=0.77~\rm cm^2g^{-1} $}  in MM1--MM3, and  $\rm \kappa_{\nu}=0.62~\rm cm^2g^{-1} $ in JetN (S)  (interpolation from \citealt{ossenkopf94}).  \\

We find peak column densities for the substructures in NGC\,7538\,S are similar $\rm (1-5)\times10^{24}~ \rm cm^{-2}$  ($\rm N_{H_{2,1}}$ in Table~\ref{source}).  
The outflow regions  have the lowest column densities which may be underestimated potentially because of the lack of short-spacing data.\\

The 1.37\,mm continuum in IRS1 is dominated by free-free emission  (e.g. \citealt{pratap92,keto08,sandell09,beuther12,beuther13}), so that  estimates of dust column density will be biassed. We note that $\rm C^{18}O$ is detected toward IRS1 and all the identified substructures of NGC\,7358 S, even though its large-scale emission  has been filtered out.  As a rare isotopologue it is reasonable to assume its line is optically thin. We therefore use $\rm C^{18}O$ to make a second, independent estimate of the $\rm H_2$ column density in these substructures. Assuming a constant abundance with respect to $\rm H_2$, $x({\rm C^{18}O}) \approx 1.64\times10^{-7}$ \citep{wilson94,giannetti14}, we can calculate the $\rm H_2$ column density, reported in the column $\rm N_{H_{2,2}}$ of Table~\ref{source}. This is further discussed in Appendix~\ref{appendix:nh2}.\\

\subsection{Molecular abundances}
To study chemical properties, we derive the molecular abundances with respect to $\rm H_2$ in each substructure with results shown in Tables \ref{col-Obearing}--\ref{col-Sbearing}. In MM1--MM3, the $\rm H_2$ column densities were derived from the 1.37\,mm dust continuum emission, while for JetN and JetS where we did not detect dust continuum emission, we estimate the $\rm H_2$ column density based on the integrated $\rm C^{18}O$ intensity ($\rm N_{H_{2,2}}$). We argue that we are not strongly biassed, provided that the $\rm C^{18}O$ abundance is close to the canonical value in the ISM in MM1 (HMC), where the outflow emanated from \citep{wilson94}. Since IRS1 and MM1 are both considered HMCs, we assume that their $\rm C^{18}O$ abundances are equal and therefore adopt $\rm N_{H_{2,2}}$ when deriving the molecular abundance of IRS1.

 \section{1-D model fit}\label{model}
 {
Despite having similar $\rm H_2$ column densities, adjacent condensations MM1--MM3 embedded in NGC\,7538\,S within {20,000} AU area, display gas temperature and spectral line properties which evolve systematically from the northeast to the southwest. We hypothesize that chemically MM1 evolves quickest, resulting in the largest abundance of COMs, consistent with a HMC phase. Non-detection of most species in MM3 indicates that it remains in an early evolutionary stage (e.g. a prestellar object). Clear detections of COMs but generally lower gas temperatures and molecular abundances than in MM1 indicating that MM2 is at an intermediate evolutionary stage between MM1 and MM3. To verify our hypothesis, we iteratively fit the observed molecular column densities using the `MUSCLE' (`MUlti Stage CLoud codE') 1-D chemical model \citep{semenov10}. This procedure has previously been successfully applied to fit the rich molecular data from several tens of HMSFRs at various evolutionary stages \citep{gerner14}, however at lower spatial resolution.

\subsection{Physical model}
\label{sec:Phys_model}
In a 1-D approximation, we assume each condensation is well described by a spherically symmetric gas condensation, with a fixed outer radius of $r_\mathrm{out}=1,100$~AU (equal to the size of the PdBI synthesized beam), and are embedded in a large-scale, low-density envelope which shields the condensation from the interstellar FUV radiation with a visual extinction of 10 mag. In addition, we define a parameter $r_{\rm {in}}$ as the inner radius, within which temperature and density are assumed to be constants. The radial density and temperature profiles are therefore described by broken power laws,

\begin{equation}
\label{eq:rho_r}
 \rho(r) =\left\{ \begin{array}{lr}  \rho_{\rm {in}}(r/r_{\rm {in}})^{-p}    &(r_{\rm {out}}\ge r\ge r_{\rm {in}}), \\
                                                 \rho_{\rm {in}}     &(0\le r<r_{\rm {in}}) \end{array} \right. 
\end{equation}

and
\begin{equation}
\label{eq:temp_r}
T(r) =\left\{ \begin{array}{lr}  T_{\rm {in}}(r/r_{\rm {in}})^{-q}    &(r_{\rm {out}}\ge r\ge r_{\rm {in}}), \\
                                                 T_{\rm {in}}     &(0\le r<r_{\rm {in}}) \end{array} \right. 
\end{equation}

\noindent Here, $p$ and $q$ are the radial indices of the density and temperature profiles, respectively. The gas and dust temperatures are assumed to be coupled. \\

In fitting each condensation model (Table. \ref{tab:bestfit1}), we adjust the parameters  $p$, $\rho_{\rm {in}}$, $T_{\rm {in}}$ and $r_{\rm {in}}$, keeping $q$ fixed and equal to 0.4 (because of a highly optically thick medium, \citealt{vandertak00}). To derive the H$_2$ column density in the modelled large-scale cloud, we adopt a grid of $40$ radial cells and integrate  Eq.~\ref{eq:rho_r} numerically.  
We note that here the physical structure is assumed static (i.e. not evolving with time) for each condensation  and is fitted iteratively to the observed chemical composition.

}

\subsection{Chemical model}
\label{sec:Chem_model}

We use the time-dependent gas-grain chemical model  `ALCHEMIC' which is described extensively in \citet{semenov10} and \citet{Albertsson_ea13}. The model is briefly summarized  as follows.\\

\noindent(1) The chemical network is based on the OSU'2007 network\footnote{http://www.physics.ohio-state.edu/~eric/research.html}, and is updated by implementing the most recent reaction rates (e.g. from the KIDA
database\footnote{http://kida.obs.u-bordeaux1.fr}, \citealt{Albertsson_ea13}).\\
(2) Both gas-phase and gas-grain interactions are included, resulting in 7907 reactions between 656 species made of 13 elements. The synthesis of COMs was included using a set of surface reactions (together with desorption energies) and photodissociation reactions of ices \citep{Garrod_Herbst06, semenov11}. \\
(3) Cosmic-ray particles (CRP,  $\zeta_{\rm CR}=5 \times 10^{-17}$~s$^{-1}$) and CRP-induced FUV radiation are considered to be the only external ionizing sources. {For these we adopt the UV dissociation and ionization photo-rates from \citet{vD06}, assuming the case corresponding to the spectral shape of the interstellar FUV radiation field.}\\
(4) Sticking : Molecules, other than H$_2$ which has a binding energy of $\sim~100$~K \citep{Lee72}, are assumed to stick to grain surfaces with 100\% probability.  \\
(5) Mobility: On the surface sites quantum tunnelling is not considered \citep{Katz_ea99}.\\
(6) Reactions: Dissociative recombination is modelled, resulting in the ion-ion neutralization and electron attachment on the grains, while chemisorption of surface molecules is not considered. Each $0.1\,\mu$m spherical olivine grain provides $\approx 1.88 \times 10^6$ surface sites for accreting gaseous species and the surface recombination proceeds solely through the Langmuir-Hinshelwood formation mechanism.\\
(7) Release: As the temperature increases, ices are released back to the gas phase by thermal, CRP, and CRP-induced UV-photodesorption ({UV yield is $3\times10^{-3}$}, e.g. \citealt{Oeberg_ea09a}). In addition, all freshly formed surface molecules are {assumed} to have a 1\% chance to leave the grain surface upon recombination.\\
(8) None of the above chemical parameters were varied during the iterative fitting of the observational data.\\

\subsection{modelling scheme}\label{sec:initi}
{Given the very close location of the  condensations MM1--MM3 to each other, it is natural to assume that they fragmented from a common parental core, as soon as the gas core became cold and dense enough.
The IRDC-like physical conditions seem to be the most appropriate for that. Thus we assume that the condensations MM1--MM3 evolved both physically and chemically from the same parental IRDC. \\

This allows us to set up the initial chemical abundances for our modelling. Following the multi-stage
MUSCLE scheme in \citet{gerner14}, we took a single-point (0-D)  IRDC physical model with the best-fit parameters by fitting the median of their observed IRDC sample.  The 0-D assumption merely reflects our lack of knowledge about the real structure of the MM progenitor core.\\

Therefore, the pre-MM IRDC is reasonably assumed to have  a  single temperature  of 15\,K, a hydrogen particle density of $2\times 10^5$~cm$^{-3}$ and a chemical age of $11,000$~years. 
Note when we model the chemical composition  with such a simplistic physical structure,  the ``chemical age'' is the physical time
moment when our chemical code describes the observed column densities most
accurately.\\

With this 0-D IRDC physical model, we use our gas-grain ALCHEMIC code to compute its chemical evolution over the chemical age of $11,000$~years, starting from the ``low metals'' initial abundances of   \citet{lee98} (see also Table~\ref{tab:inabun_MM3}). The resulting abundances are used as input  to subsequent chemical modelling of later evolutionary stages.\\

After $11,000$~years of the IRDC evolution, we assume it fragmented into  three individual MM condensations, which underwent gravitational  collapse at various speeds, due to for example, difference in initial masses or angular momenta. Using observed temperatures of the condensations MM1--MM3 as a rough estimate of how much they have evolved physically, it becomes clear that MM3 with $\rm T \sim 50\,K$ is
younger (less evolved) than MM2 with $\rm T \sim 137$~K, which is in turn younger  than MM1 
with $\rm T \sim 172$~K (see Table~\ref{source}). This also means that MM1 has likely passed through  physical conditions which were similar first to MM3 and then to MM2 phase, while MM2 has passed only 
through MM3-like evolutionary phase.  
As we do not know exactly how density and temperature of each MM condensation  evolved, we assume a jump-like transition\footnote{A temperature change occurs when simulations from one phase go to another phase: T(r) is kept constant with time during each evolutionary phase and is replaced by a new (warmer) temperature profile at each phase end.} of the physical properties from phase to phase. This is a crude approximation of more smoother, gradual increase in density and temperature  during the collapse, but
it enables us to apply our multi-stage chemical modelling approach. \\

For that, the IRDC chemical composition at $t=11,000$~years was taken as the input to the 1-D power-law physical model described in Sec.~\ref{sec:Phys_model} and  the observed column densities of the MM3 condensation were iteratively fitted. 
Then, we used the best-fit MM3 chemical structure as an input for similar iterative 1-D chemical fitting of the more evolved MM2 condensation.   
After that, we repeated the same procedure and used the best-fit MM2 chemical structure as an input to model and to fit the observed column densities in the MM1 condensation.

\begin{table}
\caption{Top 35 initial abundances for the {parental IRDC 0-D  model at 11,000~years}.
\label{tab:inabun_MM3}  }    
\centering         
\begin{tabular}{lp{1.5cm}lp{1.5cm}}        
\hline\hline
Species & Relative abundance  & Species & Relative abundance\\
\hline
H$_2$          &   $5.00 (-01)$
&H          &   $9.75 (-02)$\\
He           &   $3.77 (-04)$
&C           &   $1.79 (-05)$\\
CO          &   $2.35 (-05)$
&N           &   $1.24 (-05)$\\
O           &   $8.16 (-05)$
&CH$_4$  ice   &  $2.63 (-05)$\\
H$_2$O  ice    & $6.79 (-05)$
&NH$_3$  ice    & $1.02 (-05)$\\
C$_3$          &$1.17 (-06)$
&CH$_3$OH  ice  & $5.39 (-06)$\\
CH$_4$         &$4.07 (-07)$
&H$_2$O         &$9.43 (-07)$\\
N$_2$          &$7.75 (-07)$
&C$_3$H$_4$  ice & $3.27 (-07)$\\  
CO$_2$   ice    &$1.43 (-07)$
&N$_2$    ice    & $1.62 (-07)$\\
CH$_2$CO       &$1.43 (-08)$
&CH$_2$OH       &$2.11 (-08)$\\
CH$_3$OH       &$1.35 (-08)$
&CN          &   $1.04 (-08)$\\
CO$_2$         &$1.85 (-08)$
&e$^-$      &   $1.96 (-08)$\\
HCN         &   $2.41 (-08)$
&HNC         &   $2.36 (-08)$\\
NH$_3$         &$2.54 (-08)$
&O$_2$          &$1.63 (-08)$\\
OH          &   $1.16 (-08)$
&S           &   $8.29 (-08)$\\
CO ice        & $1.71 (-08)$
&H$_2$CO  ice   & $1.77 (-08)$\\
HCN     ice   & $1.22 (-08)$
&C$^+$          &$1.77 (-09)$\\
C$_2$H         &$2.05 (-09)$\\
\hline                
\end{tabular}
\end{table}

\subsection{Iterative fit to the data}\label{sec:fitting}
The best-fit result of the models described above was obtained via an iterative procedure. Parameters, including the inner radius $r_{\rm {in}}$, the density at the inner radius $\rho_{\rm {in}}$, the temperature at the inner radius $\rm T_{\rm {in}}$ and the density slope $p$, were adjusted  to the following constraints:\\

\begin{enumerate}
\item The density index $p$ is restricted to 1.0--2.5 \citep{guertler91,beuther02,mueller02,hatchell03}.
\item The derived $\rm H_2$ column densities set $r_{\rm {in}}$ ($1\text{--}300$~AU) and $\rho_{\rm {in}}$ ($10^{10}-10^{13}$~cm$^{-3}$)
within a factor of 3.
\item Based on our previous temperature estimates (Section~\ref{tem}), $\rm T_{\rm {in}}$ is limited to  $30-80$~K for MM3 and to $100-300$~K for the MM2 and MM1 {condensation}s, respectively. 
\end{enumerate}

\noindent By varying the above parameters and fixing the rest,  a model grid for each condensation consisting of $1,000\text{--}3,000$ parameter combinations was produced.\\

We modelled the physical and chemical structures of MM1--MM3 over a period of $10^5$~years (over 99 logarithmic time steps), and compared the modelled beam-averaged molecular column densities with the observed values. The goodness-of-fit, measured by $\chi^2$, for each model at each time step was estimated using a confidence criterion taken from Eq.~(3) in \citet{Garrod_ea07}, where the standard deviation for each detected species is taken as $\sim 0.3$~dex. For species where our observations only provided upper limits for their column densities, we manually set the goodness-of-fit to be 0 once the modelled column density is more than 10 times lower than the observed upper limit. Subsequently, a total goodness-of-fit was obtained as the mean value of the confidence criteria for all species at a given time step and with a given parameter combination. Minimizing $\chi^2$-value yielded the best-fit physical model and the best-fit chemical age for our observations. \\

Our best fit parameters for the three observed condensations are listed in Table. \ref{tab:bestfit2}. Molecular abundances with respect to $\rm H_2$ are shown in Figure \ref{model_fit}, with the black lines representing observed values and the red filled area representing the modelling range. The x-axis shows the chemical variations from MM1 to MM3. From the above fitting, we found that the time for a pre-MM3 object to chemically evolve to a MM1-like object is short\footnote{Such short best-fit chemical ages are caused by very high densities of $\ga 10^{10}$~cm$^{-3}$ and warm temperatures $\ga 50$~K in the best-fit models. Under these specific conditions, chemical evolution is driven by very rapid gas-phase processes, which are able to quickly ``reset'' the initial chemical composition from a cold  IRDC phase. More precisely, it is a steep, jump-like transition from the progenitor IRDC phase with a density of $\sim 10^5$~cm$^{-3}$  to the MM3 phase with a density of $>10^{10}$~cm$^{-3}$ that makes the resulting best-fit chemical ages so short.}  (10--450\,yrs), indicating that the fast chemical evolution in the process from MM3-like structure to MM1-like structure leads to significant chemical differences. 
 
\color{black}

% Best-fit IRDC parameters:
\begin{table*}
\caption{Parameters of the best-fit 1-D gas-grain {condensation} models  of MM1--MM3  }\label{tab:bestfit1}
\small
\begin{center}
I. Parameters of the best-fit {condensation} models. 
%{\color{red}Referee: I think it might be a good idea to give here the masses of MM1--MM3 derived using these parameters. }\\
\begin{tabular}{lllll}       
\hline\hline               
Parameter & Symbol  & MM3$^a$   &MM2$^a$   &MM1$^a$
\\    % table heading 
\hline                        % inserts single horizontal line
Inner radius & $r_{\rm {in}}$                                               & $3.7$~AU                                                    & $1.1$~AU                                               & $1.1$~AU\\
Outer radius$^b$ & $r_{\rm {out}}$                                            & $\rm 1100~AU$                                       & $\rm 1100~AU$                                & $\rm 1100~AU$  \\
Density within the inner radius & $\rho_{\rm {in}}$             & $1.01\times10^{13}$~cm$^{-3}$                  &$1.1\times10^{11}$~cm$^{-3}$             & $6.1\times10^{12}$~cm$^{-3}$\\
Density beyond the outer radius & $\rho_{\rm {out}}$        & $1.2\times10^{8}$~cm$^{-3}$                      & $1.1\times10^{8}$~cm$^{-3}$                & $4.8\times10^{7}$~cm$^{-3}$\\
Density profile & $p$                                                            & $2.0$                                                             &$1.0$                                                      & $1.7$\\
Beam-averaged density  &$\bar{\rho}$                               & $4.7\times10^{9}$~cm$^{-3}$                      & $4.1\times10^{8}$~cm$^{-3}$                & $8.7\times10^{8}$~cm$^{-3}$ \\
Temperature within the inner radius & $T_{\rm {in}}$          & $59.4$~K                                                      & $229.2$~K                                                & $278.6$~K  \\
Temperature beyond the outer radius & $T_{\rm {out}}$       & $10.0$~K                                                       & $14.5$~K                                               & $17.6$~K\\
Temperature profile & $q$                                                     & $0.4$                                                              & $0.4$                                                      & $0.4$\\
Beam-averaged temperature & $\bar{T}$                            & $14.1$~K                                                        & $20.6$~K                                               & $28.5$~K \\
Mass   &M & 2.5 $\rm M_\odot$   & 3.8 $\rm M_\odot$  & 8.3 $\rm M_\odot$ \\
\hline  
\multicolumn{5}{l}{{Note.} {\it a}.  The values listed were obtained at the end of each evolutionary stage, corresponding  to the values used in Figure~\ref{model_fit}.}\\    
\multicolumn{5}{l}{{~~~~~~~~~} {\it b}.  This value is limited by the $0.4\arcsec$ PdBI synthesized beam size used in our observations.}\\                                
\end{tabular}
\end{center}
\end{table*}

\begin{table*}
\caption{Best-fit 1-D gas-grain modelled column densities compared to the observed values in MM1--MM3  }\label{tab:bestfit2}
\small
\begin{center}                   
\scalebox{0.8}{
\begin{tabular}{l|rl|rl|rl|l}        
\hline\hline                
&\multicolumn{2}{|c|}{MM3$^f$}&\multicolumn{2}{|c|}{MM2$^f$}&\multicolumn{2}{|c|}{MM1$^f$} &\\
\cline{2-7}
Molecule$^{d,e}$ &$\rm N_{Observed}^{\it g}$ & $\rm N_{Modeled}^{\it h}$&$\rm N_{Observed}^{\it g}$ & $\rm N_{Modeled}^{\it h}$&$\rm N_{Observed}^{\it g}$ & $\rm N_{Modeled}^{\it h}$&Notes on the Observed Values \\
 & $(\rm cm^{-2})$ & $({\rm cm}^{-2})$ & $({\rm cm}^{-2})$ & $({\rm cm}^{-2})$ & $({\rm cm}^{-2})$ & $({\rm cm}^{-2})$ &
\\    
\hline                     
$\rm H_2$    &$\rm 5.25_ {\pm0.23}(24)$    &$\rm  9.9 (24)$     &$\rm 1.70_ {\pm0.22}(24)$   &$\rm  3.1 (24)$     &$\rm 1.69_ {\pm0.27}(24)$    &$\rm 2.7 (24) $  &derived from dust continuum\\
$\rm CO$    &$\rm 2.39_ {\pm0.06}(18)$    &$\rm 2.6 (18)$    &$\rm 2.21_ {\pm0.16}(19)$    &$\rm 2.2 (19)$    &$\rm 1.35_ {\pm0.16}(20)$    &$\rm 5.6 (19)$   &derived from $\rm C^{18}O$ with $\rm ^{16}O/^{18}O=607^{\it a}$\\
$\rm H_2CO$    &$\rm \le 4.95(14)$    &$\rm 7.0 (14)$    &$\rm 4.50_ {\pm0.51}(16)$    &$\rm 3.3 (16)$     &$\rm 2.22_ {\pm0.34}(16)$    &$\rm 1.6 (16)$   &derived from $\rm H_2^{13}CO$  with $\rm ^{12}C/^{13}C=73^{\it a}$\\
$\rm CH_2CO~~^{~\dag}$    &$\rm \le 3.67(13)$    &$\rm 2.6 (14)$    &$\rm 1.43_ {\pm0.12}(15)$     &$\rm 1.0 (15)$    &$\rm 4.66_ {\pm0.37}(14)$    &$\rm 1.3 (15)$   &\\
$\rm CH_3OH~~^{~\dag}$    &$\rm 2.25_ {\pm0.02}(15)$     &$\rm 4.5 (14)$    &$\rm 4.33_ {\pm0.58}(16)$   &$\rm 2.2 (16)$    &$\rm 4.61_ {\pm0.88}(16)$    &$\rm 7.2 (16)$   &\\
$\rm HCOOCH_3~~^{~\dag}$    &$\rm \le 4.95(14)$    &$\rm 2.4 (09)~^{\times}$    &$\rm 1.33_ {\pm0.13}(16)$    &$\rm 1.2 (14)~^{\times}$    &$\rm 2.91_ {\pm0.39}(16)$    &$\rm 1.2 (13)~^{\times}$   &\\
$\rm CH_3OCH_3^{~~*~\dag}$    &$\rm \le 1.96(17)$    &$\rm 4.8 (11)~^{\times}$    &$\rm 1.97_ {\pm0.19}(17)$    &$\rm 5.3 (14)~^{\times}$    &$\rm 7.62_ {\pm0.57}(17)$    &$\rm 1.1 (14)~^{\times}$   &\\

$\rm HNCO~~^{~\dag}$    &$\rm \le 2.44(13)$    &$\rm 9.1 (12)$     &$\rm 1.89{\pm1.01}(15)$     &$\rm 1.8 (16)$    &$\rm 2.64 {\pm1.35}(15)$    &$\rm 2.4 (17)~^{\times}$   &\\
$\rm HC_3N~~^{~\dag}$    &$\rm \le 6.60(12)$    &$\rm 2.8 (13)$   &$\rm 2.11_ {\pm0.03}(14)$    &$\rm 6.1 (14)$    &$\rm 6.33_ {\pm0.12}(14)$    &\color{black}$\rm 8.7 (15)~^{\times}$   &\\
$\rm CH_3CN$    &$\rm \le 1.18(15)$    &$\rm 1.3 (13)~^{\times}$    &$\rm 2.64{\pm0.96}(15)$    &\color{black}$\rm 1.3 (14)~^{\times}$    &$\rm 4.58 {\pm1.97}(16)$    &$\rm 3.8 (15)$   &derived from opacity corrected $\rm CH_3CN$ in Sec. \ref{spl}\\
$\rm CH_3CH_2CN~~^{~\dag}$    &$\rm \le 8.43(13)$     &$\rm 1.3 (10)~^{\times}$    &$\rm \le 3.77(13)$    &$\rm 1.5 (11)~^{\times}$    &$\rm 1.70_ {\pm0.13}(15)$     &$\rm 6.3 (12)~^{\times}$   &\\
$\rm NH_2CHO~~^{~\dag}$    &$\rm \le 7.11(12)$    &$\rm 2.3 (13)$    &$\rm 5.46_ {\pm0.50}(14)$    &$\rm 1.8 (14)$    &$\rm 2.91_ {\pm0.13}(14)$     &$\rm 2.9 (14)$   &\\

$\rm OCS$    &$\rm 8.25_ {\pm0.11}(16)$    &$\rm 1.3 (16)$    &$\rm 5.39_ {\pm1.71}(16)$    &$\rm 2.5 (17)$    &$\rm 6.79_ {\pm0.14}(16)$    &$\rm 4.6 (16)$   &derived from $\rm O^{13}CS$ with $\rm ^{12}C/^{13}C=73^{\it a}$\\
$\rm SO_2$    &$\rm \le 4.29(14)$    &$\rm 1.2 (16)~^{\times}$    &$\rm \le 1.58(15)$    &$\rm 1.6 (16)$    &$\rm 3.05_ {\pm0.43}(17)$    &\color{black}$\rm 2.7 (16)~^{\times}$   &derived from $\rm ^{34}SO_2$ with $\rm ^{32}S/^{34}S=22^{\it b}$\\
$\rm SO$    &$\rm \le 1.93(15)$     &$\rm 1.8 (16)$    &$\rm 3.29_ {\pm0.58}(16)$    &$\rm 3.5 (16)$  &$\rm 3.57_ {\pm0.38}(17)$    &$\rm 3.4 (17)$   &derived from $\rm ^{33}SO$  with $\rm ^{32}S/^{33}S=127^{\it b}$\\

\hline
Agreement & & 10/15 = 67\%   & & 11/15 = 73\% & & 9/15 = 60\%   &\\%Tolerence: $\rm  N_{Modelled}/N_{Observed}$= 0.1--10\\ 
\hline
%Chemical Age & & 200 yrs  & & 10 yrs & & 450 yrs  \\ 
Age $^c$ & & 11200 yrs  & & 11210 yrs & & 11660 yrs  \\

\hline  
\multicolumn{8}{l} {{\bf Note.} {\it a}. derived from galactic local isotope ratio \citep{giannetti14};}\\
\multicolumn{8}{l} {~~~~~~~~~ {\it b}. derived from isotope ratio in solar system \citep{lodders03};} \\    
\multicolumn{8}{l} {~~~~~~~~~ {\it c}. {The physical time moment when our chemical code describes the observed column densities most accurately in our assumed physical structures.}}\\ 
\multicolumn{8}{l} {~~~~~~~~~ {\it d}. {``*" denotes the tentatively identified line;} }\\     
\multicolumn{8}{l} {~~~~~~~~~ {\it e}. {``$\rm \dag$" denote the  species which do not have rare isotopologues, so we derive their values by assuming they are optically thin;}}\\
\multicolumn{8}{l} {~~~~~~~~~ {\it f}. { ``$\times$" denote the  model predicted column densities inconsistent to the observation ($\rm  N_{Modeled}/N_{Observed}<0.1$ or $>10$);} }\\  
\multicolumn{8}{l} {~~~~~~~~~ {\it g}. {Uncertainties of the observed value are determined from $\rm T_{rot}$, partition function $\rm Q(T_{rot})$, and Gaussian fit to  $\rm \int T_B(\upsilon)d\upsilon$;} }\\
\multicolumn{8}{l} {~~~~~~~~~ {\it h}. {The values listed were obtained at the end of each evolutionary stage, corresponding  to the values used in Figure~\ref{model_fit}.}}
                       
\end{tabular}
}
\end{center}
\end{table*}

%%\onecolumn
  \begin{figure*}
  
  \small
\begin{center}
 %\begin{figure}
%\centering
\begin{tabular}{llll}
&\includegraphics[width=4cm,angle=-90]{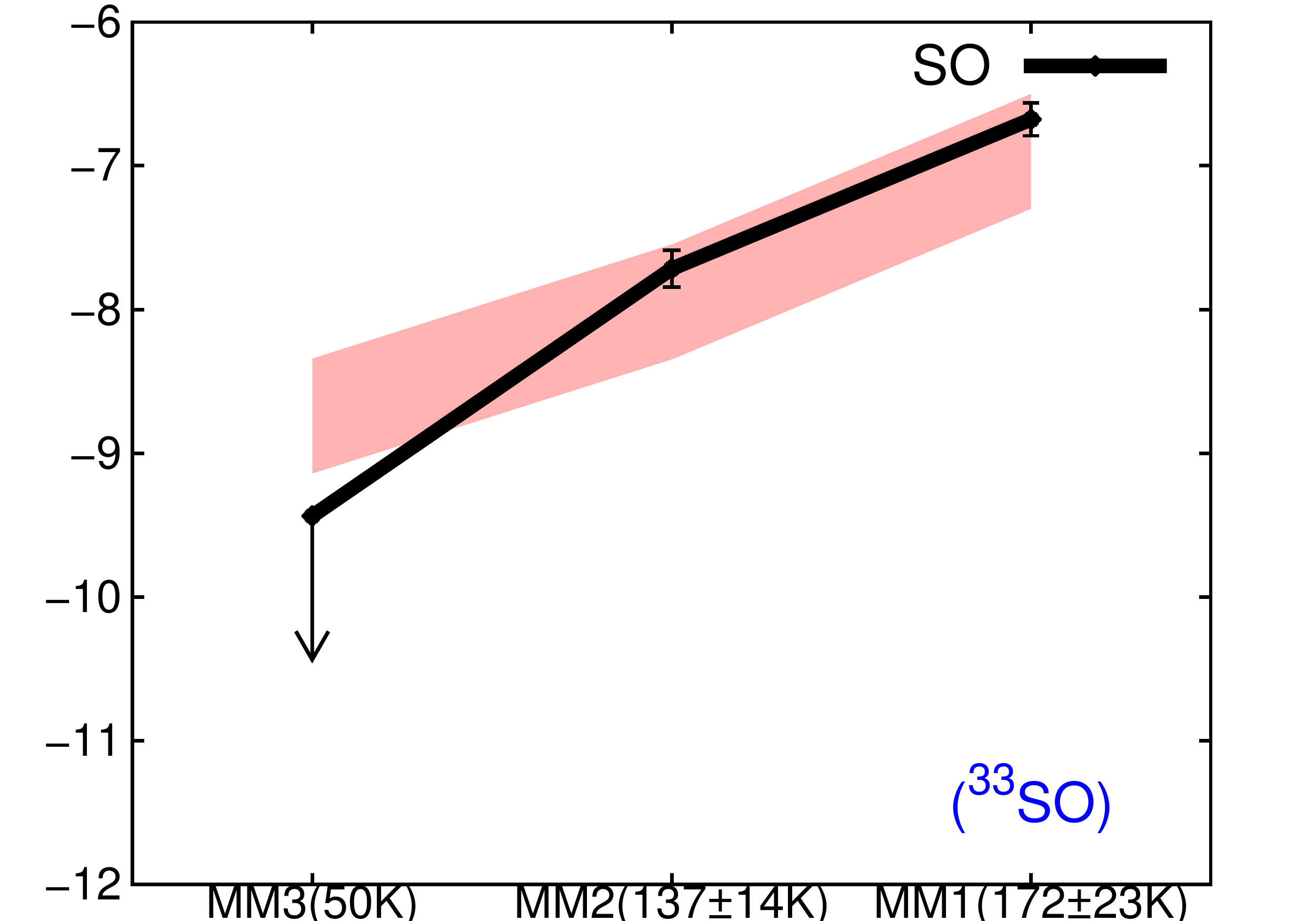}
&\includegraphics[width=4cm, angle=-90]{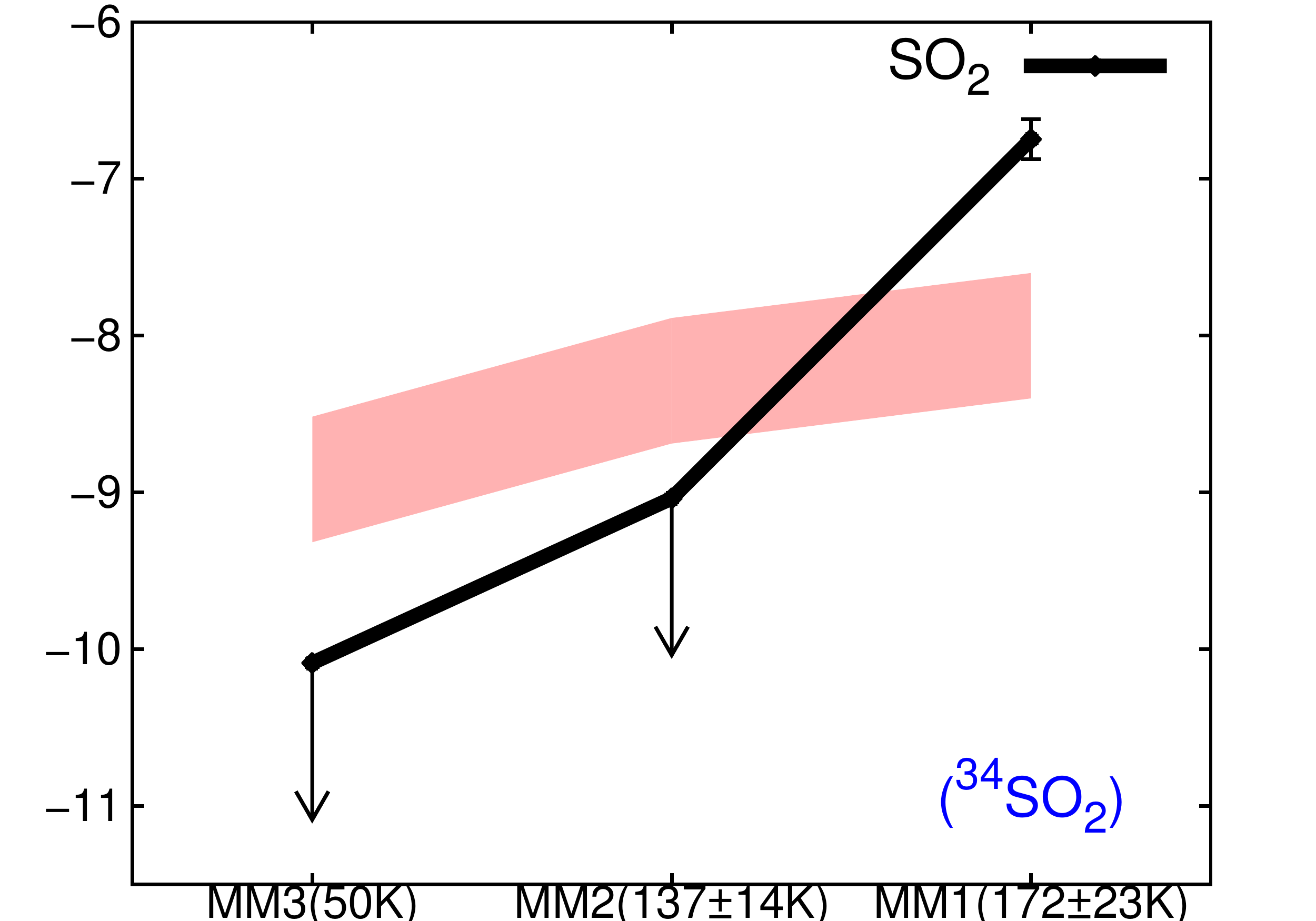}
&\includegraphics[width=4cm, angle=-90]{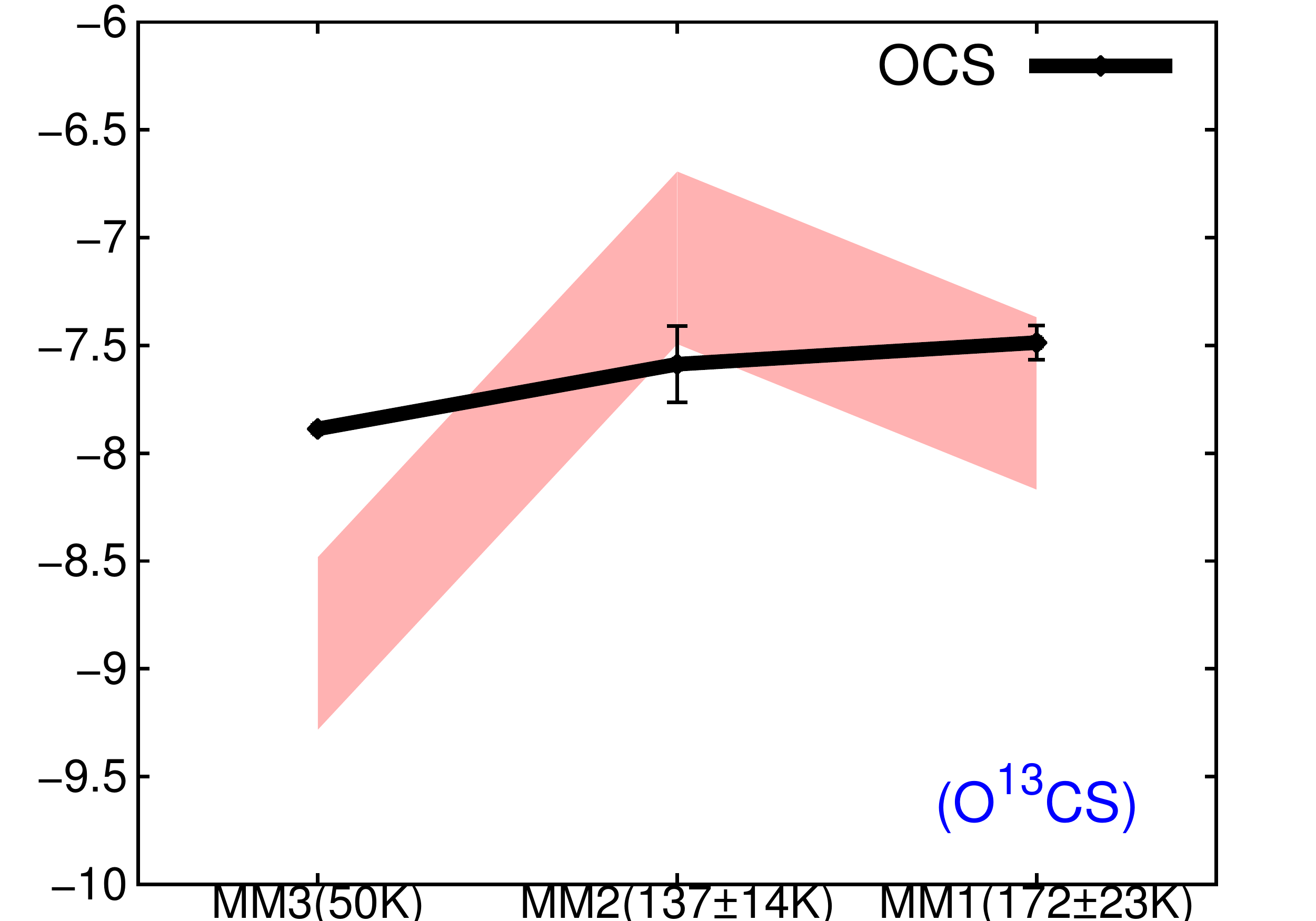}\\
\multirow{3}{*}{
 \begin{sideways}
\Large$\rm Log_{10}$  (molecular abundance with respect to $\rm H_2$)
 \end{sideways}
}
&\includegraphics[width=4cm,angle=-90]{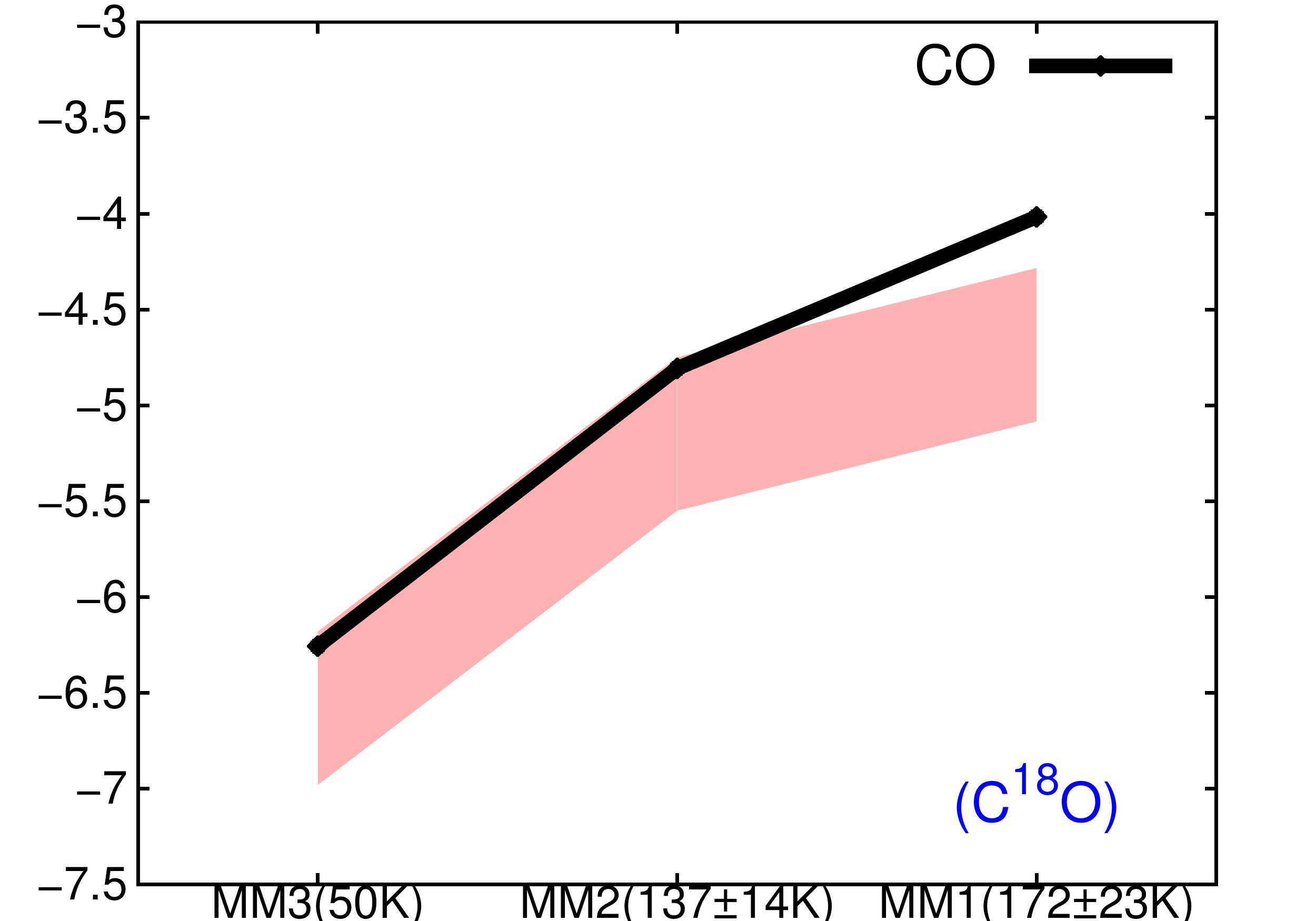}
&\includegraphics[width=4cm, angle=-90]{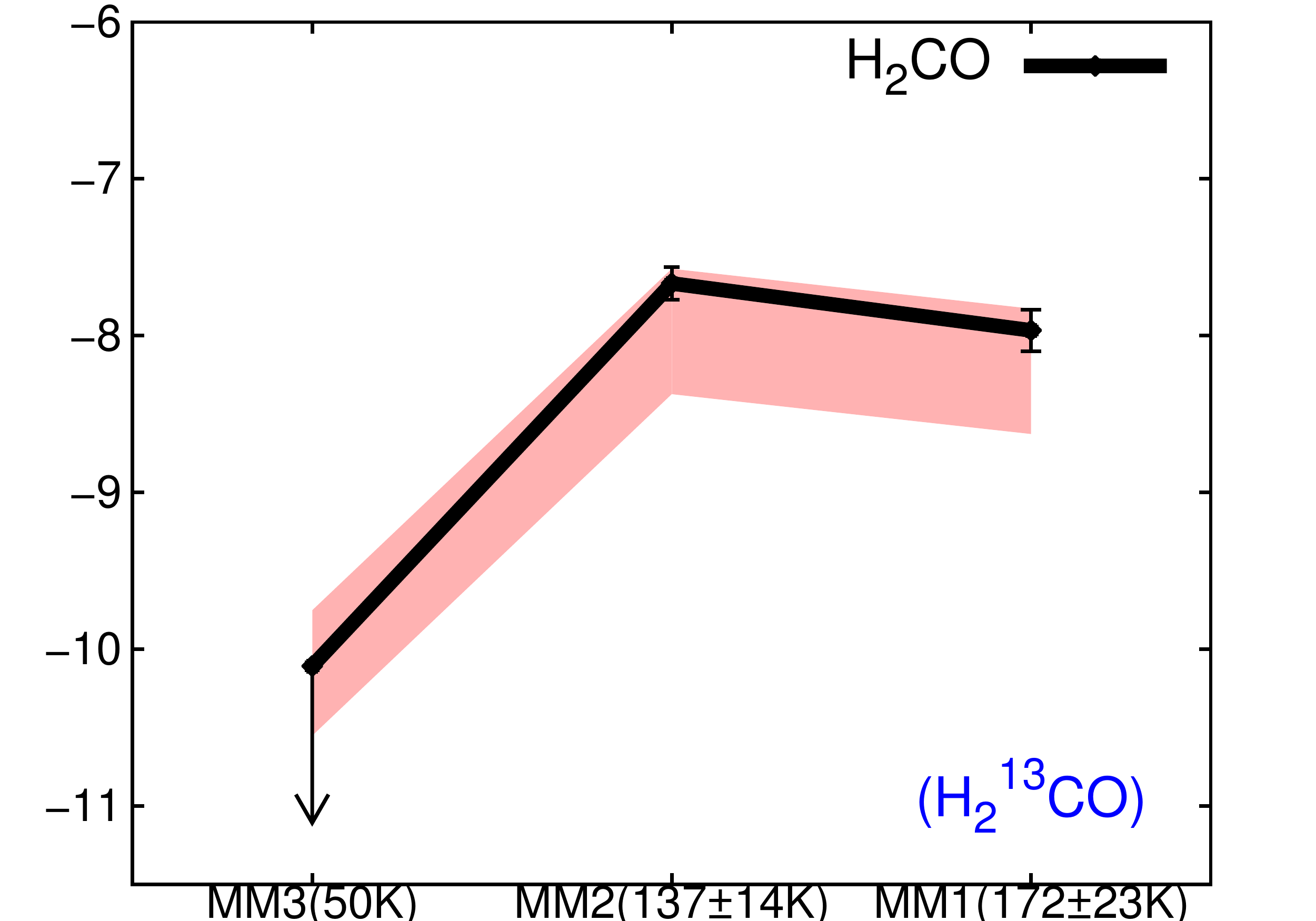}
&\includegraphics[width=4cm, angle=-90]{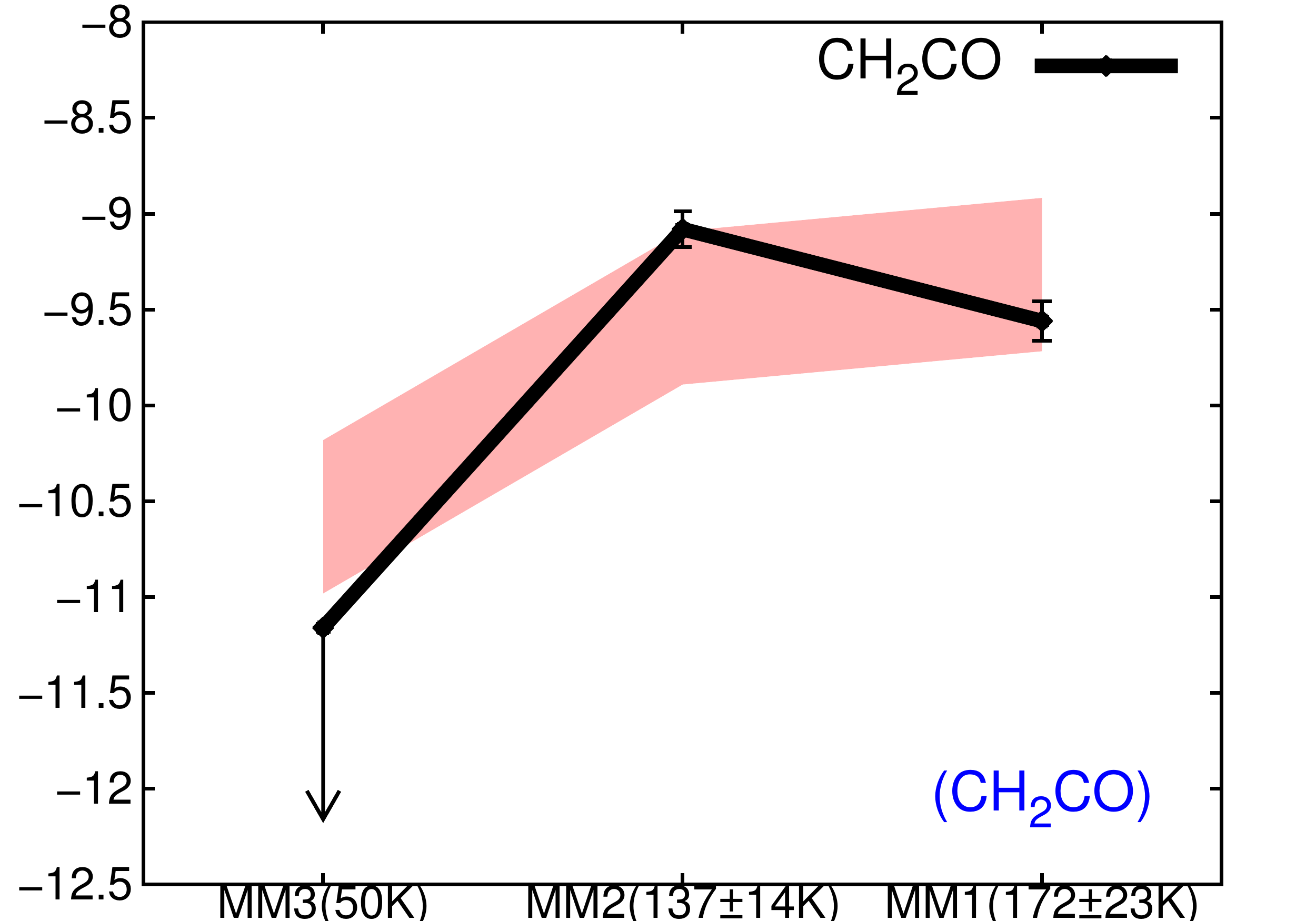}\\
&\includegraphics[width=4cm,angle=-90]{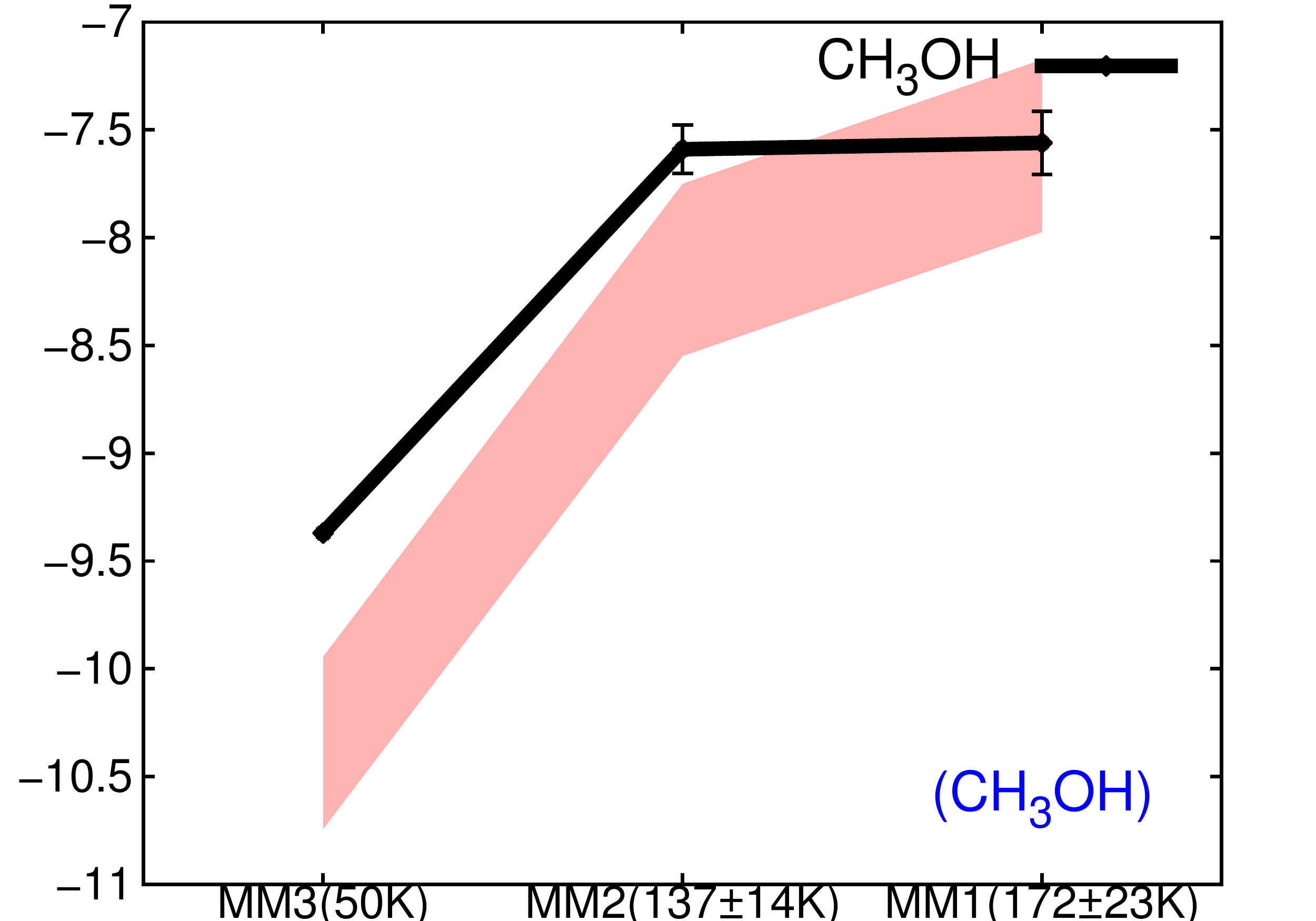}
&\includegraphics[width=4cm, angle=-90]{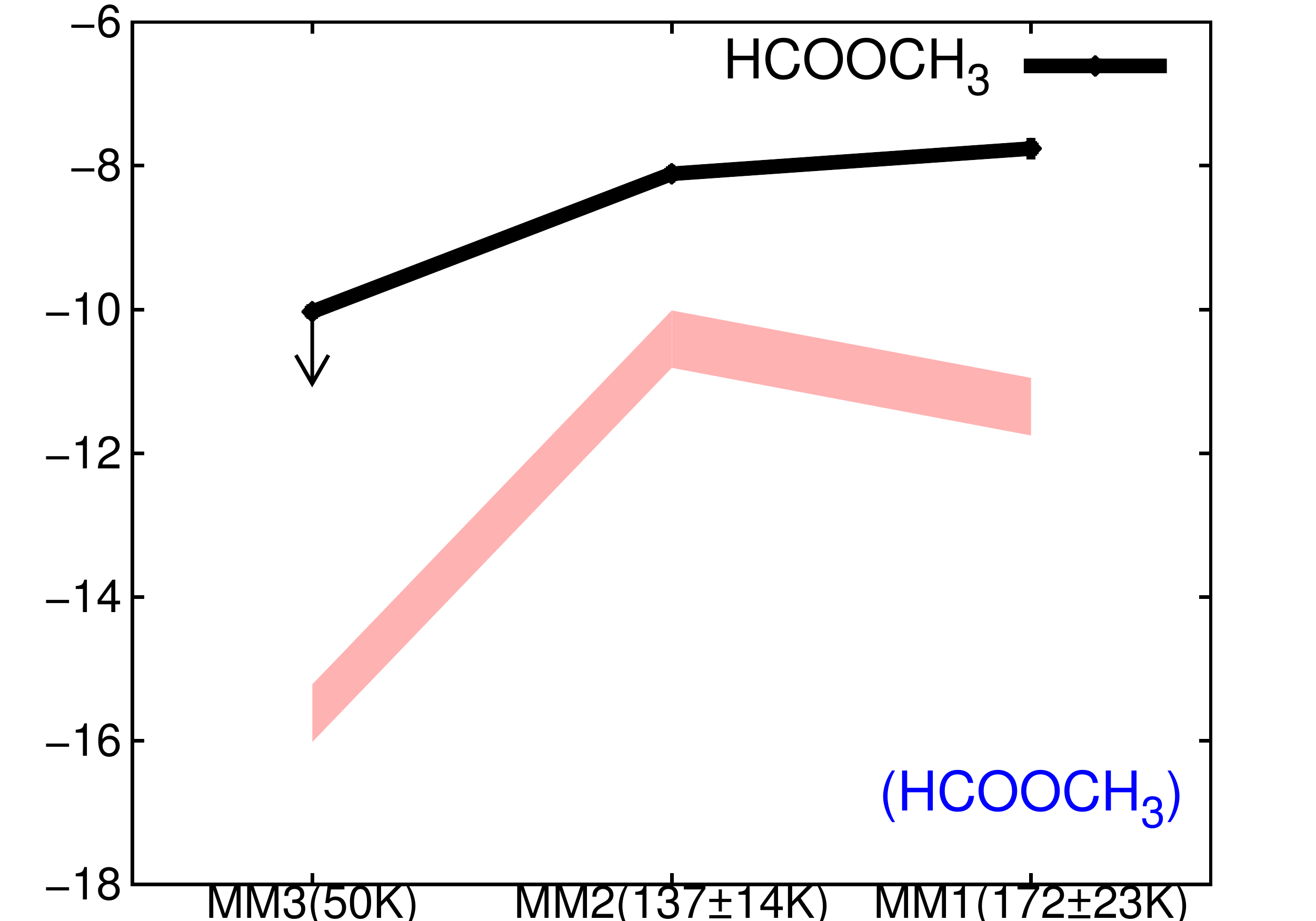}
&\includegraphics[width=4cm, angle=-90]{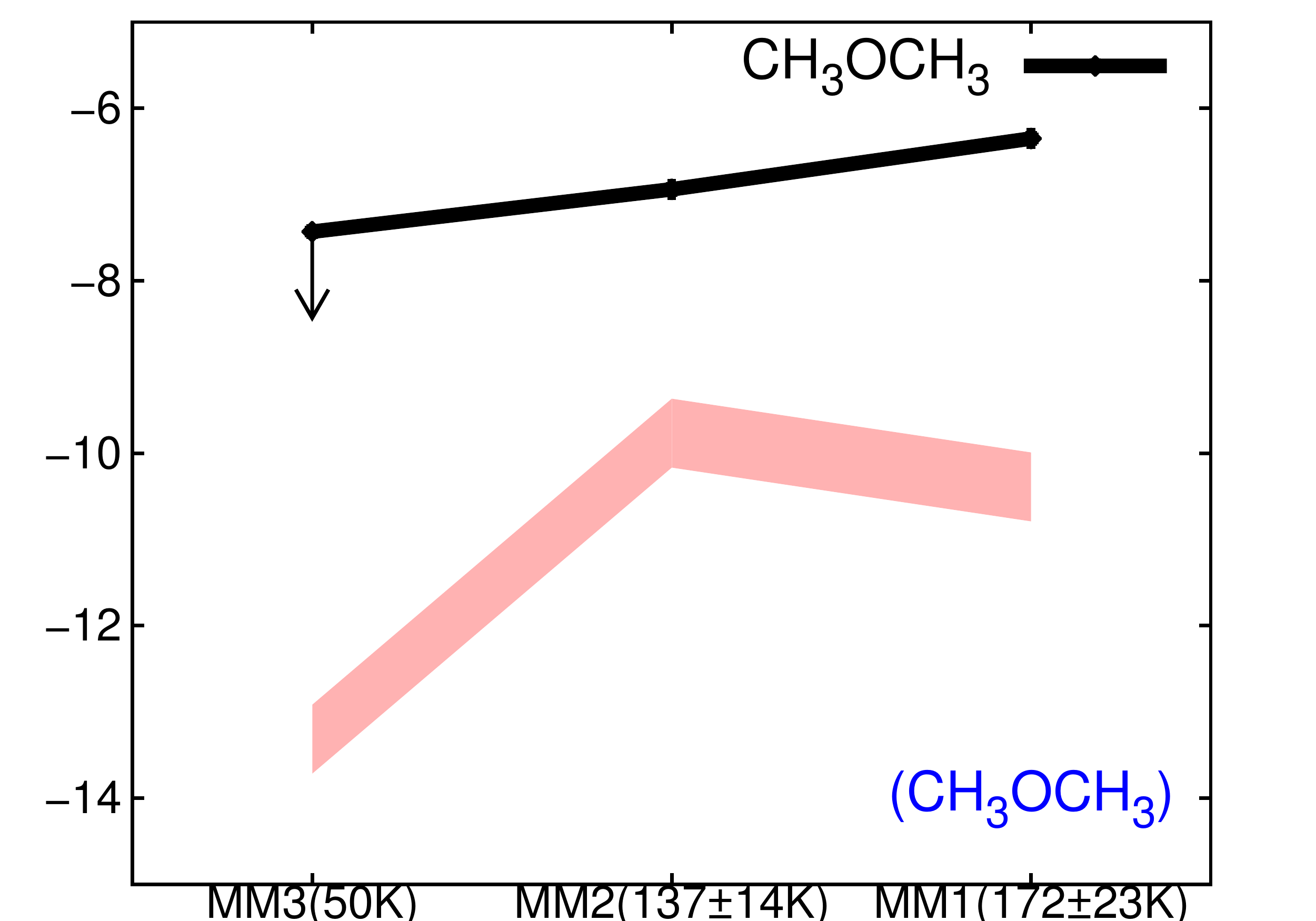}\\
&\includegraphics[width=4cm,angle=-90]{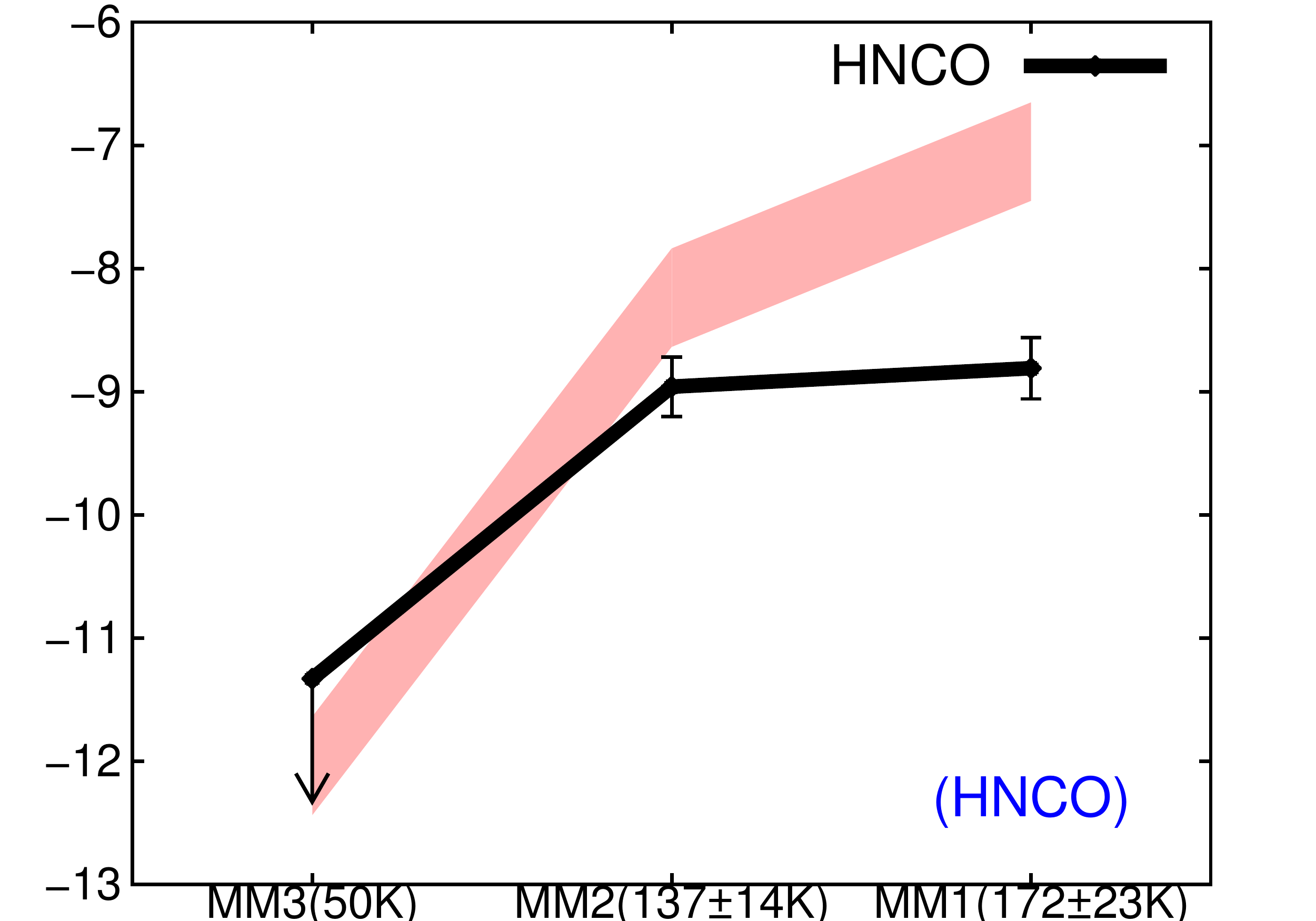}
&\includegraphics[width=4cm, angle=-90]{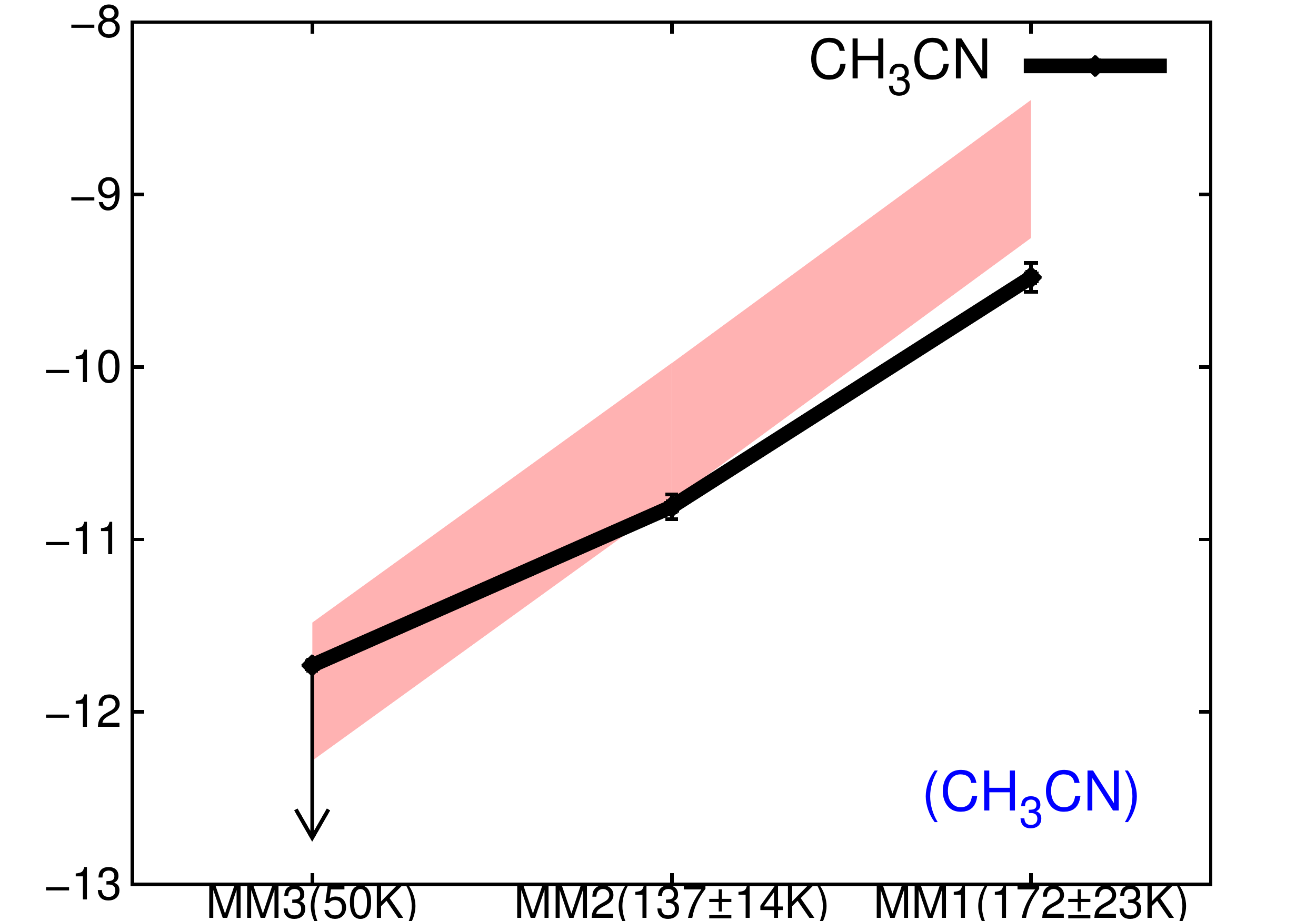}
&\includegraphics[width=4cm, angle=-90]{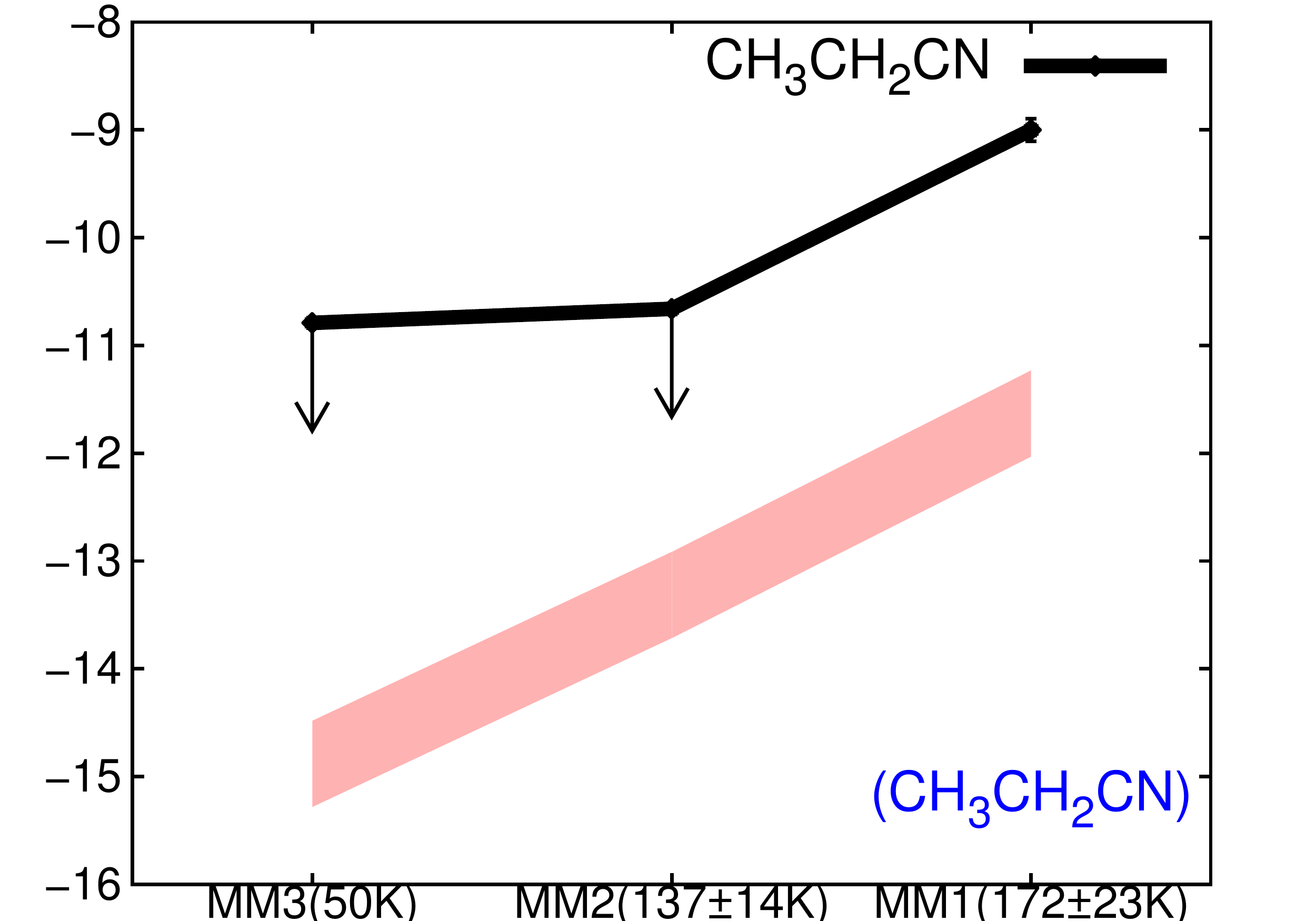}\\
&\includegraphics[width=4cm,angle=-90]{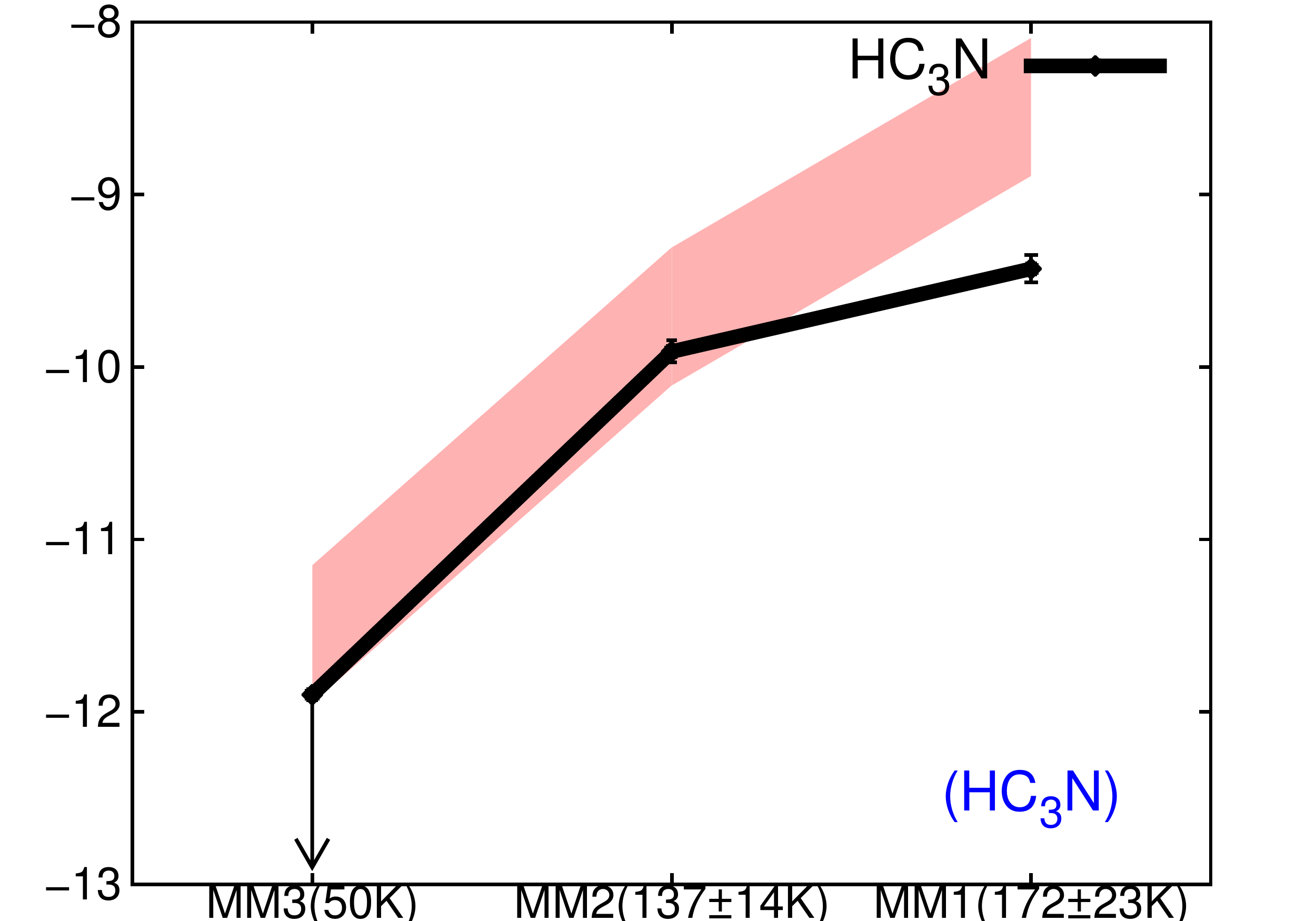}
&\includegraphics[width=4cm, angle=-90]{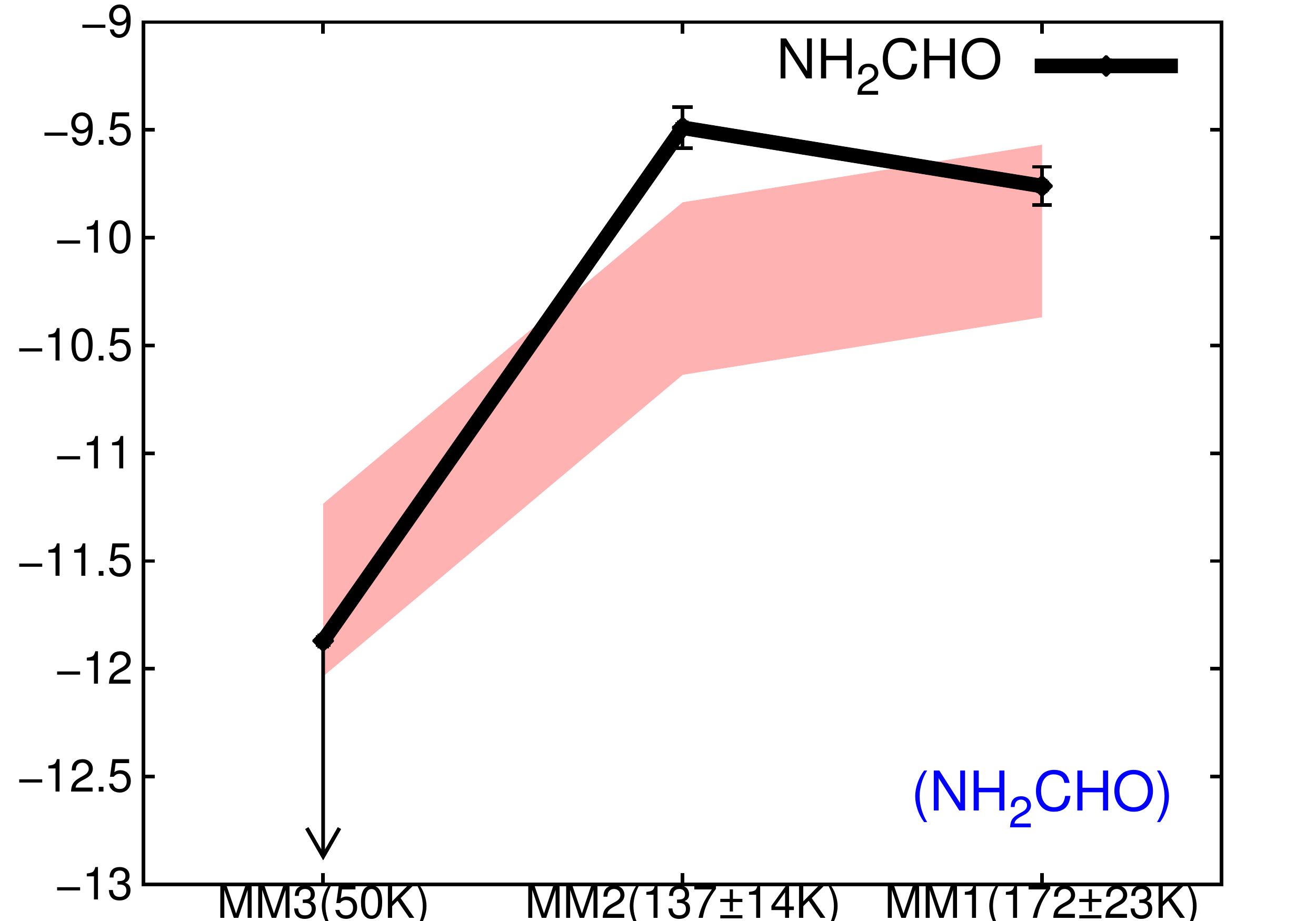}
&\includegraphics[width=4cm, angle=-90]{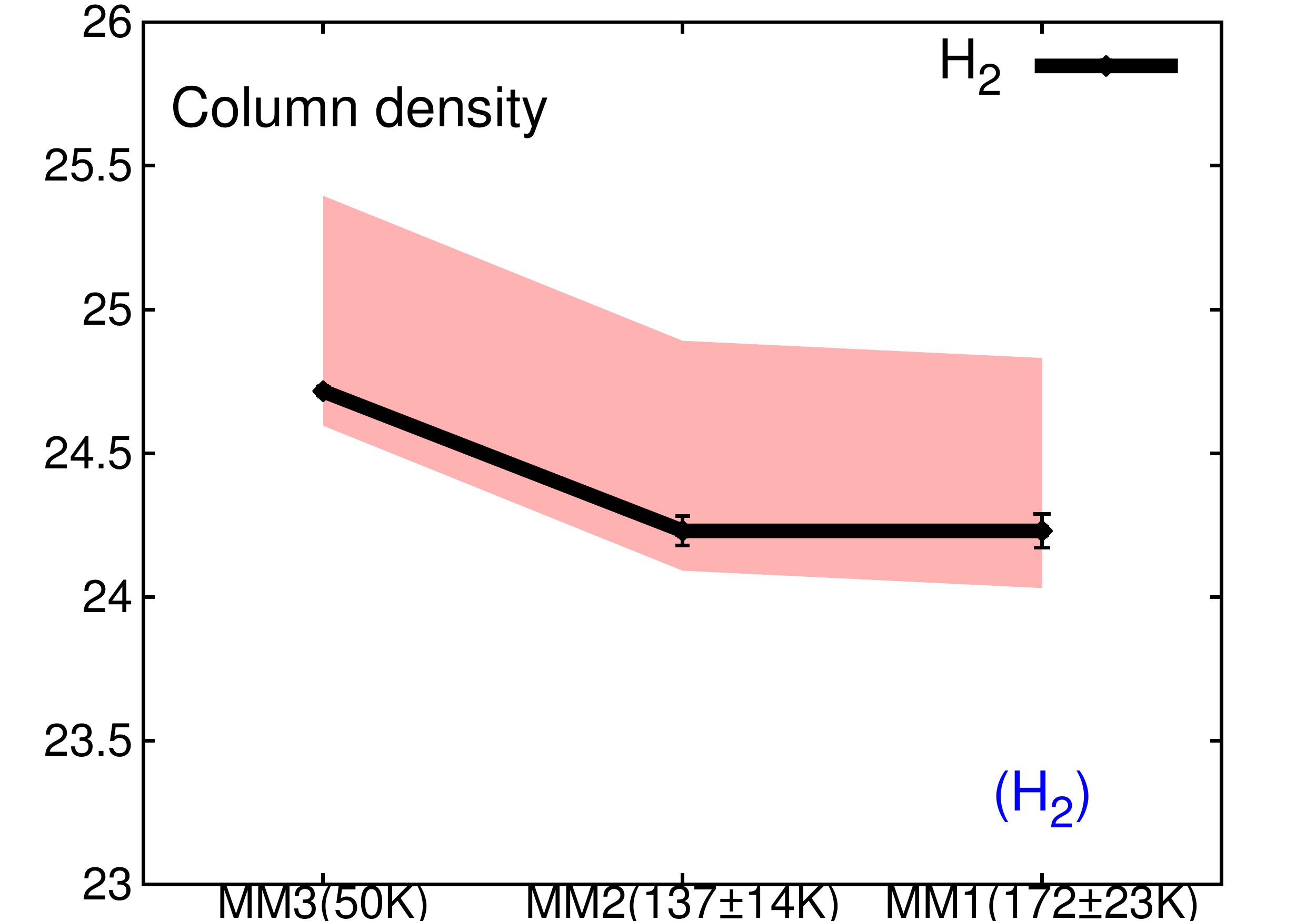}\\
%&&\Large$\triangle\alpha ['']$
\end{tabular}

\caption{Best fit of molecular abundances in MM1--MM3, with the chemical ages of 11200 yrs, 11210 yrs, and 11660 yrs, respectively. The last panel shows the best fit of $\rm H_2$ column density, whose observed values are derived from dust emission. The modelled values and tolerance  (0.1--10) are shown as a red filled {region}, while the observed values are plotted as solid black lines, with the uncertainties  determined from $\rm T_{rot}$, partition function $\rm Q(T_{rot})$, and Gaussian fit to  $\rm \int T_B(\upsilon)d\upsilon$ or scatter of transitions ($\rm CH_3CN$ and HNCO). Abundances of species (name in black at the upper-right corner of each panel) are derived  from their detected rare isotopologues, or opacity correction,  or from optically thin assumption, which are all named in blue in parentheses at the lower-right corner of each panel. 
 }\label{model_fit}

\end{center}
\end{figure*}

}
%%%%%%%%%%%%%%%%%%%%%%%%%%%%%
\section{Discussion}\label{dis}
This is the first comprehensive comparison of molecular lines observed at high spatial resolution with chemical models towards the NGC\,7538\,S and IRS1 regions. By resolving substructures we have unveiled chemical variation on a $\sim$1,100 AU scale between gas condensations which originate from the same natal gas cores. In the following section we discuss our findings.

 \subsection{Comparison with other results}
\subsubsection{Relation to the large-scale HMSFR studies}
The derived  gas column densities of the condensations (MM1--MM3) are comparable across all three condensations, $\rm (1\text{--}5)\times 10^{24}~ cm^{-2}$, and are also consistent with the derivations from the previous observations and simulations (e.g. \citealp{naranjo12}).\\

In addition, we have observed 7 species ($\rm C^{18}O$, $\rm H_2CO$, $\rm CH_3OH$, OCS, SO, HNCO, and $\rm CH_3CN$) which were previously observed in a survey of HMSFRs covering a broad range of evolutionary stages \citep{gerner14}. Due to the relatively coarse spatial resolution of the larger survey ($\rm 10^5$ AU), their observed chemical properties are potentially biassed towards the embedded, chemically most evolved gas substructures. For example, the averaged chemical and spectral line properties of the entire NGC\,7538\,S region, appear very similar to those of the condensation MM1 (a HMC).\\

Our results suggest that abundances of these species on the small scales of MM1, MM2, and MM3 all approximately agree with the abundances measured in large-scale HMCs, HMPOs, and prestellar objects, respectively. This not only supports  our hypothesis that MM3 is the youngest condensation, but also reinforces the idea of an evolutionary sequence from southwest (least evolved) to northeast (most evolved).\\

\subsubsection{Evolutionary stage of IRS1}
{In contrast to the resolved fragments in NGC\,7538\,S,  IRS1 remains an unresolved compact core at 1.37\,mm, which has a  much higher intensities of both continuum and lines than in NGC\,7538\,S. Based on the  $\rm H_2$ column density we estimated from $\rm C^{18}O$, we expect that the dust thermal continuum emission contributes to $\rm \sim0.07~Jy~beam^{-1}$ at the peak of the 1.37\,mm continuum image, roughly 3.6\% of the peak continuum intensity, 
 free-free emission may contribute to 96.4\% of the total flux (of 3.38 Jy), which is $\sim3.25$ Jy. }\\

{We estimated the $\rm CH_3CN$ rotational temperature from the off-peak position IRS1-mmS, taken this as the lower limit, and assumed the temperature from \citet{goddi15a} at higher resolution as the upper limit of IRS1-peak temperature. This gas temperature range is consistent with that in \citet{knez09, qiu11,zhu13}. Moreover, the lower limit column densities of HNCO, $\rm ^{13}CO$, and $\rm CH_3CN$  agree with the aforementioned study. Furthermore, the $\rm CH_3CN$ abundance at the IRS1-peak ($\rm \sim10^{-10}$) also agree with what are typically observed from HMCs \citep{hatchell98,chen06,zhang07}.}\\

As we have detected more COM lines in IRS1-mmS than in MM1 (Table \ref{tab:line}) and found that IRS1 has the highest luminosity in the NGC\,7538 region \citep{scoville86}, we consider IRS1 to be chemically and dynamically more evolved than MM1. This is in line with the notion that at larger scales the evolutionary sequence proceeds from the southeast (``youngest") to northwest (``oldest")  \citep{mcCaughrean91}.\\

\subsection{Hierarchical fragmentation in NGC\,7538\,S: synthesis of observations and models}\label{dif}

The giant molecular cloud NGC\,7538 fragments into at least three bright gas cores, namely IRS1, IRS9, and NGC\,7538\,S. The large-scale gas core NGC\,7538\,S continues to fragment, agreeing with the picture of ``hierarchical fragmentation". The separations of the three spatially resolved internal condensations in NGC\,7538\,S, are on the order of 3,800-5,000 AU. {Given that} these condensations are in the same gas core and  have similar projected sizes and column densities, they are likely to have similar dynamic ages. The significant variation in molecular line intensities and abundances among them indicate an evolutionary sequence from  MM3 through MM2 to MM1.

\subsubsection{MM1--MM3 evolutionary stages}
MM1 is characterized as a typical HMC owing to its high temperature ($\rm >150~ K$), high optical depth and high abundance of COMs. In fact, its presence of bright IR emission suggests that this condensation is at a chemically more evolved stage in chemical terms compared to many other HMCs (e.g. \citealt{chen06,zhang07,girart09,qiu09}). Previous VLA observations \citep{wright12} towards MM2 and MM3 did not detect radio continuum sources, suggesting that they are in early protostellar phases.  We observed that some species   in MM2 exhibit less intensity of emission than in MM1 (e.g.  $\rm H_2CO$, $\rm CH_3OH$, $\rm HCOOCH_3$, OCS), while some species exhibit the strongest emission in MM2, but they are barely detected in MM3 (e.g.   $\rm H_2^{13}CO$, $\rm CH_2CO$, $\rm NH_2CHO$, {which are evaporated to gas  at $\rm \sim40~K$ and may further interact with other radicals at higher temperature to produce the more complex molecules}, see Figure~8(d) of \citealt{garrod08}). This suggests that MM2 should be chemically more evolved than MM3. As a result, we assert MM2 could be in the HMPO stage while MM3 is potentially still a  prestellar object. {While the temperatures of MM3 around 50\,K are too high for a genuine isolated prestellar object, in the cases of the two neighboring more evolved cores, the elevated temperatures in MM3 may also be caused by radiation interactions of these neighbors.}\\

\subsubsection{Warm-up histories}
We adopt the terminology of  ``chemical warm-up timescale" from \citet{garrod08}. This timescale in our model  refers to the period over which the temperature  {within the inner radius $\rm r_{in}$} of a source increases from an initial value (e.g. 10-20 K) to a higher value (e.g. $>100$ K)\footnote{In fact, the entire process is more complicated than just the increase of temperature. When a prestellar object is forming, it may initially be warmer, $\sim$20 K, and then it cools off as it becomes more dense to $\rm T\sim$10 K (in the center). When collapse begins, T rises from 10 K to several hundreds K. 
}. 
{We found temperature rises steeply and not linearly with time in our model.}
We derive a general picture of hierarchical fragmentation in NGC\,7538\,S.  \\

While we are unable to distinguish whether MM1--MM3 are originated from a massive and unstable rotating gas torus, or are formed via fragmentation of a dusty core \citep{corder09,wright12,naranjo12}, both scenarios will lead to a similar ``dynamic age" (i.e. collapsing simultaneously) of these condensations, and therefore do not change our interpretation of chemical evolution. As our model shows, differences in the chemical age of these condensations are small: $\rm \sim10-10^2$ years. Such slight variations in timescale may be caused by the different physical environments \citep{ballesteros07}, e.g. turbulence, gravitational potential, magnetic fields, etc., which can cause different evolution speeds. In our picture, MM1 warms up faster than MM2 and MM3. At a particular physical age ($\rm \sim10^4$ years), 
when the abundances of $\rm H_2CO$, $\rm CH_2CO$, and $\rm NH_2CHO$ reach their maximum in MM2, their abundances already begin to decrease in MM1 because they are reacting to form the more complex, second-generation molecules, while MM3 is still in chemically the least evolved phase and most of the molecules are still on the grain surfaces due to its slowest speed.

%%%%%%%%%%%%%%%%%%%%%%%%%%%%%%%%%%%%%%%%%%%%
%%%%%%%%%%%%%%%%%%%%%%%%%%%%%%%%%%%%%%%%%%%%
\subsection{The absorption and emission feature in IRS1 }\label{dif}
\subsubsection{General P-Cygni and absorption profiles}
Most of the observed molecular lines at 1.37\,mm display {P-Cygni or inverse P-Cygni /pure absorption} features against the bright continuum source IRS1 within a 0.4\arcsec area. In  previous kinematics studies {the P-Cygni feature was interpreted as associating with jet(s)/outflow(s) from a disk-like structure, while the inverse P-Cygni and pure absorption features were interpreted by rotation and infalling gas \citep{wilson83,henkel84,sandell09,qiu11,beuther12,beuther13,zhu13}.}
However, whether the protostellar disk is face-on \citep{kawabe92,beuther13} or edge-on (e.g. \citealt{pestalozzi04,lugo04,de05,knez09}), whether the direction of the associated outflow is perpendicular (e.g. \citealt{de05,knez09,zhu13}) or along  (e.g. \citealt{beuther13})  our line of sight or  precessing (e.g. \citealt{kraus06}), and  whether the IRS1-peak is a single {protostar} or binary system \citep{moscadelli09,moscadelli14,goddi15a} {are still under debate}.\\

\subsubsection{Pure emission lines tentatively from $\rm CH_3OH$ and $\rm HCOOCH_3$}

\begin{figure}
\centering
\includegraphics[width=8cm]{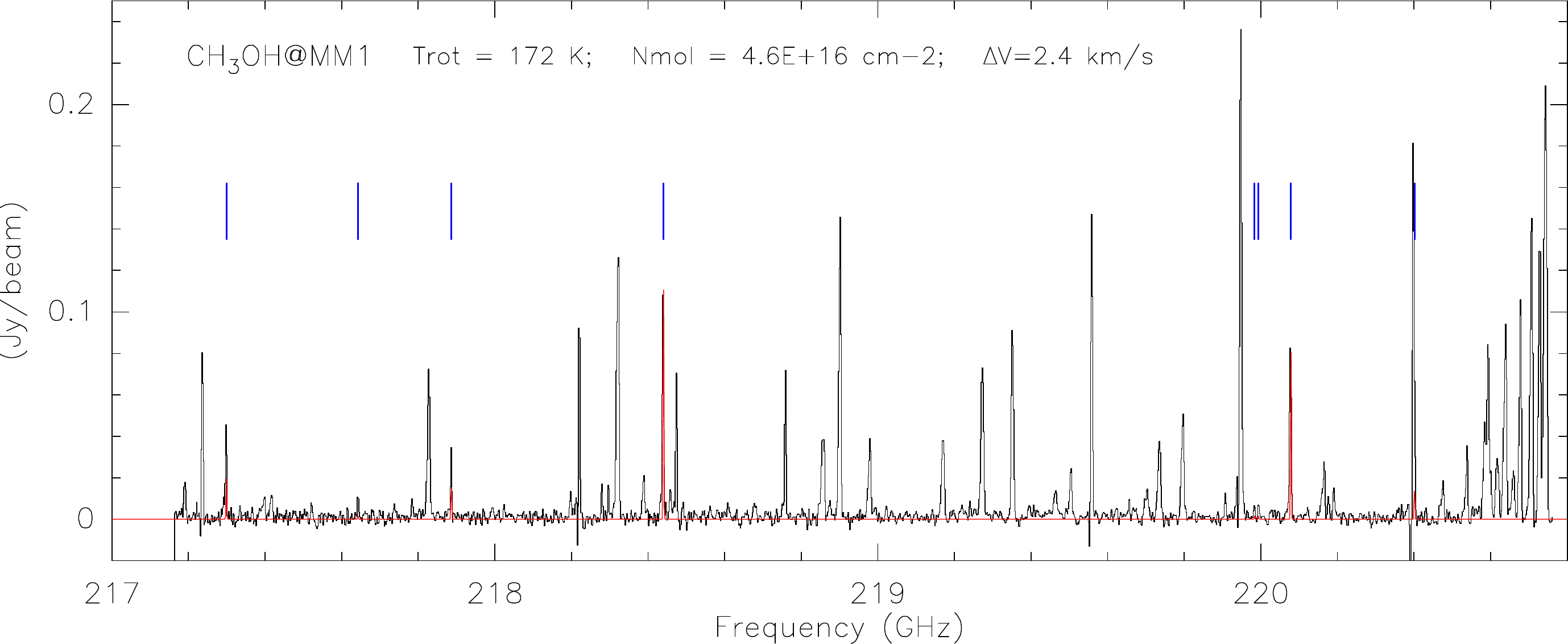}
\includegraphics[width=9cm]{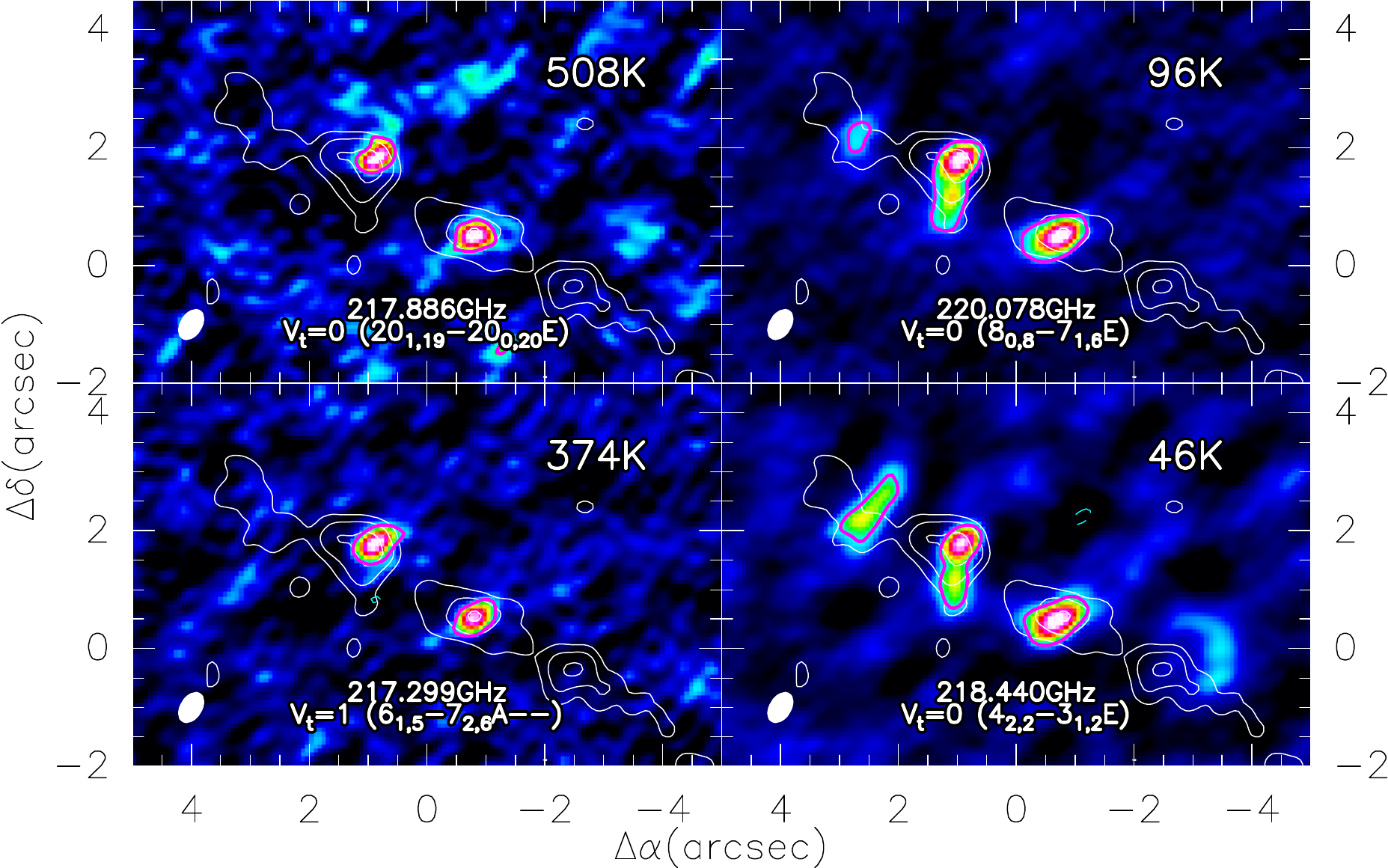}

\caption{The  upper panel shows {a synthetic  spectrum of $\rm CH_3OH$ fitted to the MM1 data}. The black histogram is the observed spectrum, overlaid with the red fit {whose parameters are listed}. Blue lines mark transition frequencies, and confirm that  lines at 217.299 GHz, 217.643 GHz, 217.886 GHz are $\rm CH_3OH$. The lower panel presents intensity  integrated maps over the velocity range from {\color{black}-65 to -50 $\rm km~ s^{-1}$}  {\color{black} ($\rm V_{lsr}\sim-60~ km~ s^{-1}$ in MM1)}, with  red contours showing the $\rm 3\sigma$ rms. White contours show the continuum (at 4$\sigma$,  14$\sigma$, and  24$\sigma$ levels).   \label{ch3oh}}
\end{figure}

\begin{figure}
\begin{center}
\begin{tabular}{ll}
\multicolumn{2}{l}{\includegraphics[width=8cm]{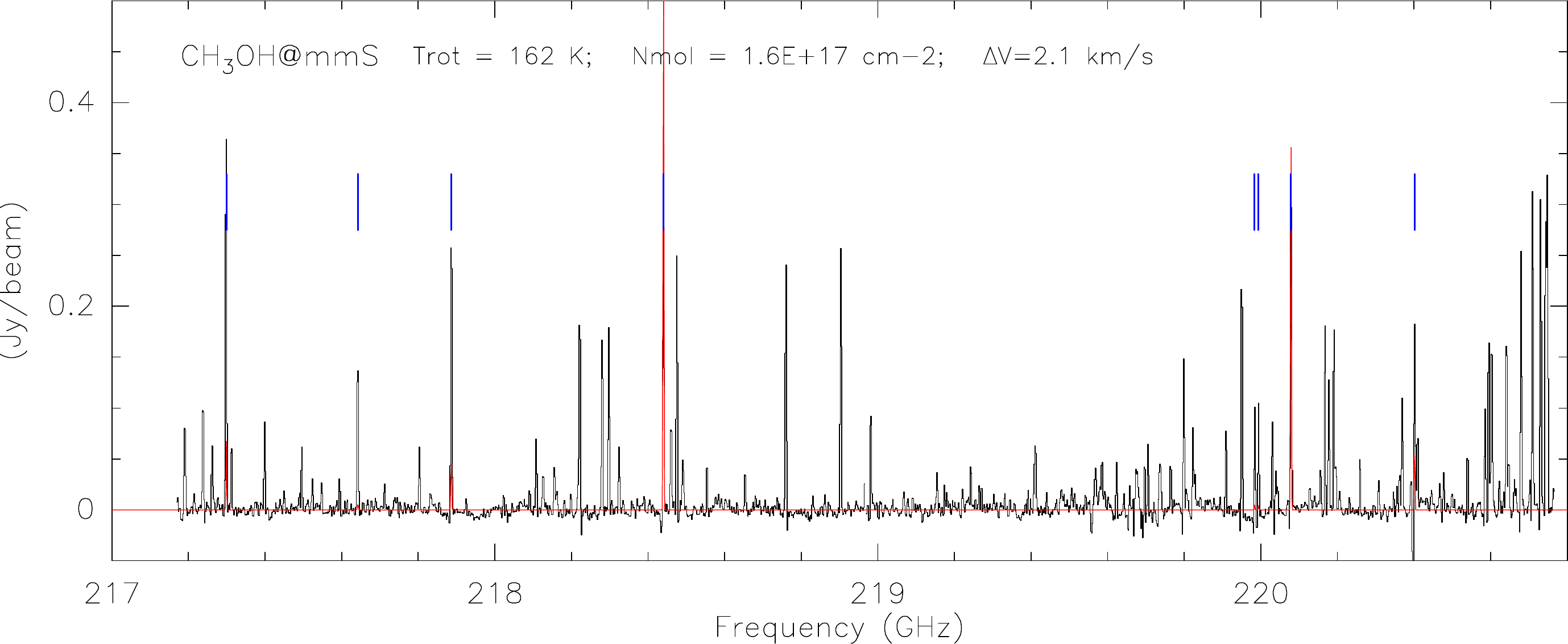}}\\
\includegraphics[width=4cm]{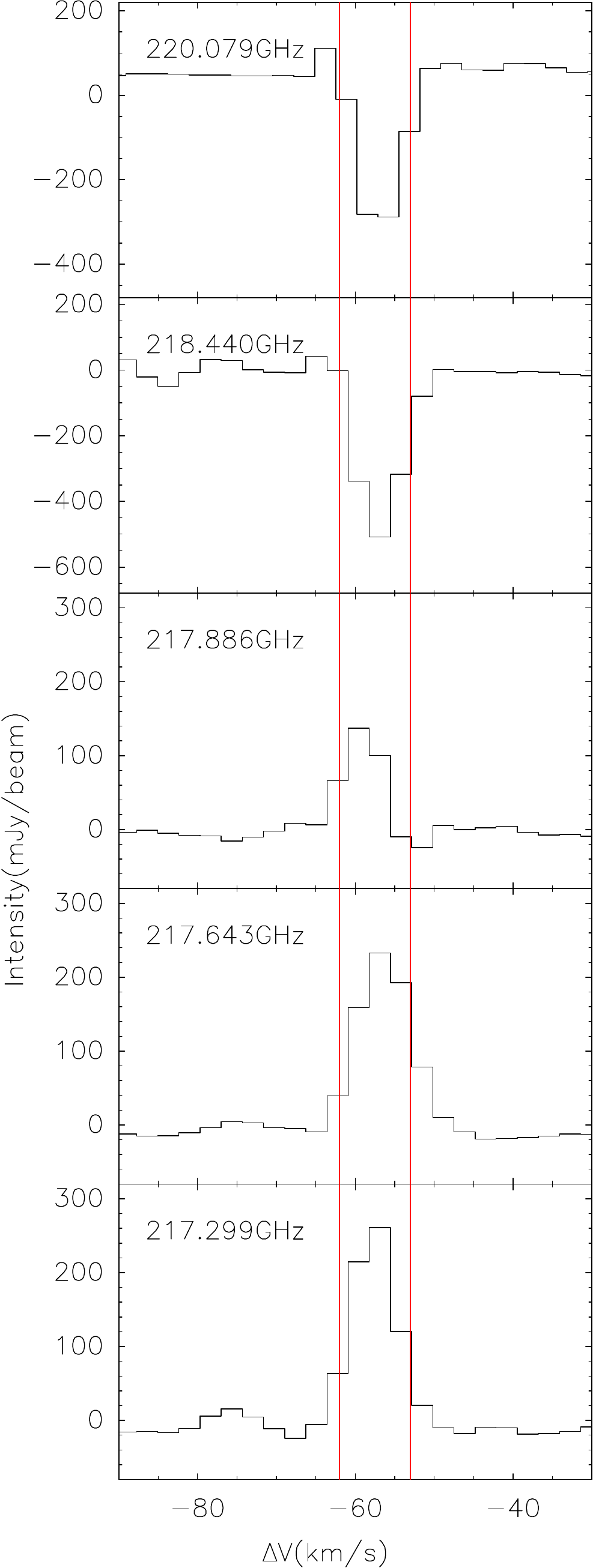}
&\includegraphics[width=2.6cm]{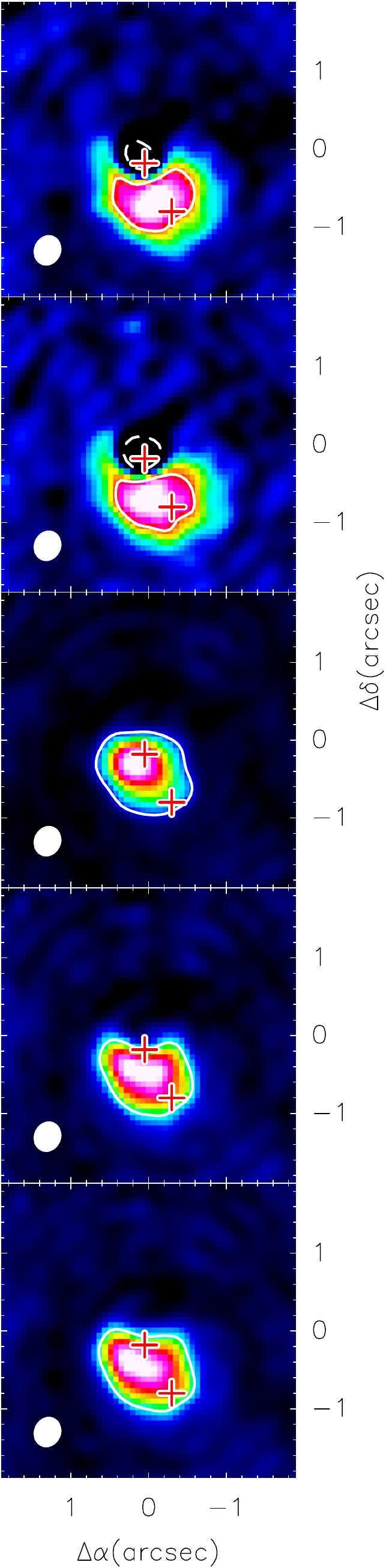}
\end{tabular}
\caption{The  upper panel shows {the fitted $\rm CH_3OH$ spectrum in IRS1-mmS}. The black histogram is the observed spectrum, overlaid with the red fit {whose parameters are listed}. Blue lines mark transition frequencies at 217.299 GHz, 217.643 GHz, 217.886 GHz. The lower {left panel shows the line profiles of $\rm CH_3OH$ transitions from the IRS1-peak; and the right} panel presents {corresponding intensity maps integrated over} the velocity range  from {\color{black}-63 to -53 $\rm km~ s^{-1}$}  {\color{black} ($\rm V_{lsr}\sim-58~ km~ s^{-1}$ at the continuum peak)}. {The red} crosses mark the IRS1-peak position and IRS1-mmS, and the yellow solid (dashed) contour {shows} $\rm \pm 4\sigma$ rms level.  \label{ch3oh-irs1}}
\end{center}
\end{figure}

At least four bright molecular lines do not  show absorption  toward the IRS1-peak (marked with ``?" in Figure~\ref{spec}). Since these lines are also detected from MM1 and IRS1-mmS, we fit them by the synthetic spectrum programme and map their spatial distribution in NGC\,7538\,S (Figure~\ref{ch3oh}) and IRS1 (Figure~\ref{ch3oh-irs1}). We propose that three of them (at  217.299 GHz, 217.643 GHz, and 217.886 GHz, {Table ~\ref{tab:ch3oh}}) are $\rm CH_3OH$ transitions with FWHM$\rm \sim2.5~km~s^{-1}$. This is supported by the emission of another $\rm CH_3OH$ transition at the IRS1-peak  (at 356.007 GHz, \citealt{beuther13}). We note that these four $\rm CH_3OH$ lines have upper state energy levels $\rm E_u/k>250~K$, 
 while the other two absorption E-type $\rm CH_3OH$  lines 
  have $\rm E_u/k<100~K$.\\

Moreover, we also note that some $\rm HCOOCH_3$ transitions also show tentatively ($\rm \sim3\sigma$ rms) pure emission toward the IRS1-peak (denoted in Table.~\ref{tab:line} and shown in Figure~\ref{hcooch3}). Compared to the $\rm CH_3OH$ lines, the emission of  $\rm HCOOCH_3$ lines are weaker.  
Eight out of ten  reliable pure emission $\rm HCOOCH_3$ lines are torsionally excited $\rm (\nu_t=1)$ with $\rm E_u/k>250~K$. It is likely that they have the same origin as the pure emission $\rm CH_3OH$  transitions. 

\begin{table}
\centering
\caption{Parameters of detected  $\rm CH_3OH$ lines from CDMS \label{tab:ch3oh}}
\begin{tabular}{lrcrcrcrcr}
\hline
\hline
\multicolumn{1}{c}{$\nu$}&\multicolumn{1}{c}{Transition}&\multicolumn{1}{c}{$\rm E_u/k$}&\\
\multicolumn{1}{c}{(MHz)}&\multicolumn{1}{c}{$\rm $}&\multicolumn{1}{c}{(K)}&\\
\hline
\hline
217299  &$\rm \nu_t=1(6_{1,5}\rightarrow7_{2,6}A--)$ &374   &emission\\
217643  &$\rm \nu_t=1(15_{6,9}\rightarrow16_{5,12}A++)$  &746   &emission\\
&$\rm \nu_t=1(15_{6,10}\rightarrow16_{5,11}A--)$  &&\\
217886  &$\rm \nu_t=0(20_{1,19}\rightarrow20_{0,20}E)$ &508   &emission\\
218440  &$\rm \nu_t=0(4_{2,2}\rightarrow3_{1,2}E)$ &46   &absorption\\
220079  &$\rm \nu_t=0(8_{0,8}\rightarrow7_{1,6}E)$ &96   &absorption\\
356007$^*$  &$\rm \nu_t=0(15_{1,14}\rightarrow15_{0,15}A-+)$ &295   &emission\\

\hline
\hline
\multicolumn{4}{l}{``*" from \citet{beuther13} }\\
\end{tabular}
\end{table}

\subsubsection{Proposed explanations for the emission lines at IRS1-peak}
One hypothesis for the simultaneous absorption and  emission  is that the outflow penetrates the envelope, producing a collimated, jet-like cavity along the line of sight. Here, transitions with higher  $\rm E_u/k$  from the inner hot core are detected in emission; the other lines of $\rm CH_3OH$ with lower  $\rm E_u/k$ are produced outside of the cavity, and are absorbed by the colder envelope.  Our observations with limited spectral resolution did not reveal a velocity offset, 
higher angular resolution cm observations ($\rm \sim200$ AU) have resolved  the IRS1-peak into two components, showing $\rm \sim3 ~km~s^{-1}$  $\rm CH_3OH$ maser radial velocity offset \citep{moscadelli09,goddi15a}. As a result, different structural components might explain the {co-existence of the  absorption and pure emission} line features.\\

{Moreover, we note from Figure~\ref{ch3oh-irs1} that the line at 217.886\,GHz is the only $\rm CH_3OH$ line that emits and peaks unambiguously at the IRS1-peak, and that the other two emission lines seem $\rm \sim0.5\arcsec$ offset southern to IRS1-peak. Therefore, a second hypothesis may be that  the $\rm CH_3OH$ absorption is due to a colder environment, while IRS1-mmS might be the foreground relative to the UCHII.}\\

In addition to the possible effects of the source geometry or {different origins of multi-transitions}, a third possibility to explain the pure emission lines are non-LTE effects or masering. The strong  and  compact continuum radiation of IRS1 can excite most molecular lines with lower $\rm E_u/k$, pumping them to higher rotational levels. 
$\rm CH_3OH$ masers are traditionally classified into two groups \citep{menten91}: collisionally pumped \citep{cragg92}  Class I and  radiatively pumped  Class II (\citealt{sobolev07} and references therein, \citealt{cragg05}). Class I masers  trace shocks from molecular outflows and jets during the earliest stages of star formation \citep{plambeck90, voronkov06,sutton04}. Class II masers are typically associated with 
hyper-compact HII regions (e.g. \citealt{minier01, ellingsen06}), where strong free-free emission  can dominate the dust emission up to frequencies as high as 300 GHz \citep{kurtz05, sobolev07}. Although there are no other observations confirming which class of masers our observed $\rm CH_3OH$ emission lines may be, cm large-scale ($\rm FWHP\sim43\arcsec$) observations of IRS1 suggest that a different torsionally excited $\rm CH_3OH$ emission line ($\rm 9_2\rightarrow10_1A+$)  a probable Class II maser \citep{wilson84}, and another  torsionally excited $\rm CH_3OH$ absorption line ($\rm 10_1\rightarrow11_2A+$) is not in LTE \citep{menten86}. Therefore,  ``quasi-masing"\footnote{A non-LTE status of the line with high $\rm E_u/k$, which can be detected in much lower temperature environment (than  $\rm E_u/k$) because of certain pumping mechanism.} may induce the emission from torsionally excited  $\rm CH_3OH$.\\

%%%%%%%%%%%%%%%%%%%%%%%%%%%%%%%%%%%%%%%%%%%%
\section{Conclusion}\label{conclusion}
We have performed high angular resolution observations of the two high-mass star-forming gas cores NGC\,7538\,S and IRS1.
 {These observations were complemented with a suite of chemical models with the aim of disentangling the chemical evolution of the individual sources.} Our main conclusions are summarized as follows.\\

\begin{enumerate}
\item NGC\,7538\,S is one of the brightest gas fragments in the NGC\,7538 molecular cloud. Our 1.37\,mm observations with  $\sim1,100$ AU (0.4\arcsec) spatial resolution have resolved this region into at least three gas condensations. Each condensation exhibits similar continuum specific intensities, but significantly different chemical complexity. We covered over 90 molecular lines from 15 species, including 22 isotopologues;  most showed emission peaks at MM1. A few ($\rm H_2^{13}CO$, $\rm CH_2CO$, and $\rm NH_2CHO$) have {higher intensity emission} in MM2
and are relatively weak or undetected in MM3. Comparing our chemical study on small scales ($\rm 10^3\,AU$) to the previous large sample statistics on large scales  ($\rm 10^5\,AU$), we found that the molecular abundances in MM1, MM2, and MM3  generally agree with the values measured in HMCs, HMPOs, and IRDCs (prestellar objects), respectively.

\item Comparing our observations with a 1-D gas-grain chemical model, we propose an evolutionary sequence along the northeast-southwest filamentary NGC\,7538\,S. MM1--MM3  may be  fragments formed from a global collapse within $\sim$ {20,000} AU. However, the heating paces from their internal protostars /prestellar objects are likely  different. Slight variations may be caused by mechanisms such as turbulence, gravitational potential, or magnetic fields. Our chemical model suggests that the protostar in MM1 ignited first, 
making it chemically the most evolved source. MM2 has {an intermediate warm-up speed} while  the central object in MM3 may still be a prestellar object. 
Although the chemical histories of MM1--MM3 are not synchronized, the differences in {their chemical ages are small (several hundred years), indicating that the chemical evolution of HMSFR is sensitive to the warm-up speed. Evolution from an MM3-like condensation to an MM1-like condensation} can be  rapid. {Given our model limitations, evolution may take longer time, so precise dating of the central condensations of HMSFRs from their chemical compositions needs our better understanding of the internal structures of the sources.}

\item  Our chemical model successfully fits the chemical variations for $>60\%$ of the observed species in MM1--MM3. However, abundances of COMs are poorly fit across the condensations.  The intrinsically weak nature of these lines (especially in MM3), the limitations of the chemical network in dealing with more complex chemistries and the unknown complicated source structure {(Appendix~\ref{error})  are likely causing these discrepancies}. To improve the modelling process, further observations, e.g. using single-dish telescopes, are required to  recover information on all scales, better description of the formation and excitation of COMs are needed in the chemical networks, and a more appropriate physical structure combined with dynamic processes have to be taken fully into account. 

\item {IRS1 is dynamically and chemically the most evolved core in the NGC\,7538 clump. We detected primarily absorption lines, but also  a few lines showing pure emission (e.g. $\rm CH_3OH$ and $\rm HCOOCH_3$). 
These emission lines may be caused by unresolved sub-entities in IRS1-peak, or non-LTE (quasi-masing) effects of certain line transitions.}

\end{enumerate}

%%%%%%%%%%%%%%%%%%%%%%%%%%%%%%%%%%%%%%%%%%%%
\acknowledgements{We would like to thank the IRAM staff for their support during the observation and data reduction process. We thank  Karin \"{O}berg, Hauyu Baobab Liu, and  Malcolm Walmsley for useful discussions. \\
This research made use of NASAÕs Astrophysics Data System. S.F. acknowledges financial support by the European Community Seventh Framework Programme [FP7/2007-2013] under grant agreement no. 238258. \\
D.S. acknowledges support by the {\it Deutsche Forschungsgemeinschaft} through SPP~1385: ``The first ten million years of the 
solar system - a planetary materials approach'' (1962/1-3).
 }

\bibliographystyle{aa}
%\bibliography{ngc_accepted.bbl}
\bibliography{../HMSFR}

\newpage
\appendix
\section{An iterative approach to obtain the opacity corrected rotational temperature of $\rm CH_3CN$}\label{ap-temp}
$\rm CH_3CN$  is  a symmetric-top molecule with no dipole moment perpendicular to the molecular axis, and its K-ladders in rotational levels  can be excited solely by collisions, so it is a good temperature tracer in dense cloud. The traditional rotation diagram method is based on the following assumption:  (1)  The collision rates of $\rm CH_3CN$  are sufficient to thermalize the rotational levels, i.e. in   local thermodynamic equilibrium (LTE); (2) All lines originate from the same parcels of gas along the line of sight; (3) All lines are optically thin; (4) All lines have the same velocity and a single temperature.  Therefore, the rotational excitation temperature $\rm T_{rot}$ is equal to  the kinetic temperature, and the level populations  are then directly proportional to the line intensities of the K components.  \citep{boucher80, hollis82,  loren84,  olmi93, wright95, goldsmith99, zhang98}.\\

 Our observation band covers the J=12-11 transitions of $\rm CH_3CN$ around 220 GHz,  with upper state energy levels $\rm E_u/k$ spanning  from $\sim70$ K to $\sim500$ K,  yielding a large range for determining temperature.  With the detected lines listed in Table ~\ref{tab:rotline}, we derive the $\rm T_{rot}$ and the column density $\rm N_T$ of $\rm CH_3CN$ in each substructure (Figure~\ref{rotation}, purple dashed lines in Panel 1--2, black fit in Panel 3--4). {The fitted synthetic spectrum (\citealp{sanchez11, palau11}, see also Section \ref{id}) deviates from the observations, especially}  at $\rm K=0,1,2,3,6$, and the reasons are {multiple}: (1) the $\rm K=3,~6$ lines are from  ortho (o-) transitions, {which has forbidden collisional transition with the other para (p-)  lines, so they are treated to be two different molecules}  \citep{andersson84}. (2)  $\rm K=0,1$ lines are blended, so estimation {of their integrated intensity derived by simply taking into account their relative weights is uncertain};   
 (3) $\rm K=0,1,2$ lines are optically thick, which is the main issue \citep{feng15}. \\

Compared to $\rm CH_3CN$, lines of its rare isotopologue $\rm CH_3^{13}CN$ should be optically thin. However,  $\rm CH_3^{13}CN~ (12_2\rightarrow11_2)$ is the only transition that is not blended with the broad line wings of $\rm CH_3CN$. We use the following iterative approach to derive the opacity corrected rotational temperature  $\rm T_{rot}$ of MM1, MM2, and IRS1-mmS (red solid fittings in Panels 3--5 of Figure~\ref{rotation}):\\

\noindent (1) Obtain the rotational temperature $\rm T_{1}$ without any opacity correction;\\
(2) Estimate the optical depth  of $\rm CH_3CN~ (12_2\rightarrow11_2)$ $\rm \tau_{K,1~ (K=2)}$ by comparing its main beam temperature with that of $\rm CH_3^{13}CN~ (12_2\rightarrow11_2)$, and calculate the column density of line $12_2\rightarrow11_2$ at $\rm T_{1}$ with the optical depth  correction. Then, estimate the total column density of $\rm CH_3CN$ as $\rm N_{1}$ assuming LTE (see Appendix \ref{ap-abun} for more details about the correction);\\
(3) With $\rm T_{1}$ and $\rm N_{1}$ as trial input parameters, estimate the optical depth for the remaining transitions in the $\rm CH_3CN$ ladder $\rm \tau_{K,1~ (K=0-8)}$ with the RADEX code package\footnote{RADEX is a one-dimensional non-LTE radiative transfer code, providing an alternative to the widely used rotation diagram method. Without requiring the observation of  many optically thin emission lines, it can be used to roughly constrain the excitation temperature in addition to the column density.} \citep{vandertak07}, then correct their column densities line by line (dots and triangles in red in Figure~\ref{rotation}) and obtain a new rotational temperature $\rm T_{2}$;\\
(4) Repeat step (2)--(3) until $\rm (|T_{n}-T_{n-1}|)/T_{n-1}\le10\%$ (n=2,3,...; see fittings in red in Figure~\ref{rotation}).\\

In reality, molecules are not distributed homogeneously in the clouds, and 
they are far from being homogeneous in density and temperature. The ultimate
best-fit  requires a  sophisticated, fully 3-D
physico-chemo long-range transport modelling, which is
beyond the scope of our paper. 

\section{Spectral line column density  estimates}\label{ap-abun}

For all the species we detected, we calculate their column densities (from both main and rare isotopologues) from the integrated spectral line intensities as below.\\

The optical depth $\rm \tau_\upsilon$  (in velocity) along the line of sight is (\citealt{zeng06},  Eq. A.25): 
\begin{equation}
\rm \tau_\upsilon=\frac{c^3}{8\pi \nu^3 } A_{ul} N_u (e^{\frac{h \nu}{k_BT_{rot}}}-1) \Phi(\upsilon). 
\end{equation}
where c is the speed of light, $\rm k_B$  the Boltzmann constant, h  the Planck constant, $\rm \nu$  the line rest frequency, $ \Phi(\upsilon)$  the distribution function of the line shape in terms of velocity $\upsilon$. 
Here, $\rm  A_{ul}$ is the average spontaneous emission rate from the upper state $\rm E_u$ into the lower state $\rm E_l$, and it is calculated from line strength $S_{ul}$ and dipole moment $\mu$ as \\
\begin{equation}
\rm  A_{ul}=\frac{64\pi^4\nu^3}{3hc^3}\frac{\it S_{ul}\mu^2}{g_u}
\end{equation}

After integrating the optical depth  over the observed  linewidth, $\rm \int \Phi(\upsilon)  d\upsilon=1$, and the column density  $\rm N_{u}$ of the line at $\rm \nu$ Hz is
\begin{equation}\label{up}
\rm N_u=\frac{8 \pi \nu^3}{c^3 A_{ul}}\frac{1}{exp(\frac{h \nu}{k_BT_{rot}})-1} \int \tau_\upsilon  d\upsilon. ~~~~~(cm^{-2})
\end{equation}

When assuming LTE,  the total column density of all transitions for a molecule $\alpha$ is 
\begin{equation}\label{NT}
\rm N_{T,\alpha}=\frac{N_u}{g_u} Q(T_{rot}) e^{\frac{E_u}{k_BT_{rot}}} ~~~~~(cm^{-2})
\end{equation}
where $\rm Q(T_{rot})$ is the partition function 
for the given  (rotational) excitation temperature of each sources, which is interpolated from a table of partition functions for fixed temperatures,  obtained from CDMS/JPL.\\

Here,  $S_{ul}\mu^2$ can be  calculated from the CDMS/JPL line intensity $\rm  \ell(T_{rot})$  \citep{pickett98},

 \begin{eqnarray}
  S_{ul}\mu^2&=&\rm \frac{3hc}{8\pi^3\nu}\ell(T_{rot})Q(T_{rot})(e^{-\frac{E_l}{k_BT_{rot}}}-e^{-\frac{E_u}{k_BT_{rot}}})^{-1}\\
&=&\rm 2.4025\times10^{10}[\frac{\nu}{Hz}]^{-1}[\frac{\ell(300\,K)}{nm^2MHz}]Q(300K) \nonumber\\
&&\rm \times[e^{-\frac{E_l}{k_B300K}}-e^{-\frac{E_u}{k_B300K}}]^{-1} Debye^2\label{sul}
 \end{eqnarray}

 The spectra from NGC\,7538\,S consist of pure emission lines, while spectrum from the IRS1-peak is a mixture of emission and absorption lines. Accordingly, we calculate the line column densities separately for these two cases.

 \begin{enumerate}
\item
For absorption lines in IRS1-peak, the optical depth can be calculated from the flux intensity ratio between a particular line $\rm F_\upsilon$ and the continuum $\rm F_{\upsilon,c}$, $\rm \tau_\upsilon=-ln(F_\upsilon/F_{\upsilon,c})$.  Then, Eq.~\ref{NT} yields:
\begin{equation}
\rm N_{T,\alpha}= \frac{8 \pi \nu}{\ell(T_{rot})\eta c^3} \int \tau_\upsilon  d\upsilon~~~~~~~~~~~(cm^{-2})~~~~~~~~{\rm (absorption)}
\end{equation}

\item 
For the emission lines, we assume the Rayleigh-Jeans approximation ($\rm \frac{h\nu}{k_BT_{rot}}\ll1$), therefore Eqs.~\ref{up} and \ref{NT} can be simplified as  (\citealt{zeng06},  Eq.  4.26, also see \citealt{wilson09}),
\begin{equation} \label{emi}
  \left. {
 \begin{array}{l}
 \rm \frac{N_u}{g_u}\cong \frac{3k_B}{8\pi^3\nu \it S_{ul}\mu^2} T_{rot} \int \tau_\upsilon  d\upsilon  ~~~~~~~~~~~~~~~~~~~~\rm(cm^{-2})\\
\rm N_{T,\alpha}=
\rm \frac{k_B}{h c} \frac{(e^\frac{h\nu}{k_BT_{rot}}-1)}{\ell(T_{rot})}T_{rot} \int \tau_\upsilon  d\upsilon  ~~~~~~~~~~\rm(cm^{-2})
\end{array}
}
\right. {\rm (emission)}
\end{equation}

Assuming that  the observed emission in each substructure is homogeneous and fills the  beam, 
the integration of measured  main beam brightness temperature within the  velocity range $\rm \int T_{B}(\upsilon)  d\upsilon$  can be  substituted for the last term in the above equation: 
\begin{equation}\label{eq:correc}
\rm T_{rot} \int \tau_\upsilon  d\upsilon \cong \frac{\tau_{\alpha,0}}{1-e^{-\tau_{\alpha,0}}}\int T_{B}(\upsilon)  d\upsilon~~~~~~~~~~~~(K~cm\,s^{-1})\\ 
\end{equation}
where $\rm \tau_{\alpha,0}$ is the optical depth at line centre, and $\rm \int T_{B}(\upsilon)  d\upsilon$ is measured from Gaussian/hyperfine structure (HFS) {fit} by using the Gildas software package.\\

 \end{enumerate}
 
The column densities and related uncertainties of species listed in Table~\ref{col-Nbearing} ($\rm CH_3CN$ and  HNCO) can be estimated directly from their rotation diagrams and the scatterings of the data points (Figure~\ref{rotation}).
For the species whose transitions are not sufficient to derive the rotation diagrams, we estimate their column densities by assuming that their transitions  have the same $\rm T_{rot}$ as those derived from the opacity-corrected $\rm CH_3CN$.\\

In general, we assume that the observed transitions are optically thin ($\tau_{\alpha,0}\le1, \frac{\tau_{\alpha,0}}{1-e^{-\tau_{\alpha,0}}} \approx1$). However, for several molecules with large abundances, we  observe the corresponding transitions of  their main and rare isotopologues. By measuring the ratio between main beam brightness temperature of the main line $\rm T_{B,~\alpha,0}$ and its rare isotopologue $\rm T_{B,~\beta,0}$, we can estimate the optical depth at line centre of a given transition $\rm \tau_{\alpha,0}$ \citep{myers83}:\\ 

  \begin{equation}\label{eq:tau}
\rm \frac{1- exp(-\tau_{\alpha,0}/\Re^\alpha)}{1-exp(-\tau_{\alpha,0})}\approx \frac{T_{B, ~\beta,0}}{T_{B, ~\alpha,0}}\\
  \end{equation}
where $\Re_\alpha$ is the intrinsic abundance of the main isotope (e.g. $\rm ^{12}C$) compared to its rare  isotope (e.g. $\rm ^{13}C$) in the ISM (Table \ref{tab:correction}, e.g. \citealt{wilson94,chin96}). \\

\section{Error budget in observed parameters and modelling fits}\label{error}
Various assumptions must be made in order to estimate observed molecular abundances and best-fits from modelling. Below we consider the main assumptions in the method and how they affect the resulting values. \\

\subsection{Rotational Temperature}
We adopt the gas temperature derived from the opacity corrected rotation diagrams of $\rm CH_3CN$ in MM1, MM2, and IRS1-mmS. To verify these $\rm T_{rot}$, we compare them with the temperatures derived from previous studies.  \\

{Our temperatures in  MM1 and MM2 are slightly higher than derived from unresolved observations, for example \citet{zheng01} ($\sim25$ K from  $\rm NH_3$) and  \citet{sandell10} ($52\pm10$ K from $\rm H^{13}CN$, DCN, and $\rm CH_3CN$). However, this is expected as we are able to resolve localized warm gas concentrations on 0.01\,pc-scale.} \\

{We also find a slightly warmer temperature for IRS1-mmS  ($\rm \sim160~ K$)  than the cm observations of \citealp{goddi15a} ($\rm \sim120~ K$), although this is likely due to their cm observations and our mm observations tracing different components (with offset $\triangle\alpha\sim1\arcsec$). {The temperature  towards IRS1-peak ($\rm \sim214\pm66~ K$) is assumed by taking the measurement towards IRS1-mmS as the lower limit and that from  \citet{goddi15a} as upper limit, so it is consistent with the temperature measured towards the entire IRS1, for example, from an LVG fit of $\rm CH_3CN$ by \citet{qiu11} ($\rm 245 \pm 25~ K$) and by \cite{zhu13} ($\rm \sim260~ K$)}. 
}

\subsection{Molecular Column Densities}\label{appendix:molcol} 
We found  that the abundance ratios {in between the main and rare isotopologues  are consistent with the isotope ratio in the local ISM (e.g. SO, $\rm ^{13}CO$,  OCS, and  $\rm CH_3CN$ in Table.~\ref{tab:correction}). }
However, there are some  uncertainties which might impact our estimation:\\

\begin{table*} 
\small
\begin{center}
\caption{Optical depths and abundance ratios for SO,  $\rm ^{13}CO$,  OCS, and $\rm \rm CH_3CN$  lines with respect to their rare isotopologue lines  in each substructure of  NGC\,7538S and IRS1-mmS
\label{tab:correction}}
\begin{tabular}{c p{0.2cm}p{0.8cm} | p{1.8cm} p{2cm} p{2cm} p{1.8cm}  p{1.8cm} |p{2cm}}\hline\hline

Species   &  &$\Re_\alpha$  & $\rm MM1$      & $\rm MM2$      & $\rm MM3 $       & $\rm JetN$       & $\rm JetS$     & $\rm IRS1-mmS$  \\
&   & &$\rm ( 172\pm 23~K)$   &$\rm ( 137\pm 14~K)$   &$\rm ( 50 ~K)$   &$\rm ( 150~K)$   &$\rm ( 150~K)$   &$\rm ( 162\pm 14~K)$\\

\hline
$\rm \tau_{\rm SO~(6_5\rightarrow5_4)}$         &     &         &44.5        &7.5                &$-$                      &$\le24$         &$\le6.5$   &10   \\
$\rm N_{SO_\tau}/N_{^{33}SO}$ &$^\dag$   &127    &$\rm 112.20_{\pm0.96}$   &$\rm 94.46_{\pm10.42}$   &$\rm --$   &$\rm \le638.08$   &$\rm \le409.18$   &$\rm 132.30_{\pm21.26}$\\
$\rm N_{SO}/N_{^{33}SO}$ &$^\star$    &  &$\rm 2.52_{\pm0.02}$   &$\rm 12.59_{\pm1.39}$   &$-$   &$\rm \ge26.59$   &$\rm \ge62.86$   &$\rm 13.23_{\pm2.13}$\\

\hline 
$\rm \tau_{\rm ^{13}CO~(2\rightarrow1)}$             &     &    &13.5             &9                          &$\ge9$                                  &8         &8    &$--$\\
$\rm N_{^{13}CO_\tau}/N_{C^{18}O}$ &$^\dag$ & 8.3    &$\rm 16.07_{\pm0.52}$   &$\rm 17.58_{\pm0.02}$   &$\rm \ge12.54$   &$\rm 13.84_{\pm0.52}$   &$\rm 16.00_{\pm0.95}$   &$--$\\
$\rm N_{^{13}CO}/N_{C^{18}O}$ &$^\star$ &      &$\rm 1.19_{\pm0.04}$   &$\rm 1.95_{\pm0.00}$   &$\rm \le1.39$   &$\rm 1.73_{\pm0.06}$   &$\rm 2.00_{\pm0.12}$   &$--$\\
\hline
$\rm \tau_{\rm OCS~(18\rightarrow17)}$               &      &     &4.5                    &5.5                  &$36$                                   &$\le38$            &15    &4\\
$\rm N_{OCS_\tau}/N_{O^{13}CS}$ &$^\dag$    &73.5   &$\rm 76.71_{\pm1.88}$   &$\rm 72.81_{\pm16.34}$   &$\rm 69.24_{\pm1.17}$   &$\rm \le159.60$   &$\rm \le72.06$   &$\rm 86.40_{\pm27.69}$\\
$\rm N_{OCS}/N_{O^{13}CS} $ &$^\star$    &   &$\rm 16.86_{\pm0.41}$   &$\rm 13.18_{\pm2.96}$   &$\rm 1.92_{\pm0.03}$   &$\ge\rm 4.20$   &$\rm \ge4.80$   &$\rm 21.20_{\pm6.80}$\\
\hline 
$\rm \tau_{\rm CH_3CN~(12_2\rightarrow11_2)}$         &        &                &16           &3            &$\le$29                          &$-$             &$-$    &7 \\
$\rm N_{CH_3CN_\tau}/N_{CH_3^{13}CN} $ &$^\dag$   &73.5    &$\rm 70.22_{\pm7.06}$   &$\rm 78.13_{\pm2.02}$   &$\rm \le119.60$   &$--$   &$--$   &$\rm 71.88_{\pm5.30}$\\
$\rm N_{CH_3CN}/N_{CH_3^{13}CN} $ &$^\star$    &    &$\rm 4.39_{\pm0.44}$   &$\rm 24.75_{\pm0.64}$   &$\rm \ge4.12$   &$--$   &$--$   &$\rm 10.26_{\pm0.76}$\\
\hline
 \hline
\end{tabular}
\end{center}
\begin{tablenotes}
\item {\bf Note.} 
\item1) $\dag$ or $\star$ mark the estimations with or without optical depth correction, {respectively}. 
\item2) $\ge$ ($\le$) come from the $\rm 3\sigma$ limit of the non-detection or partly absorption of $\rm C^{18}O$ in IRS1-mmS. 
\item3) The ratios $\Re_\alpha$ between {isotopes}  of $\rm ^{12}C/^{13}C$ and $\rm ^{16}O/^{18}O$ are provided from the Galactic ISM ratio measured in \citet{giannetti14} (with $\rm R_{GC}(NGC\,7538) \sim9.7~ kpc$), while $\rm ^{32}S/^{33}S$ is provided from the {solar system} ratio measured in \citet{lodders03}.
\item4) Uncertainties on the measured values are typically $\le10\%$ as determined from  $\rm T_{rot}$, partition function $\rm Q(T_{rot})$, and Gaussian fit to $\rm T_B(\upsilon)d\upsilon$. 

\end{tablenotes}
\end{table*}

The isotopic S-elements ratio in NGC\,7538  is assumed to be close to the ratio of the solar system (e.g.  $\rm ^{32}S/^{33}S=127$, \citealt{lodders03}).  However, solar system values might have a bias since the Galactic disk outside the solar system is affected by  isotope changes (e.g. \citealt{kobayashi11}). Observationally, $\rm ^{32}S/^{33}S$ is expected in the  range 100--210 at the galactocentric distance of  NGC\,7538 (e.g. \citealt{chin96}). Therefore, a more precise opacity correction for  SO line requires a more accurate local $\rm ^{32}S/^{33}S$ ratio.\\

Fractionation reactions can also produce changes in the main/rare isotopologue ratios, in particular via exchange reactions\footnote{Exchange reactions of $\rm ^{16}O\Leftrightarrow^{18}O$ and $\rm ^{32}S\Leftrightarrow^{33}S\Leftrightarrow^{34}S$ have been found in AGB star, PDR, and meteoritic records.} involving $\rm ^{12}C\Leftrightarrow^{13}C$ \citep{penzias80,langer92},  $\rm ^{16}O\Leftrightarrow^{18}O$, and $\rm ^{32}S\Leftrightarrow^{33}S\Leftrightarrow^{34}S$. \\

We assume that emission lines of the same species (including all isotopologues) have the same excitation temperature, which is often not the case. In these dense, warm gas condensations with strong background IR dust thermal emission, it is possible that some of the lines, especially for COMs, are weakly masing, i.e. their intensities depart from LTE (S.~Parfenov, priv. comm.). The degree of these differences are a factor of a few at most (see the scatter of the data points in Figure~\ref{rotation}). Therefore, a more accurate calculation requires proper chemical modelling and LVG calculation.\\

In brief, our analytic calculations suggest that an uncertainty in optical depth dominates the uncertainty in the derived molecular column densites. Without the opacity correction, molecular column densities are on average underestimated by a factor of 5--10 (Figure~\ref{abun_line}). For abundant species such as HNCO, $\rm HC_3N$, and $\rm SO_2$,  for which we do not observe  corresponding transitions of their rare isotopologues (marked with $\dag$ in Table. \ref{tab:bestfit2}), 
our assumption that their strongest lines are optically thin may result in an underestimation by a similar factor. Furthermore, when we derive the molecular column density from only one line, uncertainty about the level of thermalization will result in an uncertainty on the resulting column density on the order of $<$20\% (estimated from the scatter of $\rm CH_3CN$ and HNCO rotation diagrams). In addition to these, uncertainties of $\rm T_{rot}$ and line intensity integration are negligible compared to the chemical variation among the  substructures (on average $\rm <10\%$, written in subscript in Tables \ref{col-Obearing}--\ref{col-Nbearing}).\\

\subsection{$\rm N_{H_2}$ and Molecular Abundances} \label{appendix:nh2}
{$\rm H_2$ column densities are calculated using two methods: from dust continuum ($\rm N_{H_{2,1}}$) as in MM1--MM3, and from $\rm C^{18}O$ ($\rm N_{H_{2,2}}$) as in all substructures of NGC\,7538\,S and IRS1 as described in Section~\ref{continu}.}  We found $\rm N_{H_{2,2}}/ N_{H_{2,1}}$ is $\rm \sim0.98$ in MM1,  $\rm \sim0.16$ in MM2, but only 0.006 in MM3. Several reasons may lead to these differences: 
 {
\begin{enumerate}
\item Since dust properties are unknown, we simply assume that the number densities in condensations MM1--MM3 are the same ($\rm 10^8~cm^{-3}$), and that the gas distribution within each condensation is homogeneous. Therefore we expect the gas-to-dust ratio ($\sim$ 150)  and the dust opacity to be the same across all three condensations. However, grain properties, such as the thickness of the ice mantle, can vary strongly along the line of sight, resulting in differences in derived H$_2$ column densities.\\

\item Difference in grain growth between the envelopes can result in differing dust emissivities (e.g. \citealp{miotello14}).\\

\item A constant relative abundance of $\rm C^{18}O$ to H$_2$ may not be true and would strongly affect the derived $\rm N_{H_{2,2}}$. A varying value could be caused by changes in the optical depth of the C$^{18}$O line, isotopic-selective depletion effects or simply C$^{18}$O and dust tracing different regions within the condensations. \\
\end{enumerate}
}
With these in consideration, molecular abundances with respect to $\rm H_2$ in MM1--MM3 are more reliable when $\rm N_{H_{2}}$ is derived from the dust continuum.

\subsection{Short Spacings}
Without short spacings, our interferometric data can lead to underestimated column densities, especially in the extended regions JetS and JetN due to filtering out flux. Compared to the specific intensity extrapolated from the single-dish 1.2 mm MAMBO data \citep{sandell04}, we estimate that 48\% of the flux has been filtered out from IRS1 and 90\% of flux from NGC\,7538\,S \citep{beuther12}, suggesting that emission from NGC\,7538\,S is more extended than that from IRS1. Since the fraction of filtered-out flux should be lower at the center of the primary beam,  uncertainties in column densities are larger in the outflow of NGC\,7538\,S than in the dense {condensation}s MM1--MM3. \\

\subsection{Local Thermal Equilibrium}
We have assumed all molecules are in LTE, however this is likely not the case for some species. An indication of this case is that SO has differing brightness and rotational temperatures in MM1 ($\rm T_{rot}\sim170~K$ but  $\rm T_B\sim30~K$). This dichotomy could be explained by: (1) SO is absorbed in the colder mantle which has not been resolved in our observations; or (2) the liberation of SO from grains by shocks and thus be {non-LTE excited.}  
Without higher spatial resolution observations of the substructures within each condensation, we are unable to distinguish between clumpiness, non-LTE effects or the beam dilution which causes this inconsistency.

\subsection{The modelling fits}\label{sec:mouncertain}
Table. \ref{tab:bestfit2} and Figure \ref{model_fit} show that 60\%-70\% of the modelled molecular column densities agree with the observed values within the quoted uncertainties. Several sources  relate to our model fitting procedure may bring uncertainties and disagreement with the observations. \\

{
The effects of shocks have not been included in our models. The $\rm N_{observed}$ of $\rm SO_2$ is at least 1-2 magnitudes lower than the $\rm N_{modelled}$ in MM2 and MM3. MM1 shows $\rm N_{observed}$ of $\rm SO_2$ is higher than $\rm N_{modeled}$ by a factor of 10. The significantly enhanced gas phase $\rm SO_2$ in MM1 may be related to the associated stronger shocks \citep{pineau93, bachiller97,garrod13} produced by outflows detected around MM1.\\

In addition, our model contains many assumptions which may affect the resulting fit. Chemical models are hugely sensitive to initial conditions. In the above framework we have assumed that our best-fit MM3 model provides the initial conditions for the following two evolutionary stages. However, with many non-detections in MM3, the initial conditions were based on upper limits and so they may not be entirely correct. It most be noted however that the problem of initial conditions is ubiquitous among all chemical models. Furthermore, photodissociation rates for COMs can be uncertain by factors of 10 -- 100 which will greatly affect the resulting column densities. Finally, a spherically symmetric source is the limit of a 1-D model. Velocity gradients across both MM1 and MM2 indicate rotation, suggesting a disk-like morphology \citep{beuther12}. In such a scenario one would find much higher densities and therefore much shorter timescales for chemical processes.
}

\setcounter{table}{0}
\renewcommand{\thetable}{A\arabic{table}}

\onecolumn
%\scalebox{0.9}{
{\scriptsize
\begin{longtable}{lllrr|lllrr}
\caption{Identified  lines from PdBI dataset. In case of  blending, lines with stronger CDMS/JPL intensity are listed in the left column, and the possible weaker blended transitions in the right column. 
}\label{tab:line}\\ 
%\begin{tabular}
\hline \hline
Freq.$^a$ & Mol$^b$ &Trans.$^c$ & $E_u/k$  &Cmt.$^d$      &Freq.$^a$ & Mol.$^d$ (candidates) &Trans.$^c$ & $E_u/k$   &Cmt.$^d$\\
(GHz) &    &   &  (K)      &        & (GHz) &     &  & (K)     & \\
\hline
%%%%%
$\rm 217.19\pm0.01$       &UL & & &$^{2.2}$
&217.191                 &$\rm CH_3OCH_3$  &$\rm (22_{4,19}-22_{3,20})A(E)A(E)$  &253   &\\
~&&&&&217.194                 &$\rm g-(CH_2OH)_{2} (\nu=1)$  &$\rm (17_{4, 14}\rightarrow16_{3, 13})$  &83          \\
~&&&&&217.194                 &$\rm HCOOCH_{3 (\nu=0)}$  &$\rm (30_{4, 26}\rightarrow30_{3, 27})E$  &291           \\
%%%%%
217.216                 &$\rm HCOOCH_{3 (\nu=0)}$  &$\rm (32_{9, 24}\rightarrow32_{8, 25})A$  &368   &$^{1.3}$          \\
217.236$^*$                 &$\rm HCOOCH_{3 (\nu=0)}$  &$\rm (32_{9, 24}\rightarrow32_{8, 25})E$  &368   &$^{1.3}$            \\
217.239          &$\rm DCN$  &$\rm  (3\rightarrow2)$              &21  &$\dag$ \\	
%%%%
$\rm 217.3\pm0.01$       &UL & & &$^{2.2}$
&217.300                 &$\rm CH_2^{13}CHCN$  &$\rm (23_{5,19}\rightarrow22_{5,18})$  &178   &\\
~&&&&&217.301                 &$\rm CH_2^{13}CHCN$  &$\rm (23_{5,18}\rightarrow22_{5,17})$  &178   &\\
~&&&&&217.290-217.307                           &$\rm ^{13}CN$  &$\rm  (N=2-1)$              &16  &\\
~&&&&&\color{black}217.299                 &\color{black}$\rm CH_3OH_{(v_t=1)}$  &$\rm (6_{1,5}\rightarrow7_{2,6}--)$  &\color{black}374   &\\

217.313                 &$\rm HCOOCH_{3 (\nu=1)}$  &$\rm (17_{4, 13}\rightarrow16_{4, 12})A$  &290   &$^{1.2}$            \\
%%%%%
$\rm 217.64\pm0.01$       &UL & & &$^{2.2}$
&217.639                 &$\rm ^{13}CH_3CH_2CN$  &$\rm  (25_{1,24}\rightarrow24_{1,23})$  &139   &\\
~&&&&&217.642                 &$\rm CH_2CH^{13}CN$  &$\rm  (23_{4,19}\rightarrow22_{4,18})$  &160   &\\
~&&&&&217.646                 &$\rm CH_2^{13}CHCN$  &$\rm  (23_{14,9}\rightarrow22_{14,8})$  &533   &\\
~&&&&&217.646                 &$\rm CH_2^{13}CHCN$  &$\rm  (23_{14,10}\rightarrow22_{14,9})$  &533   &\\
~&&&&&217.633-217.640                           &$\rm ^{13}CN$  &$\rm   (N=2-1)$              &15  &\\
~&&&&&\color{black}217.643                 &\color{black}$\rm CH_3OH_{(v_t=1)}$  &$\rm (15_{6,10}\rightarrow16_{5,11}--)$  &\color{black}746   &\\
~&&&&&\color{black}217.643                 &\color{black}$\rm CH_3OH_{(v_t=1)}$  &$\rm (15_{6,9}\rightarrow16_{5,12}++)$  &\color{black}746   &\\
%%%%%
217.833          &$\rm ^{33}SO$  &$\rm  (6_5\rightarrow5_4)$              &35       &$\dag$\\
$\rm 217.89\pm0.005$       &UL & & &$^{2.2}$ 
&\color{black}217.886                 &\color{black}$\rm CH_3OH_{(v_t=0)}$  &$\rm (20_{1,19}\rightarrow20_{0,20})$  &\color{black} 508   & \\
~&&&&&217.889                 &$\rm CH_2CH^{13}CN$  &$\rm  (23_{14,9}\rightarrow22_{14,8})$  &544   &\\
~&&&&&217.889                 &$\rm CH_2CH^{13}CN$  &$\rm  (23_{14,10}\rightarrow22_{14,9})$  &544   &\\
~&&&&&217.895                 &$\rm ^{13}CH_2CHCN$  &$\rm  (23_{1,23}\rightarrow22_{0,22})$  &121   &\\
%%%%%
218.108                &$\rm HCOOCH_{3 (\nu=1)} $  &$\rm (17_{4, 13}\rightarrow16_{4, 12})E$  &290   &$^{1.2}$            \\
218.198$^*$         &$\rm O^{13}CS$  &$\rm  (18\rightarrow17)$           & 99   &$^{3}$ $\dag$\\
218.222         &$\rm H_{2}CO$  &$\rm  (3_{0,3}\rightarrow2_{0,2})$           & 21     &\\
218.260                 &$\rm HCOOCH_{3 (\nu=0)}$  &$\rm  (31_{9, 23}\rightarrow31_{8, 24})A$  &349   &$^{1.3}$           \\
218.281                 &$\rm HCOOCH_{3 (\nu=0)}$  &$\rm  (17_{3, 14}\rightarrow16_{3, 13})E$  &100   &$^{2.2}$           \\
218.298                 &$\rm HCOOCH_{3 (\nu=0)}$  &$\rm (17_{3, 14}\rightarrow16_{3, 13})A$  &100   &$^{2.2}$           \\
218.325         &$\rm HC_{3}N_{(\nu=0)}$  &$\rm  (24\rightarrow23)$            &131   &$\dag$\\
218.390         &$\rm CH_3CH_2CN_{(\nu=0)}$  &$\rm  (24_{3,21}\rightarrow23_{3,20})$          &140        &$^{3}$ $\dag$\\
218.440         &$\rm CH_{3}OH_{(\nu=0)}$  &$\rm  (4_{2,2}\rightarrow3_{1,2})$           &46    &$\dag$ \\
218.460         &$\rm NH_{2}CHO$  &$\rm  (10_{1,9}\rightarrow9_{1,8})$           &61 &$\dag$\\
218.476         &$\rm H_{2}CO$  &$\rm  (3_{2,2}\rightarrow2_{2,1})$            &68     &$\dag$\\
218.585$^*$                 &$\rm HCOOCH_{3 (\nu=0)}$  &$\rm  (36_{9, 28}\rightarrow36_{8, 29})A$  &450   &$^{2.2}$           \\
218.593$^*$                 &$\rm HCOOCH_{3 (\nu=0)}$  &$\rm  (27_{7, 21}\rightarrow27_{5, 22})A$  &258   &$^{2.2}$           \\
218.607$^*$                 &$\rm HCOOCH_{3 (\nu=0)}$  &$\rm  (36_{9, 28}\rightarrow36_{8, 29})E$  &450   &$^{2.2}$           \\
218.633$^*$                 &$\rm HCOOCH_{3 (\nu=0)}$  &$\rm  (27_{7, 21}\rightarrow27_{5, 22})E$  &258   &$^{2.2}$           \\
218.655                 &$\rm HCOOCH_{3 (\nu=1)}$  &$\rm  (18_{16, 2}\rightarrow17_{16, 1})E$  &460   &$^{1.3}$           \\
218.738                 &$\rm HCOOCH_{3 (\nu=1)}$  &$\rm  (18_{14, 4}\rightarrow17_{14, 3})E$  &419   &$^{2.3}$           \\
218.760         &$\rm H_{2}CO$  &$\rm  (3_{2,1}\rightarrow2_{2,0})$           &68  &\\
218.831                 &$\rm HCOOCH_{3 (\nu=1)}$  &$\rm  (18_{13, 5}\rightarrow17_{13, 4})E$  &401   &$^{1.3}$           \\
218.861         &$\rm HC_3N_{(v_7=1)}$  &$\rm  (24\rightarrow23, l=1e)$           &452  &$\dag$\\
218.903          &$\rm OCS_{(\nu=0)}$  &$\rm  (18\rightarrow17)$           &100   &$\dag$\\
218.966                 &$\rm HCOOCH_{3 (\nu=1)}$  &$\rm  (18_{12, 6}\rightarrow17_{12, 5})E$  &384   &$^{1.3}$           \\
218.981         &$\rm HNCO_{(\nu=0)}$  &$\rm   (10_{1, 10}\rightarrow9_{1, 9})$           &101 &$\dag$\\
219.068$^*$                 &$\rm HCOOCH_{3 (\nu=1)}$  &$\rm  (18_{17, 2}\rightarrow17_{17, 1})E$  &481   &$^{2.2}$           \\
219.079$^*$                 &$\rm HCOOCH_{3 (\nu=1)}$  &$\rm  (28_{3, 25}\rightarrow28_{2, 26})E$  &434   &$^{2.2}$           \\
219.090$^*$                 &$\rm HCOOCH_{3 (\nu=0)}$  &$\rm  (34_{7, 28}\rightarrow34_{5, 29})E$  &388   &$^{2.2}$           \\
219.109$^*$                 &$\rm HCOOCH_{3 (\nu=0)}$  &$\rm  (34_{7, 28}\rightarrow34_{5, 29})A$  &388   &$^{2.3}$           \\
219.153$^*$                 &$\rm HCOOCH_{3 (\nu=1)}$  &$\rm  (10_{4, 6}\rightarrow9_{3, 6})E$  &230   &$^{2.3}$           
&219.155$^*$                 &$\rm HCOOCH_{3 (\nu=1)}$  &$\rm  (18_{11, 7}\rightarrow17_{11, 6})E$  &369   &$^{2.3}$           \\
219.174         &$\rm HC_3N_{(\nu_7=1)}$  &$\rm  (24\rightarrow23)$         &452 &\\
219.195$^*$                 &$\rm HCOOCH_{3 (\nu=1)}$  &$\rm  (18_{16, 3}\rightarrow17_{16, 2})E$  &459   &$^{2.3}$           \\
219.264$^*$                 &$\rm HCOOCH_{3 (\nu=0)}$  &$\rm  (36_{6, 30}\rightarrow36_{6, 31})A$  &429   &$^{2.2}$           \\
219.276         &$\rm SO_{2 (\nu=0)}$  &$\rm    (22_{7, 15}\rightarrow23_{6, 18}) $             &352 &$\dag$\\
219.331$^*$                 &$\rm HCOOCH_{3 (\nu=1)}$  &$\rm  (18_{15, 4}\rightarrow17_{15, 3})E$  &438   &$^{2.3}$           \\
219.355          &$\rm ^{34} SO_{2 (\nu=0)}$  &$\rm     (11_{1, 11}\rightarrow10_{0, 10})$           &60  &$^{3}$ $\dag$\\
219.412                 &$\rm HCOOCH_{3 (\nu=1)}$  &$\rm  (18_{10, 8}\rightarrow17_{10, 7})E$  &355   &           \\
219.417$^*$                 &$\rm HCOOCH_{3 (\nu=0)}$  &$\rm  (30_{5, 26}\rightarrow30_{4, 27})E$  &292   &$^{2.3}$           \\
219.479                 &$\rm HCOOCH_{3 (\nu=1)}$  &$\rm  (18_{14, 5}\rightarrow17_{14, 4})E$  &419   &$^{2.3}$           \\
219.484$^*$                 &$\rm HCOOCH_{3 (\nu=0)}$  &$\rm  (30_{5, 26}\rightarrow30_{4, 27})A$  &292   &$^{2.3}$           \\
219.506         &$\rm CH_{3}CH_{2}CN_{(\nu=0)}$  &$\rm   (24_{2, 22}\rightarrow23_{2, 21})$             &136   &$^{2.3}$ \\
%219.547         &$\rm HNCO (10_{4,6}\rightarrow9_{4,5})$           &485  & \\
219.560          &$\rm C^{18}O$  &$\rm  (2\rightarrow1)$             &16       &$\dag$\\

219.566                 &$\rm HCOOCH_{3 (\nu=1)}$  &$\rm  (18_{14, 4}\rightarrow17_{14, 3})A$  &419   &           
&219.566                 &$\rm HCOOCH_{3 (\nu=1)}$  &$\rm  (18_{14, 5}\rightarrow17_{14, 5})A$  &419   &           \\
~&&&&&219.566                 &$\rm HCOOCH_{3 (\nu=1)}$  &$\rm   (18_{15, 4}\rightarrow17_{15, 3})A$  &438   &            \\
~&&&&&219.566                 &$\rm HCOOCH_{3 (\nu=1)}$  &$\rm   (18_{15, 3}\rightarrow17_{15, 3})A$  &438   &           \\
~&&&&&219.571                 &$\rm HCOOCH_{3 (\nu=1)}$  &$\rm   (18_{16, 2}\rightarrow17_{16, 1})A$  &459   &           \\
~&&&&&219.571                 &$\rm HCOOCH_{3 (\nu=1)}$  &$\rm   (18_{16, 3}\rightarrow17_{16, 2})A$  &459   &           \\

219.579                 &$\rm HCOOCH_{3 (\nu=0)}$  &$\rm   (28_{9, 19}\rightarrow28_{8, 20})A$  &295   &$^{2.3}$           \\
219.584                &$\rm HCOOCH_{3 (\nu=1)}$  &$\rm   (18_{13, 5}\rightarrow17_{13, 4})A$  &401   &$^{2.3}$           
&219.584                &$\rm HCOOCH_{3 (\nu=1)}$  &$\rm   (18_{13, 6}\rightarrow17_{13, 5})A$  &401   &$^{2.3}$           \\
219.592                &$\rm HCOOCH_{3 (\nu=0)}$  &$\rm   (28_{9, 19}\rightarrow28_{8, 20})E$  &295   &$^{1.3}$           \\
219.600                &$\rm HCOOCH_{3 (\nu=0)}$  &$\rm   (30_{9, 22}\rightarrow30_{8, 23})E$  &330   &$^{1.1}$           \\
219.607                &$\rm HCOOCH_{3 (\nu=0)}$  &$\rm   (30_{5, 26}\rightarrow30_{3, 27})E$  &292   &           \\
219.623                &$\rm HCOOCH_{3 (\nu=1)}$  &$\rm   (18_{12, 6}\rightarrow17_{12, 5})A$  &384   &$^{1.1}$           
&219.623                &$\rm HCOOCH_{3 (\nu=1)}$  &$\rm   (18_{12, 7}\rightarrow17_{12, 6})A$  &384   &$^{1.1}$           \\
219.642                &$\rm HCOOCH_{3 (\nu=1)}$  &$\rm   (18_{13, 6}\rightarrow17_{13, 5})E$  &400   &$^{1.1}$           \\
219.657$^*$          &$\rm HNCO$  &$\rm  (10_{3, 8}\rightarrow9_{3, 7})$            &447  & 
&219.657          &$\rm HNCO$  &$\rm  (10_{3, 7}\rightarrow9_{3, 6})$            &447  & \\
219.696                &$\rm HCOOCH_{3 (\nu=1)}$  &$\rm   (18_{11, 8}\rightarrow17_{11, 7})A$  &369   &$^{1.3}$           
&219.696                &$\rm HCOOCH_{3 (\nu=1)}$  &$\rm   (18_{11, 7}\rightarrow17_{11, 6})A$  &369   &$^{1.3}$           \\
219.705                &$\rm HCOOCH_{3 (\nu=1)}$  &$\rm   (18_{4, 15}\rightarrow17_{4, 14})A$  &299   &$^{2.2}$           \\
219.734$^*$          &$\rm HNCO_{(\nu=0)}$  &$\rm   (10_{2, 9}\rightarrow9_{2, 8})$            &231  &
&219.737          &$\rm HNCO_{(\nu=0)}$  &$\rm   (10_{2, 8}\rightarrow9_{2, 7})$            &231  & \\
219.764                &$\rm HCOOCH_{3 (\nu=1)}$  &$\rm   (18_{9, 9}\rightarrow17_{9, 8})E$  &342   &$^{1.2}$           \\
219.798           &$\rm HNCO_{(\nu=0)}$  &$\rm   (10_{0, 10}\rightarrow9_{0, 9})$            &58     &\\
219.822               &$\rm HCOOCH_{3 (\nu=1)}$  &$\rm   (18_{10, 9}\rightarrow17_{10, 8})A$  &355   &$^{1.2}$           
&219.822               &$\rm HCOOCH_{3 (\nu=1)}$  &$\rm   (18_{10, 8}\rightarrow17_{10, 7})A$  &355   &$^{1.2}$           \\
219.827$^*$               &$\rm HCOOCH_{3 (\nu=1)}$  &$\rm   (18_{12, 7}\rightarrow17_{12, 6})E$  &384   &$^{2.3}$           \\
219.908         &$\rm H_{2} ^{13}CO$  &$\rm  (3_{1, 2}\rightarrow2_{1, 1})$          &33    &$\dag$ \\
219.949         &$\rm SO_{(\nu=0)}$  &$\rm   (6_{5}\rightarrow5_{4})$           &35     &$\dag$ \\
220.030        &$\rm HCOOCH_{3 (\nu=1)}$  &$\rm   (18_{9, 10}\rightarrow17_{9, 9})A$           &342       &
&220.030        &$\rm HCOOCH_{3 (\nu=1)}$  &$\rm   (18_{9, 9}\rightarrow17_{9, 8})A$           &342       &\\
220.043        &$\rm HCOOCH_{3 (\nu=1)}$  &$\rm   (18_{11, 8}\rightarrow17_{11, 7})E$           &368       &$^{2.3}$\\
220.078        &$\rm CH_{3}OH_{(\nu_t=0)}$  &$\rm   (8_{0, 8}\rightarrow7_{1, 6})$           &97       &\\
220.167        &$\rm HCOOCH_{3 (\nu=0)}$  &$\rm   (17_{4, 13}\rightarrow16_{4, 12})E$           &103       &$\dag$\\
220.178         &$\rm CH_{2}CO$  &$\rm  (11_{1, 11}\rightarrow10_{1, 10})$           &76   &$\dag$\\
220.190         &$\rm HCOOCH_{3 (\nu=0)}$  &$\rm   (17_{4, 13}\rightarrow16_{4, 12})A$           &103   &\\
220.258          &$\rm HCOOCH_{3 (\nu=1)}$  &$\rm   (24_{2, 23}\rightarrow24_{1, 24})E$           &355   &$^{1.3}$\\
220.296         &$\rm CH_{3}^{13}CN$  &$\rm  (12_{9}\rightarrow11_{9})$           &646      &$^{1.3}$  \\
220.307          &$\rm HCOOCH_{3 (\nu=1)}$  &$\rm   (18_{10, 9}\rightarrow17_{10, 8})E$           &354   &$^{1.2}$\\
220.368          &$\rm HCOOCH_{3 (\nu=1)}$  &$\rm   (18_{8, 11}\rightarrow17_{8, 10})A$           &331   &$^{2.2}$
&220.370          &$\rm HCOOCH_{3 (\nu=1)}$  &$\rm   (18_{8, 10}\rightarrow17_{8, 9})A$           &331   &$^{2.2}$\\
~&&&&&220.366          &$\rm HCOOCH_{3 (\nu=0)}$  &$\rm   (33_{5, 28}\rightarrow33_{5, 29})E$           &357   &$^{2.2}$\\
220.399         &$\rm ^{13}CO$  &$\rm  (2\rightarrow1)$           &16       &$\dag$\\

220.409          &$\rm HCOOCH_{3 (\nu=1)}$  &$\rm   (18_{4, 15}\rightarrow17_{4, 14})E$           &299   &$^{1.3}$\\
220.476        &$\rm CH_{3}CN_{(\nu=0)}$  &$\rm    (12_{8}\rightarrow11_{8})$            &526   &\\
220.539       &$\rm CH_{3}CN_{(\nu=0)}$  &$\rm   (12_{7}\rightarrow11_{7})$           &419     &\\
220.570        &$\rm CH_{3}^{13}CN$  &$\rm   (12_{4}\rightarrow11_{4})$            &183        &\\
220.594        &$\rm CH_{3}CN_{(\nu=0)}$  &$\rm   (12_{6}\rightarrow11_{6})$           &326   &\\
220.621         &$\rm CH_{3}^{13}CN$  &$\rm  (12_{2}\rightarrow11_{2})$            &97        &$\dag$\\
220.641        &$\rm CH_{3}CN_{(\nu=0)}$  &$\rm    (12_{5}\rightarrow11_{5})$            &248    &\\
220.647          &$\rm HCOOCH_{3 (\nu=1)}$  &$\rm   (18_{9, 10}\rightarrow17_{9, 9})E$  &342   &$^{2.2}$ \\
220.661         &$\rm CH_{3}CH_{2}CN_{(\nu=0)}$  &$\rm    (25_{2, 24}\rightarrow24_{2, 23})$            &143         &$^{2.3}$ \\
220.679        &$\rm CH_{3}CN_{(\nu=0)}$  &$\rm    (12_{4}\rightarrow11_{4})$            &183    &\\
220.709       &$\rm CH_{3}CN_{(\nu=0)}$  &$\rm    (12_{3}\rightarrow11_{3})$            &133       &\\
220.730        &$\rm CH_{3}CN_{(\nu=0)}$  &$\rm    (12_{2}\rightarrow11_{2})$            &98     &$\dag$\\
220.743       &$\rm CH_{3}CN_{(\nu=0)}$  &$\rm    (12_{1}\rightarrow11_{1})$            &76     &\\
220.747        &$\rm CH_{3}CN_{(\nu=0)}$  &$\rm    (12_{0}\rightarrow11_{0})$            &69      &\\

%%%%%%%%%%%%%%%%%%%%%%%%%%%%%

\hline \hline
\multicolumn{9}{l}{{\bf Note.} {\it a}. Tentative detections (intensity $\rm \sim3\sigma$ rms) are marked with ``*"; }\\
\multicolumn{9}{l}{~~~~~~~~~~{\it b}. Unidentified  lines are denoted as ``UL", and possible transitions are given; }\\
\multicolumn{9}{l}{~~~~~~~~~~{\it c}. ``A(E)A(E)" indicates 4 types of transitions are possible for a certain line: AA, EE, AE, EA;}\\
\multicolumn{9}{l}{~~~~~~~~~~{\it d}. Lines with  $\dag$ are imaged in Figure~\ref{into};}\\
\multicolumn{9}{l}{~~~~~~~~~~~~~ Without any digital annotation indicates that the   lines are detected  in both IRS1 and MM1, especially showing absorption or an (inverse) P-Cygni profile towards IRS1-peak;}\\
\multicolumn{9}{l}{~~~~~~~~~~~~~ ``$^{1.1}$"   denote the lines being detected in IRS1 only, especially showing  absorption or an (inverse) P-cygni profile towards IRS1-peak. }\\
\multicolumn{9}{l}{~~~~~~~~~~~~~ ``$^{1.2}$"   denote the lines being detected in IRS1 only, especially showing  pure emission towards IRS1-peak. }\\
\multicolumn{9}{l}{~~~~~~~~~~~~~ ``$^{1.3}$"  denote the lines being detected in IRS1-mmS only, but without clear detection in IRS1-peak. }\\

\multicolumn{9}{l}{~~~~~~~~~~~~~ ``$^{2.2}$"  denote the lines being detected in both  IRS1 and MM1, especially showing  pure emission towards IRS1-peak. }\\
\multicolumn{9}{l}{~~~~~~~~~~~~~ ``$^{2.3}$"   denote the lines being detected in both  IRS1 and MM1, but without clear detection in IRS1-peak. }\\
\multicolumn{9}{l}{~~~~~~~~~~~~~ ``$^{3}$"   denote the lines being detected only in MM1. }\\

\end{longtable}

}
\newpage

\begin{landscape}
\begin{table}
%  \begin{threeparttable}

\caption{Integrated intensity $\rm \int T_B(\upsilon)d\upsilon~ (K km~s^{-1}$) for  NGC\,7538S and IRS1-mmS,$\rm \int \tau_(\upsilon)d\upsilon~ (km~s^{-1}$) for IRS1-peak,corresponding to the transitions shown in Figure~\ref{velpro}. \label{tab:lineprofile} }
\small
\begin{center}

\begin{tabular}{c||p{1.6cm} |p{2cm}  p{2cm} p{2cm} p{2cm} p{2cm} ||p{2cm} |p{2cm} }\hline\hline

Species   &Freq (GHz)   &$\rm MM1$   &$\rm MM2$   &$\rm MM3$   &$\rm JetN$   &$\rm JetS$   &$\rm IRS1-mmS$   &$\rm IRS1-peak$\\
\hline
$\rm C^{18}O$     & 219.56     &$\rm 120.82_{\pm0.42}$     &$\rm 24.22_{\pm0.47}$     &$\rm 5.80_{\pm0.15}$     &$\rm 20.96_{\pm0.09}$     &$\rm 27.52_{\pm0.26}$     &$\nabla\nabla$     &$\rm 1.26_{\pm0.02}$\\ 
$\rm ^{13}CO$     & 220.399     &$\rm 144.74_{\pm6.38}$     &$\rm 47.60_{\pm0.88}$     &$\rm \ge8.13$     &$\rm 36.47_{\pm1.53}$     &$\rm 55.36_{\pm3.84}$     &$\nabla\nabla$     &$\rm 11.82_{\pm0.18}$\\ 
$\rm H_2CO$     & 218.476     &$\rm 43.20_{\pm1.89}$     &$\rm 51.35_{\pm0.38}$     &$\rm \le1.04$     &$\rm 24.68_{\pm0.55}$     &$\rm 11.89_{\pm0.59}$     &$\rm 225.81_{\pm12.56}$     &$\rm 2.16_{\pm0.00}$\\ 
$\rm H_2^{13}CO$     & 219.909     &$\rm 9.24_{\pm0.13}$     &$\rm 24.95_{\pm0.31}$     &$\rm \le0.81$     &$\rm \le0.54$     &$\rm 7.87_{\pm0.02}$     &$\rm 72.24_{\pm1.31}$     &$\rm 0.07_{\pm0.00}$\\ 
$\rm CH_2CO$     & 220.178     &$\rm 5.77_{\pm0.36}$     &$\rm 22.09_{\pm0.17}$     &$\rm \le0.97$     &$\rm 1.78_{\pm0.06}$     &$\rm \le0.61$     &$\rm 117.79_{\pm1.30}$     &$\rm \le0.03$\\ 
$\rm CH_3OH$     & 218.44     &$\rm 76.22_{\pm0.69}$     &$\rm 101.27_{\pm0.89}$     &$\rm 16.25_{\pm0.13}$     &$\rm 7.59_{\pm0.72}$     &$\rm 12.87_{\pm0.35}$     &$\rm 316.60_{\pm10.35}$     &$\rm 1.84_{\pm0.03}$\\ 
$\rm HCOOCH_3$     & 220.167     &$\rm 24.08_{\pm0.36}$     &$\rm 14.13_{\pm0.08}^{~\S}$     &$\rm \le0.78$     &$\rm \le0.65$     &$\rm 0.25_{\pm0.04}^{~\S}$     &$\rm 112.05_{\pm20.48}$     &$\rm \le0.06$\\ 
$\rm ?CH_3OCH_3^{~*}$     & 217.191     &$\rm 17.80_{\pm0.07}$     &$\rm 5.10_{\pm0.51}$     &$\rm \le1.14$     &$\rm \le0.94$     &$\rm \le0.93$     &$\rm 131.87_{\pm1.37}$     &$\nabla\nabla$\\ 
$\rm DCN$     & 217.239     &$\rm 65.97_{\pm0.18}$     &$\rm 5.40_{\pm0.65}$     &$\rm \le0.94$     &$\rm \le0.83$     &$\rm 5.73_{\pm0.82}$     &$\rm 98.58_{\pm6.93}$     &$\rm 0.92_{\pm0.00}$\\ 
$\rm HNCO$     & 218.981     &$\rm 43.66_{\pm0.35}$     &$\rm 26.13_{\pm0.06}$     &$\rm \le0.57$     &$\rm \le0.72$     &$\rm \le0.39$     &$\rm 83.03_{\pm3.19}$     &$\rm 0.62_{\pm0.05}$\\ 
$\rm HC_3N$     & 218.325     &$\rm 163.85_{\pm0.96}$     &$\rm 56.58_{\pm0.75}$     &$\rm \le0.92$     &$\rm \le0.63$     &$\rm 10.42_{\pm0.53}$     &$\rm 63.66_{\pm1.58}$     &$\rm 0.46_{\pm0.01}$\\ 
$\rm HC_3N(v_7=1)$     & 218.861     &$\rm 54.90_{\pm0.56}$     &$\rm 7.84_{\pm0.27}$     &$\rm \le0.91$     &$\rm \le0.63$     &$\rm \le1.51$     &$\rm 10.35_{\pm1.24}$     &$\rm 0.18_{\pm0.01}$\\ 
$\rm CH_3CN$     & 220.73     &$\rm 136.57_{\pm1.93}$     &$\rm 46.87_{\pm1.40}$     &$\rm 3.97_{\pm0.02}$     &$\rm \le0.38$     &$\rm \le1.28$     &$\rm 280.88_{\pm3.92}$     &$\rm 0.91_{\pm0.03}$\\ 
$\rm CH_3^{13}CN$     & 220.621     &$\rm 10.43_{\pm1.00}$     &$\rm 0.60_{\pm0.03}$     &$\rm \le0.29$     &$\rm \le0.22$     &$\rm \le0.21$     &$\rm 9.03_{\pm0.66}$     &$\rm \le0.04$\\ 
$\rm CH_3CH_2CN$     & 218.39     &$\rm 21.32_{\pm0.10}$     &$\rm \le0.54$     &$\rm \le0.92$     &$\rm \le0.63$     &$\rm \le1.28$     &$\rm 10.13_{\pm0.79}$     &$\rm \le0.09$\\ 
$\rm NH_2CHO$     & 218.46     &$\rm 8.30_{\pm0.89}$     &$\rm 19.77_{\pm0.28}$     &$\rm \le0.54$     &$\rm 0.97_{\pm0.41}$     &$\rm \le0.69$     &$\rm 87.89_{\pm2.66}$     &$\rm 0.19_{\pm0.01}$\\ 
$\rm OCS$     & 218.903     &$\rm 147.54_{\pm0.97}$     &$\rm 98.99_{\pm0.24}$     &$\rm 17.04_{\pm0.53}$     &$\rm 3.49_{\pm0.06}$     &$\rm 5.28_{\pm0.60}$     &$\rm 250.85_{\pm3.88}$     &$\rm \le0.14$\\ 
$\rm O^{13}CS$     & 218.199     &$\rm 8.72_{\pm0.66}$     &$\rm 7.48_{\pm2.62}$     &$\rm 8.86_{\pm0.12}^{~\S}$     &$\rm \le0.83$     &$\rm \le1.10$     &$\rm 11.78_{\pm6.42}$     &$\rm \le0.14$\\ 
$\rm SO_2$     & 219.276     &$\rm 98.07_{\pm0.26}$     &$\rm \le0.67$     &$\rm \le0.58$     &$\rm \le0.53$     &$\rm \le0.88$     &$\rm 30.03_{\pm5.17}$     &$\rm 0.25_{\pm0.01}$\\ 
$\rm ^{34}SO_2$     & 219.355     &$\rm 102.32_{\pm0.58}$     &$\rm \le0.68$     &$\rm \le0.39$     &$\rm \le0.47$     &$\rm \le1.29$     &$\rm 12.80_{\pm0.18}$     &$\rm 0.03_{\pm0.00}$\\ 

$\rm SO$     & 219.949     &$\rm 223.73_{\pm2.95}$     &$\rm 123.71_{\pm4.35}$     &$\rm \le1.37$     &$\rm 22.98_{\pm0.51}$     &$\rm 132.20_{\pm3.98}$     &$\rm 229.06_{\pm0.66}$     &$\rm 1.90_{\pm0.06}$\\ 
$\rm ^{33}SO$     & 217.833     &$\rm 26.83_{\pm0.08}$     &$\rm 2.97_{\pm0.83}$     &$\rm \le0.33$     &$\rm \le0.26$     &$\rm \le0.64$     &$\rm 5.23_{\pm1.94}$     &$\rm \le0.11$\\ 

 \hline
\end{tabular}
  \begin{tablenotes}
{\bf Note.} \\
 1). Most species are measured from Gaussian fits, and uncertainties (written {in} subscript) on the measured intensities are typically $\le10\%$.\\
2). For those marked with  ``$\S$", {integration has been performed down to the noise level due to their non-Gaussian profile. Uncertainties area however derived from the difference with the Gaussian fit. } \\
3). For species which are not detected, an upper limit derived from  $\rm 3\sigma$ rms is given. \\
4).  ``$\nabla\nabla$" {indicates no meaningful area can be derived because} of obvious absorption   in IRS1-mmS, orobvious emission in IRS1-peak.\\
5). {A lower limit  is given to species which has partly absorption in NGC\,7538\,S because of the missing short space}.\\
6). ``*" { indicates the tentative detected species because  only one unblended line has been detected, which has not been observed in our source previously}. 
    \end{tablenotes}
    
\end{center}
%  \end{threeparttable}

\end{table}
\end{landscape}

%%%%%%%%%%%%%%%%%%%%%%%%
%%%%%%%%%%%%%%%%%%%%%%%%%
%%%%%%%%%%%%%%%%%%%%%%%%%

\newpage
\begin{landscape}
\begin{table}
\caption{Laboratory parameters of $\rm CH_3CN$ and HNCO and the intensity integrated  over the width of each line $\rm \int T_{B}({\upsilon})  d{\upsilon}$, which is used for rotation diagrams in  Figure~\ref{rotation}. Uncertainties on the measured intensities are typically $\le10\%$ as determined from Gaussian {fit}. 
  \label{tab:rotline} 
 }
\small
\begin{center}

\begin{tabu}{cc| p{0.2cm} p{1cm} p{1cm} p{1cm}p{0.7cm}|p{2cm} p{2cm} p{2cm}|p{1cm}}\hline\hline

Mol. &Freq    &K  &$\rm I(300K)$    &$\rm E_{L}$    &$S_{ul}\mu^2$       &o-/p-  &$\rm MM1$                      & $\rm MM2$                     & $\rm IRS1-mmS$                    &Note\\
       &(GHz)    &  &$\rm Log_{10}$    &$\rm (cm^{-1})$     &$\rm (D^2)$                       &     &$\rm (K~km~s^{-1})$     &$\rm (K~km~s^{-1})$            &$\rm (K~km~s^{-1})$       \\
\hline
\multirow{9}*{$\rm CH_3CN~(J=12\rightarrow11)$}
    &220.476   &         8   &    -3.098  &   357.938  &   205.033 &p-     &$\rm 24.76\pm 2.50$	&$\rm 5.53\pm 0.09$		&$\rm 38.15\pm 0.92$      &\multirow{9}*{\it  1, 2, 3}\\

   &220.539   &         7   &    -2.868  &   283.608  &   243.499 &p-      &$\rm 32.99\pm 2.60$	&$\rm 13.07\pm 0.16$		&$\rm 46.61\pm 4.02$    \\

   &220.594   &         6   &    -2.377  &   219.154  &   553.690 &o-     &$\rm 129.85\pm 10.78$		&$\rm 60.20\pm 0.08$		&$\rm 166.07\pm 2.04$   \\

   &220.641   &         5   &    -2.522  &   164.592  &   305.047 &p-   &$\rm 109.82\pm 4.97$		&$\rm 36.07\pm 0.18$		&$\rm 165.53\pm 11.26$  \\

   &220.679   &         4   &    -2.397  &   119.933  &   328.188 &p-   &$\rm 105.81\pm 3.19$		&$\rm 39.42\pm 0.22$		&$\rm 225.55\pm 3.57$  \\

   &220.709   &         3   &    -2.001  &    85.187  &   692.234 &o-    &$\rm 173.43\pm 7.99$		&$\rm 66.98\pm 0.72$		&$\rm 287.13\pm 2.31$ \\

   &220.730   &         2   &    -2.234  &    60.363  &   358.988 &p-   &$\rm 136.57\pm 1.93$		&$\rm 46.87\pm 1.40$		&$\rm 280.88\pm 3.92$  \\

   &220.743   &         1   &    -2.194  &    45.467  &   366.614 &p-    &$\rm 143.27\pm 31.70$		&$\rm 54.19\pm 0.70$		&$\rm 301.46\pm 4.56$ \\

  & 220.747   &         0   &    -2.180  &    40.501  &   369.277 &p-   &$\rm 155.20\pm 31.70$		&$\rm 58.70\pm 0.70$		&$\rm 326.57\pm 4.56$  \\
\hline

\multirow{5}*{HNCO}

&218.981   &   & -2.677  &    62.949  &    70.596  &  &$\rm35.32\pm1.11$	&$\rm17.38\pm0.19$	&$\rm88.86\pm3.25$
   &\multirow{6}*{\it 1, 2, 4, 5}\\
   &219.656  & &   -3.232   &    293.596  &59.063  &  &$\rm3.41\pm0.04$	&$\rm3.86\pm0.03$	     &$\rm8.49\pm0.49$\\
   %   &219.734  & &    -2.887  &   151.337  &    66.112  \\
   &219.737  & &    -2.887  &   151.337  &    66.110  &  &$\rm25.47\pm0.12^{~*}$	&$\rm14.07\pm0.09^{~*}$     &$\rm46.96\pm0.24^{~*}$\\
   &219.798  & &    -2.602  &    32.994  &    72.127  &  &$\rm65.23\pm0.60$ 	&$\rm36.89\pm0.18$     &$\rm121.44\pm6.06$\\

\hline
 \hline
%    \multicolumn{7}{l}{* None detection are not given}\\
\multicolumn{11}{l}{{\bf Notes:} {\it 1.} Integrated intensities are  obtained by Gildas ``Gaussian"  {fit}s.}\\
\multicolumn{11}{l}{~~~~~~~~~~~~{\it 2.}  Laboratory parameters are  from CDMS. }\\
\multicolumn{11}{l}{~~~~~~~~~~~~{\it 3.}  Q(300K, {\scriptsize{$\rm CH_3CN$}})=14683.6324 } \\
\multicolumn{11}{l}{~~~~~~~~~~~~{\it 4.}  Q(300K, {\scriptsize{$\rm HNCO$}})=7785.741 } \\
\multicolumn{11}{l}{~~~~~~~~~~~~{\it 5.}  ``*" { denotes the line with blending multiplets due to our spectral resolution, and the values are giving by simply taking into account their relative weights}. } \\
\end{tabu}

\end{center}
\end{table}

\end{landscape}

%%%%%%%%%%%%%%%%%%%%%%%%column density
%%%%%%%%%%%%%%%%%%%%%%%%column density
%%%%%%%%%%%%%%%%%%%%%%%%column density

\newpage
\renewcommand{\baselinestretch}{1.0} 
\begin{landscape}
\begin{table}
\caption{{Beam averaged} column densities and abundances for O-bearing molecules  from different substructures in NGC\,7538S and IRS1  denoted in Figure~\ref{into}.}
\label{col-Obearing} 
\small
\begin{center}
\begin{tabu}{c||p{2cm} p{2cm} p{2cm} p{2cm} p{2cm} || p{2cm}| p{2.2cm}}\hline\hline

Species    &$\rm MM1$  &$\rm MM2$  &$\rm MM3$	&$\rm JetN$	&$\rm JetS$	&$\rm IRS1-mmS$	&$\rm IRS1-peak$\\
      &$\rm ( 172\pm 23~K)$   &$\rm ( 137\pm 14~K)$   &$\rm ( 50 ~K)$   &$\rm ( 150~K)$   &$\rm ( 150~K)$   &$\rm ( 162\pm 14~K)$   &$\rm ( 214\pm 66~K)$\\

\hline
\multicolumn{8}{c}{I. Column densities (in the form of $\rm x\pm y (z) =(x\pm y) \times 10^z  cm^{-2}$)}    \\
\hline
$\rm C^{18}O$   &$\rm 2.71_ {\pm0.31}(17)$   &$\rm 4.43_ {\pm0.32}(16)$   &$\rm 4.80_ {\pm0.13}(15)$   &$\rm 4.16_ {\pm0.02}(16)$   &$\rm 5.46_ {\pm0.05}(16)$   &$\rm --$   &$\rm 7.23_ {\pm5.17}(17)$\\
%$\rm ^{13}CO$   &$\rm 3.22_ {\pm0.26}(17)$   &$\rm 8.65_ {\pm0.63}(16)$   &$\rm \ge 6.68(15)$   &$\rm 7.19_ {\pm0.30}(16)$   &$\rm 1.09_ {\pm0.08}(17)$   &$\rm --$   &$\rm 3.91_ {\pm0.58}(18)$\\
$\rm ^{13}CO^{~\dagger}$  &$\rm 4.35_ {\pm0.35}(18)$   &$\rm 7.79_ {\pm0.56}(17)$   &$\rm \ge 6.01(16)$   &$\rm 5.76_ {\pm0.24}(17)$   &$\rm 8.74_ {\pm0.61}(17)$   &$--$  &$\rm 6.71_ {\pm4.81}(18)$\\
$\rm H_2CO$   &$\rm 6.76_ {\pm0.69}(15)$   &$\rm 6.34_ {\pm0.58}(15)$   &$\rm \le 6.82(13)$   &$\rm 3.34_ {\pm0.08}(15)$   &$\rm 1.61_ {\pm0.08}(15)$   &$\rm 3.31_ {\pm0.13}(16)$  &$\rm 9.07_ {\pm7.53}(16)$\\
$\rm H_2^{13}CO$   &$\rm 2.50_ {\pm0.39}(14)$   &$\rm 5.06_ {\pm0.57}(14)$   &$\rm \le 5.56(12)$   &$\rm \le 1.22(13)$   &$\rm 1.79_ {\pm0.00}(14)$   &$\rm 1.81_ {\pm0.17}(15)$   &$\rm 5.30_ {\pm5.34}(14)$\\
$\rm CH_2CO$   &$\rm 4.66_ {\pm0.37}(14)$   &$\rm 1.43_ {\pm0.12}(15)$   &$\rm \le 3.67(13)$   &$\rm 1.25_ {\pm0.04}(14)$   &$\rm \le 4.28(13)$   &$\rm 8.95_ {\pm0.68}(15)$   &$\rm \le 5.96(14)$\\
$\rm CH_3OH$   &$\rm 4.61_ {\pm0.88}(16)$   &$\rm 4.33_ {\pm0.58}(16)$   &$\rm 2.25_ {\pm0.02}(15)$   &$\rm 3.72_ {\pm0.35}(15)$   &$\rm 6.30_ {\pm0.17}(15)$   &$\rm 1.74_ {\pm0.18}(17)$   &$\rm 3.36_ {\pm3.98}(17)$\\
$\rm HCOOCH_3$   &$\rm 2.91_ {\pm0.39}(16)$   &$\rm 1.33_ {\pm0.13}(16)$   &$\rm \le 4.95(14)$   &$\rm \le 6.71(14)$   &$\rm 2.57_ {\pm0.45}(14)$   &$\rm 1.27_ {\pm0.09}(17)$   &$\rm \le 1.92(16)$\\
$\rm ?CH_3OCH_3^{~*}$   &$\rm 7.62_ {\pm0.57}(17)$   &$\rm 1.97_ {\pm0.19}(17)$   &$\rm \le 1.96(17)$   &$\rm \le 3.71(16)$   &$\rm \le 3.67(16)$   &$\rm 5.43_ {\pm0.19}(18)$   &$\rm --$\\

\hline
\multicolumn{8}{c}{II. Abundances (in the form of $\rm x\pm y (z) =(x\pm y) \times 10^z $)}    \\
\hline
$\rm C^{18}O$   &$\rm 1.60_ {\pm0.50}(-7)$    &$\rm 2.60_ {\pm0.58}(-8)$    &$\rm 9.14_ {\pm0.64}(-10)$   &$\rm 1.64(-7)$   &$\rm 1.64(-7)$   &$\rm--$   &$\rm 1.64(-7)$\\
%$\rm ^{13}CO$   &$\rm 1.91_ {\pm0.51}(-7)$    &$\rm 5.07_ {\pm1.14}(-8)$    &$\rm \ge 1.27(-9)$    &$\rm 2.84_ {\pm0.12}(-7)$    &$\rm 3.28_ {\pm0.23}(-7)$    &$\rm--$    &$\rm 1.52_ {\pm1.09}(-6)$\\
$\rm ^{13}CO^{~\dagger}$   &$\rm 1.33_ {\pm0.41}(-6)$    &$\rm 2.16_ {\pm0.48}(-7)$    &$\rm \ge7.58(-9)$    &$\rm 1.36_ {\pm0.01}(-6)$    &$\rm 1.36_ {\pm0.03}(-6)$  &$\rm--$    &$\rm 1.52_ {\pm1.09}(-6)$\\
 
$\rm H_2CO$   &$\rm 4.00_ {\pm1.18}(-9)$    &$\rm 3.72_ {\pm0.91}(-9)$    &$\rm \le 1.30(-11)$    &$\rm 1.32_ {\pm0.03}(-8)$    &$\rm 4.84_ {\pm0.24}(-9)$    &$\rm--$    &$\rm 2.06_ {\pm1.71}(-8)$\\
$\rm H_2^{13}CO$   &$\rm 1.48_ {\pm0.53}(-10)$    &$\rm 2.97_ {\pm0.80}(-10)$    &$\rm \le 1.06(-12)$    &$\rm \le 4.81(-11)$    &$\rm 5.37_ {\pm0.01}(-10)$    &$\rm--$    &$\rm 1.20_ {\pm1.21}(-10)$\\
$\rm CH_2CO$   &$\rm 2.76_ {\pm0.74}(-10)$    &$\rm 8.36_ {\pm1.99}(-10)$    &$\rm \le 6.98(-12)$    &$\rm 4.94_ {\pm0.17}(-10)$    &$\rm \le 1.29(-10)$    &$\rm--$    &$\rm \le 1.35(-10)$\\
$\rm CH_3OH$   &$\rm 2.73_ {\pm1.09}(-8)$    &$\rm 2.54_ {\pm0.75}(-8)$    &$\rm 4.29_ {\pm0.22}(-10)$    &$\rm 1.47_ {\pm0.14}(-8)$    &$\rm 1.89_ {\pm0.05}(-8)$    &$\rm--$    &$\rm 7.63_ {\pm9.03}(-8)$\\
$\rm HCOOCH_3$   &$\rm 1.72_ {\pm0.58}(-8)$    &$\rm 7.77_ {\pm1.96}(-9)$    &$\rm \le 9.43(-11)$    &$\rm \le 2.65(-9)$    &$\rm 7.71_ {\pm1.35}(-10)$    &$\rm--$    &$\rm \le 4.35(-9)$\\
$\rm ?CH_3OCH_3^{~*}$   &$\rm 4.51_ {\pm1.19}(-7)$    &$\rm 1.15_ {\pm0.29}(-7)$    &$\rm \le 3.74(-8)$    &$\rm \le 1.46(-7)$    &$\rm \le 1.10(-7)$    &$\rm--$ &$\rm--$\\

\hline\hline
\end{tabu}
  \begin{tablenotes}
{\bf Note.} \\
1). ``*" { indicates the tentative detected species because  only one unblended line has been detected which has not been observed in our source previously}.\\
2).  {Abundances with respect to $\rm H_2$ were computed using $\rm N_{H_2}$ derived from dust continuum for MM1--MM3, and from $\rm C^{18}O$  for the other substructures (assuming $x({\rm C^{18}O}) \approx 1.64\times10^{-7}$; see Table \ref{fig:conti})} \\
3). All species  are assumed to have the same temperature as $\rm CH_3CN$  in MM1, MM2, and IRS1-mmS.\\
4). Values of all species in MM3, the outflow regions (JetN and JetS), and IRS1-peak  are derived under assumed temperatures (Section~\ref{tem}).\\
5).  Molecular column densities are  obtained from {the strongest transition of each isotopologue} by assuming LTE with the  excitation temperatures {listed} in the table head.\\
6). Uncertainties for the values derived from one transition   (written {in} subscript) are determined from $\rm T_{rot}$, partition function $\rm Q(T_{rot})$, and  fit to  $\rm \int T_B(\upsilon)d\upsilon$ (or $\rm \int \tau_(\upsilon)d\upsilon$ in IRS1-peak). \\
7). For species which are not detected, an upper limit derived from  $\rm 3\sigma$ rms is given. \\
8).  ``$--$" {indicates that the line area could not be computed because it has obvious  absorption in IRS1-mmS, or obvious emission in IRS1-peak, or non-detection of both main and rare isotopologues}.\\
9). {A lower limit  is given to the  species which has  partial absorption in S because of the missing short spacing. }\\
10). $``\dagger"$  mark the values which are { obtained from likely optically thick lines  after we we did the optical depth correction in each substructures except for IRS1-peak.  }
  \end{tablenotes}
\end{center}
\end{table}
\end{landscape}
%\vspace{1em}

%%%%%%%%%%%%%%%%%%%%!!!!!!!!!!!!!!!!!!!!!!!!!!!!!!!!!!
%%%%%%%%%%%%%%%%%%%%!!!!!!!!!!!!!!!!!!!!!!!!!!!!!!!!!!NNNNNNNNNNNNNNNNNNNNNN
%%%%%%%%%%%%%%%%%%%%!!!!!!!!!!!!!!!!!!!!!!!!!!!!!!!!!!
\newpage
\renewcommand{\baselinestretch}{1.0} 
\begin{landscape}
\begin{table}
\caption{{Beam averaged} column densities and abundances for N-bearing molecules  from different substructures in NGC\,7538S and IRS1  denoted in Figure~\ref{into}.}
\label{col-Nbearing} 
\small
\begin{center}
\begin{tabu}{c||p{2cm} p{2cm} p{2cm} p{2cm} p{2cm} || p{2cm}| p{2.2cm}}\hline\hline

Species    &$\rm MM1$  &$\rm MM2$  &$\rm MM3$	&$\rm JetN$	&$\rm JetS$	&$\rm IRS1-mmS$	&$\rm IRS1-peak$\\
      &$\rm ( 172\pm 23~K)$   &$\rm ( 137\pm 14~K)$   &$\rm ( 50 ~K)$   &$\rm ( 150~K)$   &$\rm ( 150~K)$   &$\rm ( 162\pm 14~K)$   &$\rm ( 214\pm 66~K)$\\

\hline
\multicolumn{8}{c}{I. Column densities (in the form of $\rm x\pm y (z) =(x\pm y) \times 10^z  cm^{-2}$)}    \\
\hline
$\rm DCN$    &$\rm 2.15_ {\pm0.26}(14)$    &$\rm 1.43_ {\pm0.03}(13)$    &$\rm \le 1.19(12)$    &$\rm \le 2.38(12)$    &$\rm 1.65_ {\pm0.24}(13)$    &$\rm 3.04_ {\pm0.05}(14)$    &$\rm 7.82_ {\pm6.14}(14)$\\
$\rm HNCO$    &$\rm \bf 2.64\pm1.35(15)$    &$\rm \bf 1.89\pm1.01(15)$    &$\rm \le 2.44(13)$    &$\rm \le 4.13(13)$    &$\rm \le 2.27(13)$    &$\rm \bf 5.73\pm1.45(15)$    &$\rm 1.04_ {\pm0.96}(16)$\\
$\rm HC_3N$    &$\rm 6.33_ {\pm0.12}(14)$    &$\rm 2.11_ {\pm0.03}(14)$    &$\rm \le 6.60(12)$    &$\rm \le 2.39(12)$    &$\rm 3.93_ {\pm0.20}(13)$    &$\rm 2.43_ {\pm0.03}(14)$   &$\rm 3.95_ {\pm2.03}(14)$\\
$\rm HC_3N(v_7=1)$    &$\rm 1.38_ {\pm0.42}(15)$    &$\rm 3.08_ {\pm1.08}(14)$    &$\rm \le 4.04(15)$    &$\rm \le 2.03(13)$    &$\rm \le 4.85(13)$    &$\rm 2.88_ {\pm0.96}(14)$    &$\rm 6.89_ {\pm0.69}(14)$\\
$\rm CH_3CN^{~\dagger}$    &$\rm \bf 4.58\pm1.97(16)$    &$\rm \bf 2.64\pm0.96(15)$    &$\rm \le1.18(15)$    &$\rm--$    &$\rm --$    &$\rm \bf 3.87\pm1.13(16)$    &$\rm 4.61_ {\pm5.14}(15)$\\
$\rm CH_3^{13}CN$    &$\rm 5.58_ {\pm0.18}(14)$    &$\rm 2.67_ {\pm0.09}(13)$    &$\rm \le 9.85(12)$    &$\rm \le 1.04(13)$    &$\rm \le 1.00(13)$    &$\rm 4.58_ {\pm0.03}(14)$    &$\rm  \le 1.87(14)$\\
$\rm CH_3CH_2CN$    &$\rm 1.70_ {\pm0.13}(15)$    &$\rm \le 3.77(13)$    &$\rm \le 8.43(13)$    &$\rm \le 4.61(13)$    &$\rm \le 9.41(13)$    &$\rm 7.77_ {\pm0.17}(14)$    &$\rm \le 1.76(15)$\\
$\rm NH_2CHO$    &$\rm 2.91_ {\pm0.13}(14)$    &$\rm 5.46_ {\pm0.50}(14)$    &$\rm \le 7.11(12)$    &$\rm 2.96_ {\pm1.25}(13)$    &$\rm \le 2.11(13)$    &$\rm 2.89_ {\pm0.18}(15)$    &$\rm 1.80_ {\pm1.68}(15)$\\

\hline
\multicolumn{8}{c}{II. Abundances (in the form of $\rm x\pm y (z) =(x\pm y) \times 10^z $)}    \\
\hline
$\rm DCN$   &$\rm 1.27_ {\pm0.40}(-10)$    &$\rm 8.41_ {\pm1.40}(-12)$    &$\rm \le 2.27(-13)$    &$\rm \le 9.38(-12)$    &$\rm 4.95_ {\pm0.71}(-11)$    &$\rm--$    &$\rm 1.77_ {\pm1.39}(-10)$\\
$\rm HNCO$   &$\rm \bf 1.57\pm1.21(-9)$    &$\rm \bf 1.11\pm0.83(-9)$    &$\rm \le 4.64(-12)$    &$\rm \le 1.63(-10)$    &$\rm \le 6.80(-11)$    &$\rm--$    &$\rm 2.35_ {\pm2.17}(-9)$\\
$\rm HC_3N$    &$\rm 3.74_ {\pm0.74}(-10)$    &$\rm 1.24_ {\pm0.20}(-10)$    &$\rm \le 1.26(-12)$    &$\rm \le 9.43(-12)$    &$\rm 1.18_ {\pm0.06}(-10)$    &$\rm--$    &$\rm 8.96_ {\pm4.60}(-11)$\\
$\rm HC_3N(v_7=1)$   &$\rm 8.18_ {\pm4.39}(-10)$    &$\rm 1.81_ {\pm0.98}(-10)$    &$\rm \le 7.69(-10)$    &$\rm \le 8.01(-11)$    &$\rm \le 1.46(-10)$    &$\rm--$    &$\rm 1.56_ {\pm0.16}(-10)$\\
$\rm CH_3CN^{~\dagger}$   &$\rm \bf 2.71\pm1.85(-8)$    &$\rm \bf 1.55\pm0.86(-9)$   &$ \le2.24(-10)$   &$\rm--$   &$\rm --$   &$\rm--$  &$\rm 1.05_ {\pm1.17}(-9)$\\
$\rm CH_3^{13}CN$   &$\rm 3.30_ {\pm0.71}(-10)$    &$\rm 1.57_ {\pm0.28}(-11)$    &$\rm \le 1.88(-12)$    &$\rm \le 4.09(-11)$    &$\rm \le 3.02(-11)$    &$\rm--$    &$\rm \le 4.24(-11)$\\
$\rm CH_3CH_2CN$  &$\rm 1.01_ {\pm0.27}(-9)$    &$\rm \le 2.21(-11)$    &$\rm \le 1.61(-11)$    &$\rm \le 1.82(-10)$    &$\rm \le 2.83(-10)$    &$\rm--$    &$\rm \le 4.00(-10)$\\
$\rm NH_2CHO$   &$\rm 1.72_ {\pm0.39}(-10)$    &$\rm 3.20_ {\pm0.79}(-10)$    &$\rm \le 1.35(-12)$    &$\rm 1.17_ {\pm0.49}(-10)$    &$\rm \le 6.32(-11)$    &$\rm--$    &$\rm 4.08_ {\pm3.81}(-10)$\\

\hline\hline
\end{tabu}
  \begin{tablenotes}
{\bf Note.} \\
1). ``*" { indicates the tentative detected species because  only one unblended line has been detected which has not been observed in our source previously}.\\
2).  {Abundances with respect to $\rm H_2$ were computed using $\rm N_{H_2}$ derived from dust continuum for MM1--MM3, and from $\rm C^{18}O$  for the other substructures (assuming $x({\rm C^{18}O}) \approx 1.64\times10^{-7}$; see Table \ref{fig:conti})} \\
3). Values and uncertainties in bold face are obtained directly  from   the rotation diagram fits in Figure~\ref{rotation}, and all the rest species except for HNCO are assumed to have the same temperature as $\rm CH_3CN$   in MM1, MM2, and IRS1-mmS.\\
4). Values of all species in MM3, the outflow regions (JetN and JetS), and IRS1-peak  are derived under assumed temperatures (Section~\ref{tem}).\\
5).  Molecular column densities are  obtained from {the strongest transition of each isotopologue} by assuming LTE with the  excitation temperatures {listed} in the table head.\\
6). Uncertainties for the values derived from one transition   (written {in} subscript) are determined from $\rm T_{rot}$, partition function $\rm Q(T_{rot})$, and  fit to  $\rm \int T_B(\upsilon)d\upsilon$ (or $\rm \int \tau_(\upsilon)d\upsilon$ in IRS1-peak). \\
7). For species which are not detected, an upper limit derived from  $\rm 3\sigma$ rms is given. \\
8).  ``$--$" {indicates that the line area could not be computed because it has obvious  absorption in IRS1-mmS, or obvious emission in IRS1-peak, or non-detection of both main and rare isotopologues}.\\
9). {A lower limit  is given to the  species which has  partial absorption in S because of the missing short spacing. }\\
10). $``\dagger"$  mark the values which are { obtained from likely optically thick lines  after we we did the optical depth correction in each substructures except for IRS1-peak.  }

  \end{tablenotes}
\end{center}
\end{table}

\end{landscape}
%\vspace{1em}

%%%%%%%%%%%%%%%%%%%%!!!!!!!!!!!!!!!!!!!!!!!!!!!!!!!!!!
%%%%%%%%%%%%%%%%%%%%!!!!!!!!!!!!!!!!!!!!!!!!!!!!!!!!!!SSSSSSSSSSSSSSSSSSSSSSSS
%%%%%%%%%%%%%%%%%%%%!!!!!!!!!!!!!!!!!!!!!!!!!!!!!!!!!!
\newpage
\renewcommand{\baselinestretch}{1.0} 
\begin{landscape}
\begin{table}
\caption{{Beam averaged}  column densities and abundances for S-bearing molecules  from different substructures in NGC\,7538S and IRS1  denoted in Figure~\ref{into}.}
\label{col-Sbearing} 
\small
\begin{center}
\begin{tabu}{c||p{2cm} p{2cm} p{2cm} p{2cm} p{2cm} || p{2cm}| p{2.2cm}}\hline\hline

Species    &$\rm MM1$  &$\rm MM2$  &$\rm MM3$	&$\rm JetN$	&$\rm JetS$	&$\rm IRS1-mmS$	&$\rm IRS1-peak$\\
      &$\rm ( 172\pm 23~K)$   &$\rm ( 137\pm 14~K)$   &$\rm ( 50 ~K)$   &$\rm ( 150~K)$   &$\rm ( 150~K)$   &$\rm ( 162\pm 14~K)$   &$\rm ( 214\pm 66~K)$\\

\hline
\multicolumn{8}{c}{I. Column densities (in the form of $\rm x\pm y (z) =(x\pm y) \times 10^z  cm^{-2}$)}    \\
\hline
%$\rm OCS$    &$\rm 1.29_ {\pm0.06}(16)$    &$\rm 7.98_ {\pm0.18}(15)$    &$\rm 1.78_ {\pm0.05}(15)$    &$\rm 2.89_ {\pm0.05}(14)$    &$\rm 4.38_ {\pm0.50}(14)$    &$\rm 2.14_ {\pm0.04}(16)$    &$\rm \le 1.90(15)$\\
$\rm OCS^{~\dagger}$  &$\rm 5.86_ {\pm0.26}(16)$   &$\rm 4.41_ {\pm0.10}(16)$   &$\rm 6.42_ {\pm0.20}(16)$   &$\rm \le1.10(16)$   &$\rm \le 6.57(15)$   &$\rm 8.71_ {\pm0.14}(16)$     & $\rm \le 2.87(15)$\\
$\rm O^{13}CS$    &$\rm 7.63_ {\pm0.15}(14)$    &$\rm 6.05_ {\pm1.92}(14)$    &$\rm 9.27_ {\pm0.13}(14)$    &$\rm \le 6.88(13)$    &$\rm \le 9.11(13)$    &$\rm 1.01_ {\pm0.50}(15)$    &$\rm \le 4.15(15)$\\
$\rm SO_2$    &$\rm 1.93_ {\pm0.20}(17)$    &$\rm \le 1.57(15)$    &$\rm \le 2.65(16)$    &$\rm \le 1.13(15)$    &$\rm \le 1.89(15)$    &$\rm 6.12_ {\pm1.56}(16)$    &$\rm 9.71_ {\pm4.18}(16)$\\
$\rm ^{34}SO_2$    &$\rm 1.38_ {\pm0.20}(16)$    &$\rm \le 7.17(13)$    &$\rm \le 1.95(13)$    &$\rm \le 5.41(13)$    &$\rm \le 1.50(14)$    &$\rm 1.62_ {\pm0.13}(15)$    &$\rm 1.17_ {\pm1.20}(15)$\\

%$\rm SO$    &$\rm 7.09_ {\pm0.70}(15)$    &$\rm 3.26_ {\pm0.16}(15)$    &$\rm \le 1.89(13)$    &$\rm 6.52_ {\pm0.14}(14)$    &$\rm 3.75_ {\pm0.11}(15)$    &$\rm 6.91_ {\pm0.47}(15)$    &$\rm 8.98_ {\pm1.15}(15)$\\
$\rm SO^{~\dagger}$     &$\rm 3.16_ {\pm0.31}(17)$   &$\rm 2.45_ {\pm0.12}(16)$   &$--$   &$\rm \le1.56(16)$   &$\rm \le2.44(16)$   &$\rm 6.91_ {\pm0.47}(16)$  &$\rm 1.52_ {\pm1.10}(16)$\\
$\rm ^{33}SO$    &$\rm 2.81_ {\pm0.30}(15)$    &$\rm 2.59_ {\pm0.46}(14)$    &$\rm \le 1.52(13)$    &$\rm \le 2.45(13)$    &$\rm \le 5.96(13)$    &$\rm 5.22_ {\pm1.43}(14)$    &$\rm \le 9.85(14)$\\

\hline
\multicolumn{8}{c}{II. Abundances (in the form of $\rm x\pm y (z) =(x\pm y) \times 10^z $)}    \\
\hline
%$\rm OCS$   &$\rm 7.22_ {\pm1.65}(-9)$   &$\rm 4.44_ {\pm0.74}(-9)$   &$\rm 3.22_ {\pm0.24}(-10)$   &$\rm 1.14_ {\pm0.02}(-9)$   &$\rm 1.31_ {\pm0.15}(-9)$   &$\rm--$   &$\rm \le 7.40(-10)$\\
$\rm OCS^{~\dagger}$   &$\rm 3.47_ {\pm0.79}(-8)$    &$\rm 2.58_ {\pm0.43}(-8)$    &$\rm 1.22_ {\pm0.09}(-8)$    &$\rm \le4.33(-8)$    &$\rm \le 1.97(-8)$  &$\rm--$    &$\rm \le 6.52(-10)$\\
$\rm O^{13}CS$   &$\rm 4.52_ {\pm0.90}(-10)$    &$\rm 3.55_ {\pm1.79}(-10)$    &$\rm 1.77_ {\pm0.10}(-10)$    &$\rm \le 2.71(-10)$    &$\rm \le 2.74(-10)$    &$\rm--$    &$\rm \le 9.42(-10)$\\
$\rm SO_2$    &$\rm 1.14_ {\pm0.34}(-7)$    &$\rm \le 9.21(-10)$    &$\rm \le 5.05(-9)$    &$\rm \le 4.47(-9)$    &$\rm \le 5.67(-9)$    &$\rm--$    &$\rm 2.20_ {\pm0.95}(-8)$\\
$\rm ^{34}SO_2$   &$\rm 8.19_ {\pm2.80}(-9)$    &$\rm \le 4.21(-11)$    &$\rm \le 3.71(-12)$    &$\rm \le 2.13(-10)$    &$\rm \le 4.51(-10)$    &$\rm--$    &$\rm 2.67_ {\pm2.71}(-10)$\\

%$\rm SO$    &$\rm 4.20_ {\pm1.22}(-9)$    &$\rm 1.91_ {\pm0.37}(-9)$    &$\rm \le 3.60(-12)$    &$\rm 2.57_ {\pm0.06}(-9)$    &$\rm 1.13_ {\pm0.03}(-8)$    &$\rm--$    &$\rm 3.45_ {\pm2.49}(-9)$\\
$\rm SO^{~\dagger}$    &$\rm 2.11_ {\pm0.64}(-7)$    &$\rm 1.93_ {\pm0.66}(-8)$     &$--$   &$\rm \le 1.23(-8)$    &$\rm \le 2.27(-8)$   &$\rm--$    &$\rm 3.45_ {\pm2.49}(-9)$\\
$\rm ^{33}SO$   &$\rm 1.66_ {\pm0.50}(-9)$    &$\rm 1.52_ {\pm0.52}(-10)$    &$\rm \le 2.90(-12)$    &$\rm \le 9.66(-11)$    &$\rm \le 1.79(-10)$    &$\rm--$    &$\rm \le 2.23(-10)$\\

\hline\hline
\end{tabu}
  \begin{tablenotes}
{\bf Note.} \\
1). ``*" { indicates the tentative detected species because  only one unblended line has been detected which has not been observed in our source previously}.\\
2).  {Abundances with respect to $\rm H_2$ were computed using $\rm N_{H_2}$ derived from dust continuum for MM1--MM3, and from $\rm C^{18}O$  for the other substructures (assuming $x({\rm C^{18}O}) \approx 1.64\times10^{-7}$; see Table \ref{fig:conti})} \\
3). All species  are assumed to have the same temperature as $\rm CH_3CN$  in MM1, MM2, and IRS1-mmS.\\
4). Values of all species in MM3, the outflow regions (JetN and JetS), and IRS1-peak  are derived under assumed temperatures (Section~\ref{tem}).\\
5).  Molecular column densities are  obtained from {the strongest transition of each isotopologue} by assuming LTE with the  excitation temperatures {listed} in the table head.\\
6). Uncertainties for the values derived from one transition   (written {in} subscript) are determined from $\rm T_{rot}$, partition function $\rm Q(T_{rot})$, and  fit to  $\rm \int T_B(\upsilon)d\upsilon$ (or $\rm \int \tau_(\upsilon)d\upsilon$ in IRS1-peak). \\
7). For species which are not detected, an upper limit derived from  $\rm 3\sigma$ rms is given. \\
8).  ``$--$" {indicates that the line area could not be computed because it has obvious  absorption in IRS1-mmS, or obvious emission in IRS1-peak, or non-detection of both main and rare isotopologues}.\\
9). {An lower limit  is given to the  species which has  partial absorption in S because of the missing short spacing. }\\
10). $``\dagger"$  mark the values which are { obtained from likely optically thick lines  after we we did the optical depth correction in each substructures except for IRS1-peak.  }
  \end{tablenotes}
\end{center}
\end{table}
\end{landscape}
%\vspace{1em}

\setcounter{figure}{0}
\renewcommand{\thefigure}{A\arabic{figure}}

\newpage

%\begin{longtable}{c}
%\begin{figure*} [htb]
%\small
%\begin{center}
\begin{figure*}[htb]
\begin{center}
\begin{tabular}{ll}
\includegraphics[width=8cm]{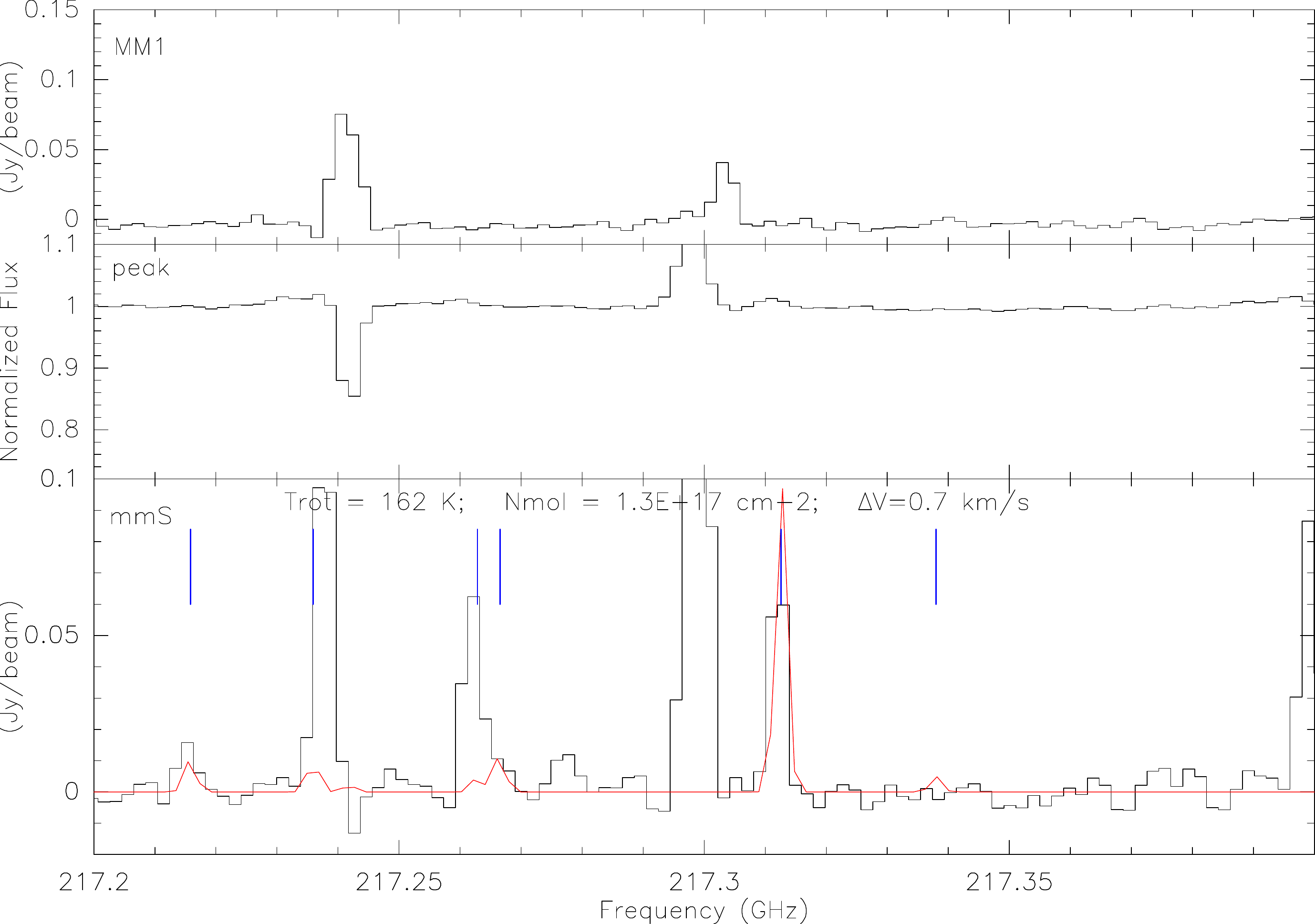}
&\includegraphics[width=8cm]{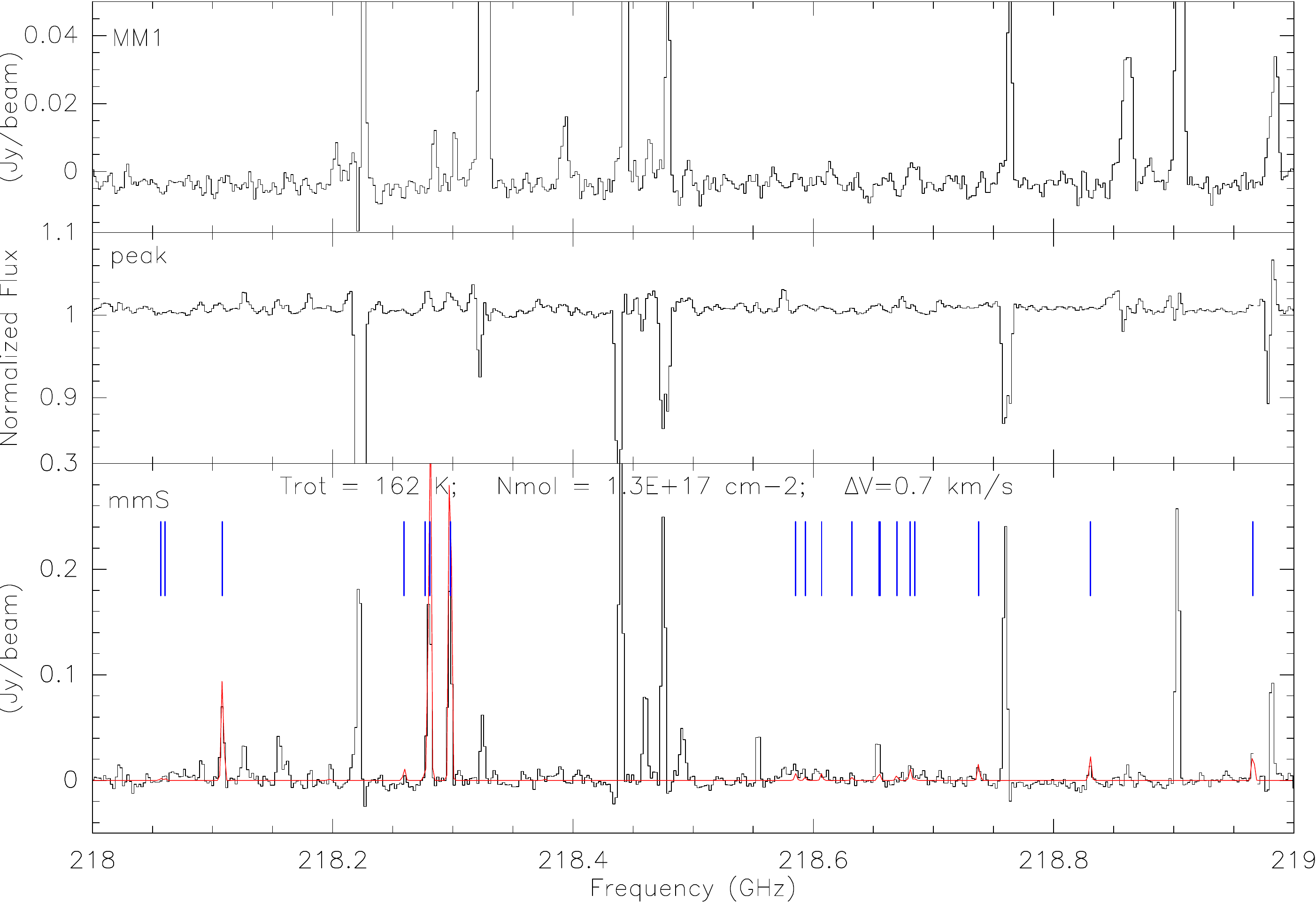}\\
\includegraphics[width=8cm]{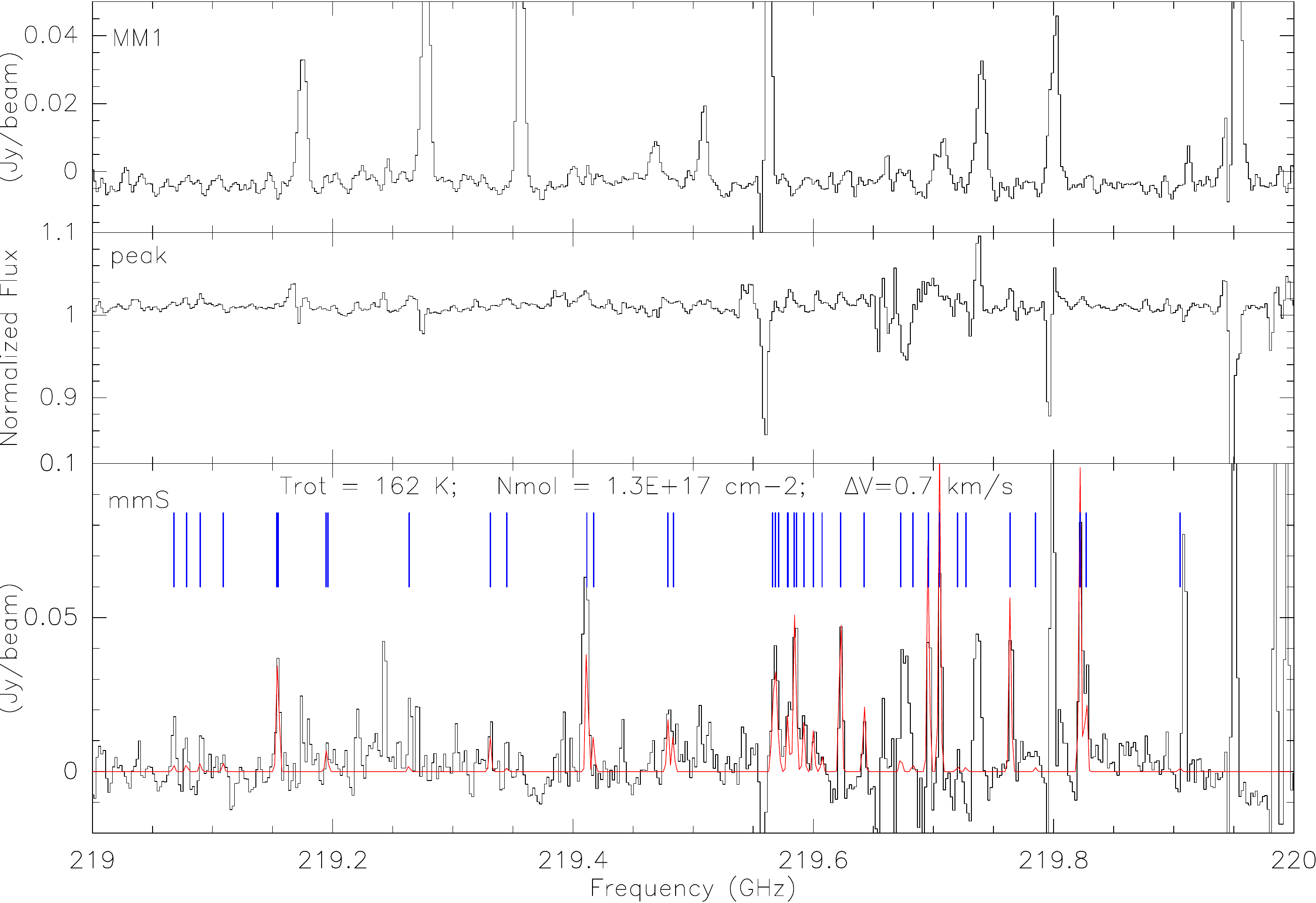}
&\includegraphics[width=8cm]{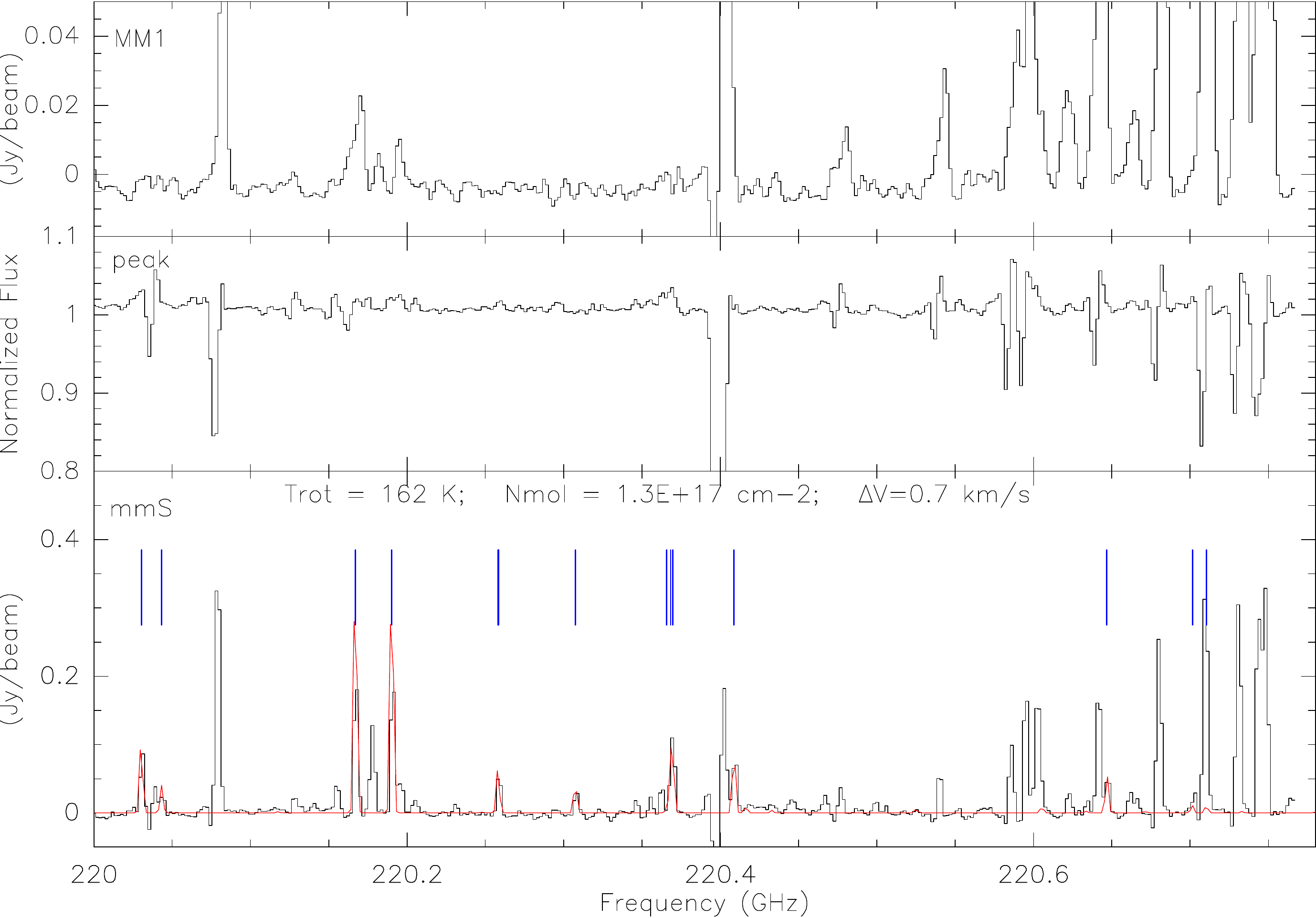}
\end{tabular}
\end{center}
\caption{Synthetic fits of $\rm HCOOCH_3$ in NGC\,7538 IRS1 (mmS and peak) and MM1. Black lines represent the observed spectra while red lines show the best-fit model based on optically thin assumption.}\label{hcooch3}
\end{figure*}

\newpage

%\begin{longtable}{c}
%\begin{figure*} [htb]
%\small
%\begin{center}

\begin{figure*}[htb]
\begin{center}
\includegraphics[angle=270,scale=0.75] {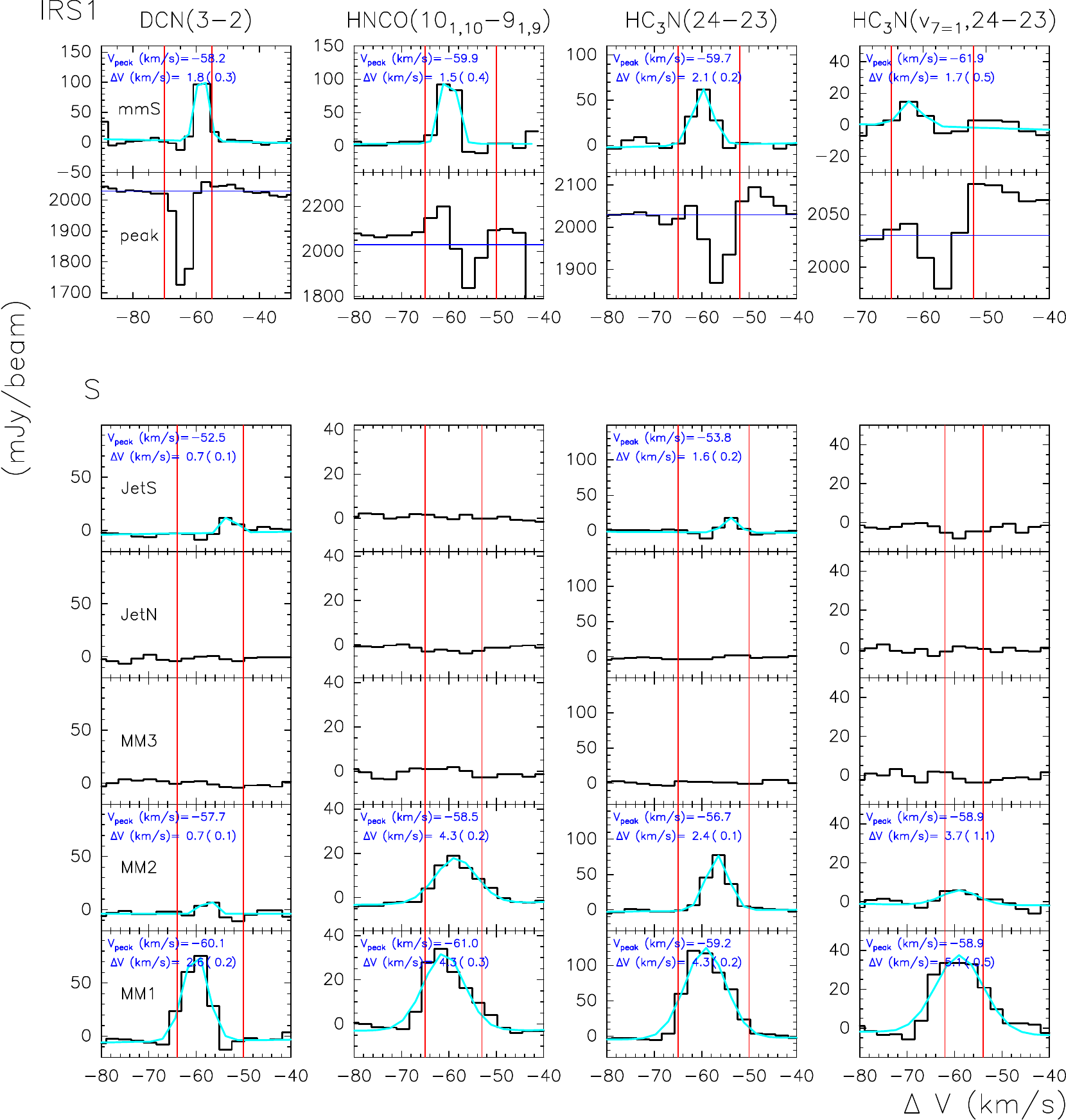}\\
~~~~~~~~~~~~~\includegraphics[angle=270,scale=0.75] {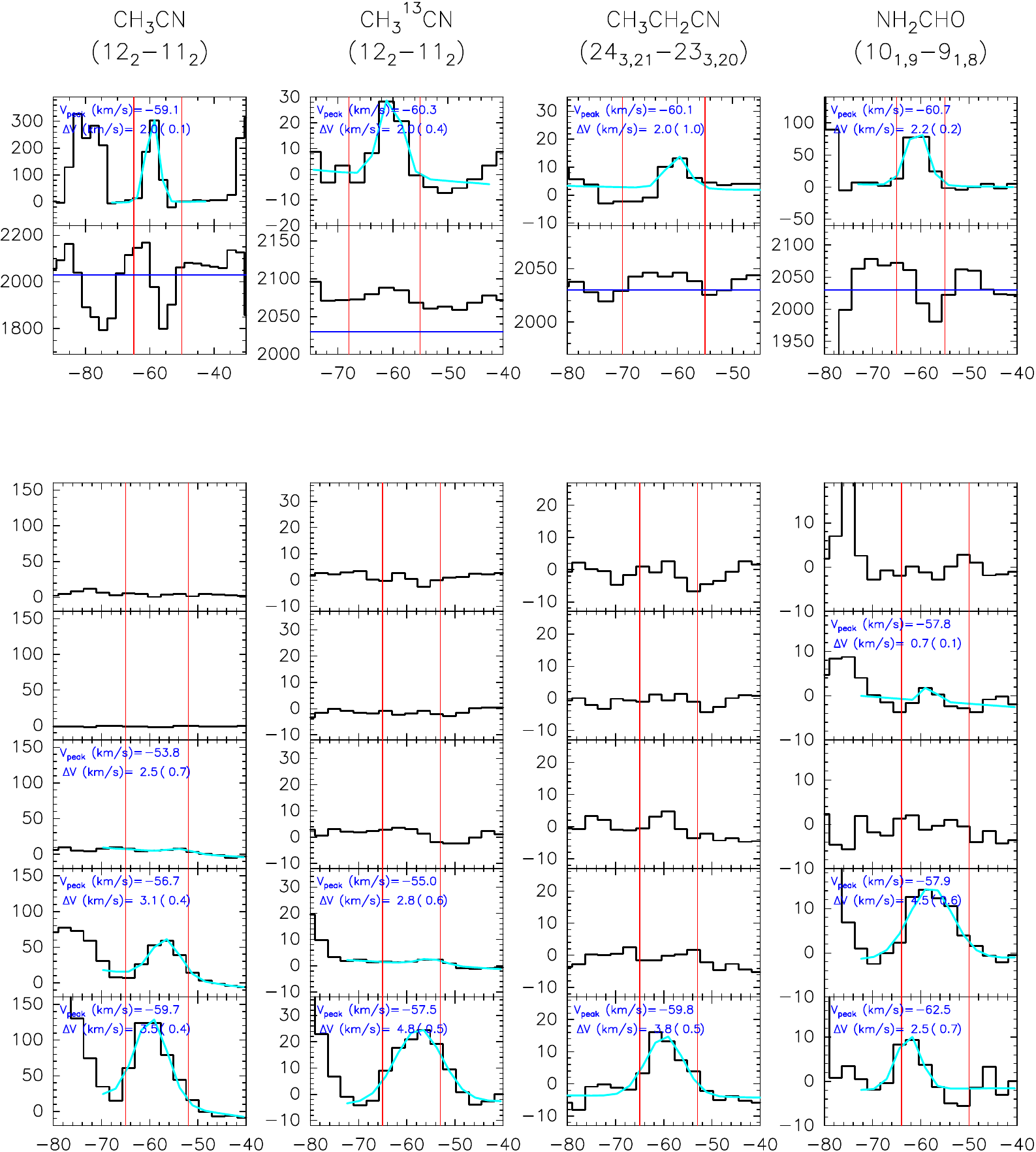}
%\end{figure}
\end{center}
\caption{Line profiles of identified species (in black) and single Gaussian {fits} (in blue) toward the  {continuum substructures identified in  this paper } (given in Table~\ref{source}). Continuum is not subtracted in IRS1. Two red vertical lines in each panel mark the velocity range we use to integrate the intensity and obtain the {spatial} distribution maps (Figure~\ref{into}). Lines marked with ``**" {has $\rm >4\sigma$ detection but could not be fitted with a single } Gaussian.  ``$\nabla\nabla$" denotes the {lines having $\rm >4\sigma$ detection but } showing obvious absorption   in IRS1-mmS or obvious emission in IRS1-peak.
}\label{velpro}
\end{figure*} 

\begin{figure*}[htb]
\ContinuedFloat
\begin{center}
\includegraphics[angle=270,scale=0.8] {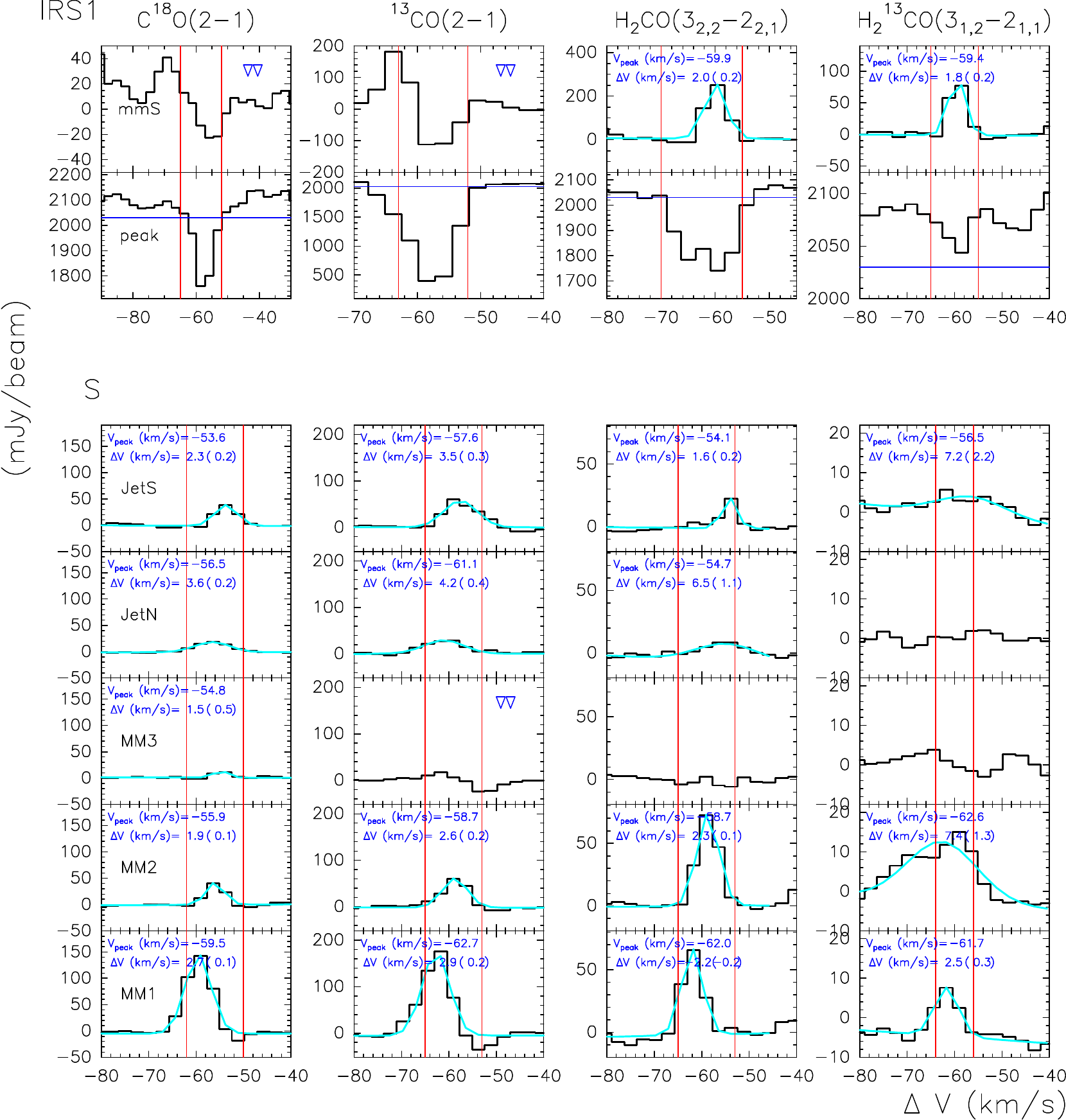}\\
~~~~~~~~~~~\includegraphics[angle=270,scale=0.8] {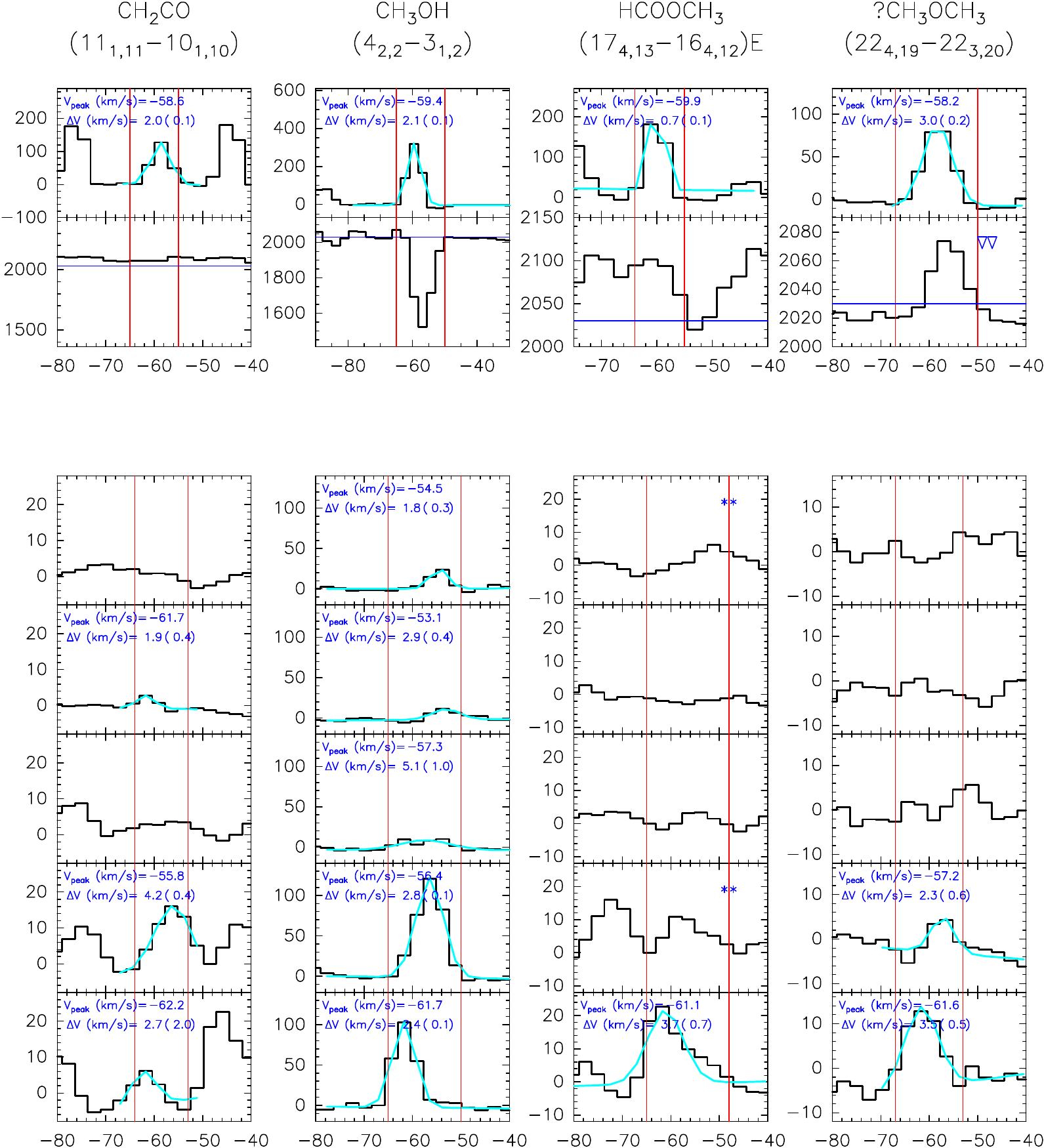}
%\end{figure}
\caption{(continued)}
\end{center}
\end{figure*} 

\begin{figure*}[htb]
\ContinuedFloat
\begin{center}
\includegraphics[angle=270,scale=0.8] {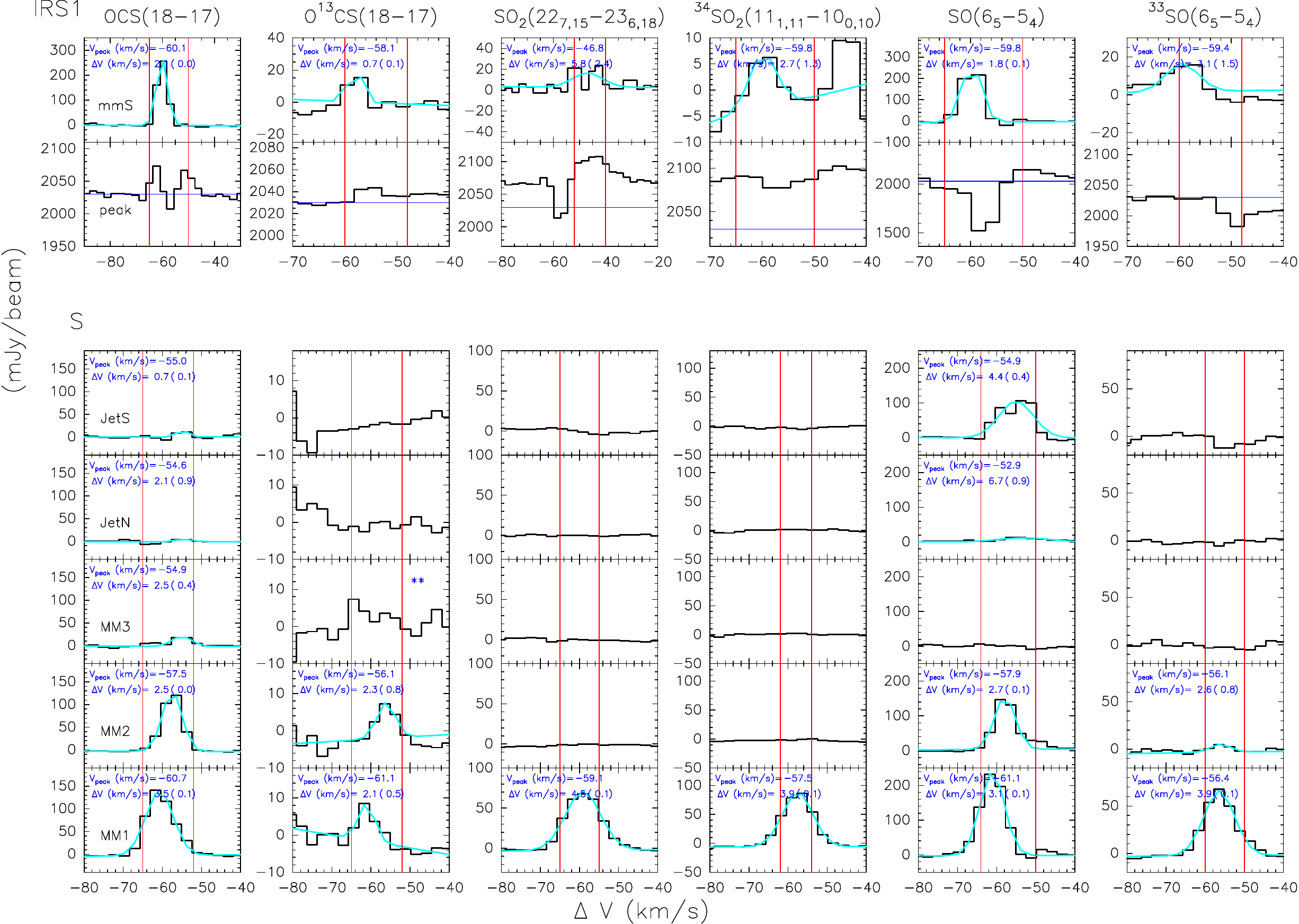}
%\end{figure}
\caption{(continued)}
\end{center}

\end{figure*}

\end{document}